\RequirePackage{fix-cm}
\documentclass[smallextended]{svjour3}  
\smartqed
 
\usepackage{packages}
\usepackage{macros}

\title{
A Rewriting Logic Approach to Specification, Proof-search, and
Meta-proofs in Sequent Systems}

\author{
  Carlos Olarte \and
  Elaine Pimentel \and
  Camilo Rocha
}

\institute{
 Carlos Olarte and Elaine Pimentel \at Universidade Federal do Rio Grande do Norte, Natal, Brazil
  \and
  Camilo Rocha \at 
  Pontificia Universidad Javeriana, Cali, Colombia
}

\date{Received: date / Accepted: date}

\begin{document}

\maketitle

\begin{abstract}
This paper develops an algorithmic-based approach for proving
inductive properties of propositional sequent systems such as
admissibility, invertibility, cut-elimination, and identity expansion.
Although undecidable in general, these structural properties are
crucial in proof theory because they can reduce the proof-search
effort and further be used as scaffolding for obtaining other
meta-results such as consistency.
The algorithms --which take advantage of the rewriting logic
meta-logical framework, and use rewrite- and narrowing-based
reasoning-- are explained in detail and illustrated with examples
throughout the paper.
They have been fully mechanized in the \toolname, thus offering both a
formal specification language and off-the-shelf mechanization of the
proof-search algorithms coming together with semi-decision
procedures for proving theorems and meta-theorems of the object
system.
As illustrated with case studies in the paper, the \toolname\,
achieves a great degree of automation when used on several
propositional sequent systems, including single conclusion and
multi-conclusion intuitionistic logic, classical logic, classical
linear logic and its dyadic system, intuitionistic linear logic, and
normal modal logics.

\end{abstract}

\tableofcontents

\section{Introduction}
\label{sec.intro}

Gentzen's sequent systems~\cite{gentzen35} have proved to be a flexible and robust logical
formalism for specifying proof systems. With the advent of new needs
in the form of techniques and technologies (e.g., to reason in
temporal, epistemic, and modal domains), the past decades have
witnessed a vivid scene where novel sequent-based calculi have been
proposed (see, e.g.,
\cite{DBLP:conf/tableaux/Lellmann15,DBLP:journals/corr/abs-1805-09437}. It
is fair to say that one of the most important aims of these new
systems is the ultimate goal of having both a formal specification
language and a mechanization of proof-search algorithms coming
together with decision procedures --whenever possible and feasible.

Cut-free sequent systems are key for this latter purpose. Intuitively,
the cut-rule expresses the mathematical use of lemmas in proofs: if
$A$ follows from $B$ and $C$ follows from $A$, then $C$ follows from
$B$. That is, one can cut the intermediate lemma $A$.  Cut-elimination
theorems then state that any judgment that possesses a proof making
use of the cut-rule, also possesses a proof that does not make use of
it. This is of practical convenience for proof-search purposes because
it is often much easier to find proofs when the need to mechanically
come up with auxiliary lemmas can be avoided.  Moreover, there are
important applications of cut-elimination theorems, such as the
reduction of consistency proofs and ``subformula properties'' to mere
structural syntactic checks.  However, structural properties of a
logical system such as cut-freeness are actually inductive
meta-properties and thus undecidable to check in general. In practice,
proof-theoretic approaches to establish them usually require
combinatorial search heuristics based on the structure of formulas of
each particular system. These heuristics, in turn, may depend on the
ability to prove other inductive properties. For
instance, 
Gentzen-style proofs
of cut-elimination heavily
depends on checking admissibility 
of other structural rules (e.g., weakening and contraction)  and showing that some of 
the sequent rules defining the system are invertible. 

The main goal of this article is to present the  $\toolname$\footnote{\url{https://carlosolarte.github.io/L-framework/}},
a tool for algebraically specifying and analyzing propositional
sequent systems. It provides both a specific scaffolding of constructs
usually found in a sequent system, and off-the-shelf algorithms for
proving inductive properties such as admissibility, invertibility,
cut-elimination, and identity expansion.  The proposed framework is
\textit{generic} in the sense that only mild restrictions are imposed
on the formulas of the sequent system, and \textit{modular} since
cut-elimination, admissibility, invertibility, and identity expansion
can be proved incrementally.

The algorithms and their implementation in the \toolname\, are based
on the expressive power of rewriting logic \cite{meseguer-rltcs-1992},
both as a logical and a reflective meta-logical
framework~\cite{marti-oliet-rlframework-2002,basin-reflmetlogframe-2004}. A
sequent system is specified as a rewriting logic theory $\scal$ and a
sequent as a formula $\varphi$ in the syntax of $\scal$. In this
setting, the process of building a proof of $\varphi$ in $\scal$ is
cast as a reachability goal of the form $\scal \vdash \varphi
\stackrel{*}{\rightarrow} \top$, where $\top$ denotes the empty
sequent. Therefore, as a logical framework, rewriting logic directly
serves the purpose of providing a proof-search algorithm at the
\textit{object} level of $\scal$. Since rewriting logic is reflective,
a universal rewrite theory $\ucal$ is further used to prove
meta-logical properties (e.g., cut-freeness) of the system
$\scal$. More precisely, such proofs are specified as reachability
goals of the form $\ucal \vdash
\pair{\overline{\scal}}{\overline{\Gamma}} \stackrel{*}{\rightarrow}
\pair{\overline{\scal}}{\overline{\top}}$, where $\Gamma$ is a set of
sequents representing the proof-obligations to be discharged, and
$\overline{\scal}$ and $\overline{\Gamma}$ are the meta-representation
of $\scal$ and $\Gamma$ in the syntax of the universal theory $\ucal$,
respectively. In this way, starting with the specification of the
propositional sequent $\scal$ in rewriting logic, the framework
presented in this paper enables both the proof of theorems and
inductive meta-theorems of $\scal$.

The framework comprises a \textit{generic} rewrite theory offering
sorts, operators, and standard procedures for proof-search at the
object- and meta-level. The sorts are used to represent different
elements of the syntactic structure of a  sequent
system such as formulas, sets of formulas, sequents, and collections
of sequents.  The inference rules of a sequent system are specified as
reversed rewrite rules, so that an inference step of the mechanized
system is carried out by matching a sequent to the conclusion of an
inference rule, and replacing the latter by the corresponding
substitution instance of the rule's premises. Conceptually, a proof at
the object level is found when all proof-obligations are conclusions
of inference rules without premises (i.e., axioms).

Once a sequent system $\scal$ is specified, proofs of
its cut-freeness, admissibility, invertibility, and identity expansion
properties can be searched for. The process of obtaining proofs for
each one of these properties follows a similar pattern. Namely, the
rules of $\scal$ are used by pair-wise comparison and narrowing to
generate the inductive hypothesis $\hcal$ and the proof obligations
$\Gamma$. Therefore, if for each pair $(\hcal, \Gamma)$ the
(meta-)reachability goal $\ucal \vdash \pair{\overline{\scal \cup
    \hcal}}{\overline{\Gamma}} \stackrel{*}{\rightarrow}
\pair{\overline{\scal \cup \hcal}}{\top}$ can be answered positively,
then the corresponding property of the sequent system $\scal$
holds. The universal theory $\ucal$ contains the rewrite- and
narrowing-based heuristics that operate over $\overline{\scal \cup
  \hcal}$ and $\overline{\Gamma}$. Much of the effort in obtaining the
proposed framework was devoted designing such meta-level heuristics
and fully implementing them in Maude~\cite{clavel-maudebook-2007}, a
high-performance reflective language and system supporting rewriting
logic.

The case studies presented in this paper comprise several
propositional sequent systems, encompassing different
proof-theoretical aspects. The chosen systems include: single
conclusion ($\Gtip$) and multi-conclusion intuitionistic logic
($\mLJ$), classical logic ($\Gtcp$), classical linear logic ($\LLM$)
and its dyadic system ($\LLD$), intuitionistic linear logic ($\ILL$),
and normal modal logics ($\kRule$ and $\Sfour$).  Beyond advocating
for the use of rewriting logic as a meta-logical framework, the novel
algorithms presented here are able to automatically discharge many
proof obligations and ultimately obtain the expected results.

This paper is an extended version of the
work~\cite{DBLP:conf/wrla/OlartePR18}, bringing not only new results
and analyses, but also new uses of the $\toolname$ tool.  In
particular:

\begin{itemize}

\item \emph{Properties and the spectrum of logics:} the procedures
  have been updated and extended to automatically check invertibility
  of rules in a larger class of logical systems, including modal
  logics and variants of  linear logic. Moreover, the current
  analyses prove stronger properties: invertibility lemmas and
  admissibility of structural rules are shown to be
  \emph{height-preserving} (a fact needed in the proof of
  cut-elimination). For that, the definitions and necessary conditions
  in Section \ref{sec.check} were refined.
  
 \item \emph{New reasoning techniques:} cut-elimination is probably
   the most important property of sequent systems. This paper shows
   how to use the proposed framework to automatically discard most of
   the cases in proofs of cut-elimination. In some cases, the proposed
   algorithms are able to complete entire cut-elimination proofs. The
   cases of failure are also interesting, since they tell much about
   the reasons for such failure.  Some are fixable by proving and adding
   invertibility and admissibility lemmas. Others can be used in order
   to shed light on the reason for the failure.  Finally, this work
   also addresses the dual property of identity expansion. Together
   with cut-elimination, this property is core on designing harmonical
   systems~\cite{DBLP:journals/jphil/FrancezD12}.

 \item \emph{Pretty printing and output: } the previous implementation
   has been updated to generate proof terms that can be used, e.g., to
   produce \LaTeX\ files with the proof trees in the meta-proofs. This
   allows for generating documents with complete and detailed proofs
   of several results in the literature, as well as identifying
   dependencies among the different theorems.
\end{itemize}

\paragraph{Outline.} The rest of the paper is organized as
follows.  Section~\ref{sec.proof} introduces the structural properties
of sequent systems that are considered in this work and
Section~\ref{sec.prelim} presents order-sorted rewriting logic and its
main features as a logical framework. Section~\ref{sec.framework}
presents the machinery used for specifying sequent systems as rewrite
theories.  Then, Section~\ref{sec.check} establishes how to prove the
structural properties based on a rewriting approach.
Section~\ref{sec:cut} addresses cut-elimination and identity
expansion.  The design principles behind the $ \toolname$ are described in
Section~\ref{sec.rew}.  Section~\ref{sec.case} presents different
sequent systems and properties that can be proved with the
approach. Finally, Section~\ref{sec.concl} concludes the paper and
presents some future research directions.

\section{Structural Properties of Sequent-Based Proof Systems}
\label{sec.proof}

This section presents and illustrates with examples the properties of {\em admissibility} and {\em
  invertibility} of rules in sequent
systems~\cite{troelstra96bpt,negri01spt}. Additional notation and
standard definitions are established to make the text self-contained.

\begin{definition}[Sequent]\label{def:seq}
  Let $\mathcal{L}$ be a formal language consisting of well-formed
  formulas.  A {\em sequent} is an expression of the form
  $\Gamma\vdash \Delta$ where $\Gamma$ (the \emph{antecedent}) and
  $\Delta$ (the \emph{succedent}) are finite multisets of formulas in
  $\mathcal{L}$, and $\vdash$ is the 
  \emph{consequence} symbol.  If the succedent of a sequent contains at most
  one formula, it is called {\em single-conclusion}; otherwise, it is
  called {\em multiple-conclusion}.
\end{definition}

\begin{definition}[Sequent System]\label{def:seqsys}
  A \emph{sequent system} $\scal$ is a set of inference rules.  Each
  inference
  rule is a {\em rule schema} and a rule instance is a {\em rule
    application}. An inference rule has the form
  $$
  \infer[r]{S}{S_1\quad\cdots\quad S_n}
  $$
  where the sequent $S$ is the {\em conclusion} inferred from the {\em
    premise} sequents $S_1,\ldots,S_n$ in the rule $r$.  If the set of
  premises is empty, then $r$ is an {\em axiom}.  In a rule
  introducing a connective, the formula with that connective in the
  conclusion sequent is the {\em principal formula}.  
  When sequents are restricted to having empty antecedents (hence being of the shape $\vdash \Delta$), the system is called {\em one-sided};
otherwise it is
  called {\em two-sided}.
\end{definition}

As an example, Fig.~\ref{fig:G3ip} presents the two-sided
single-conclusion propositional intuitionistic sequent system
$\Gtip$~\cite{troelstra96bpt}, with formulas built from the grammar:
\[
F,G ::= p \mid \top  \mid  \bot \mid F \vee G \mid F \wedge G \mid F \cimp G
\]
where $p$ ranges over atomic propositions. In this system, for instance,
the formula $F\vee G$ in the conclusion sequent of the inference rule
$\vee_L$ is the principal formula.

\begin{figure}
\resizebox{\textwidth}{!}{
$
\begin{array}{c}
\infer[I]{\Gamma, p \vdash p}{}
\qquad
\infer[\top_R]{\Gamma \vdash \top}{}
\quad
\infer[\top_L]{\Gamma, \top \vdash C}{\Gamma \vdash C}
\quad
\infer[\bot_L]{\Gamma, \bot \vdash C}{}
\\
\infer[\vee_L]{\Gamma, F \vee G \vdash C }{
 \deduce{\Gamma, F \vdash C}{}
 &
 \deduce{\Gamma, G \vdash C}{}
}
\qquad
\infer[\vee_{R_i}]{\Gamma \vdash F_1 \vee F_2}{\Gamma \vdash F_i}
\qquad
\infer[\wedge_L]{\Gamma, F \wedge G \vdash C}
{\Gamma, F , G \vdash C}
\qquad
\infer[\wedge_R]{\Gamma  \vdash F \wedge G }{
 \deduce{\Gamma  \vdash F}{}
 &
 \deduce{\Gamma  \vdash G}{}
}
\\
\infer[\cimp_L]{\Gamma, F\cimp G \vdash C}{
\deduce{\Gamma, F\cimp G \vdash F}{}
&
\deduce{\Gamma,  G \vdash C}{}
}
\quad
\infer[\cimp_R]{\Gamma \vdash F \cimp G}{\Gamma , F\vdash  G}
\end{array}
$
}
  \caption{System $\Gtip$ for propositional 
    intuitionistic logic. In the $I$ axiom, $p$ is atomic.
  }
  \label{fig:G3ip}
\end{figure}

\begin{definition}[Derivation]\label{def.deriv}
A {\em derivation} in a sequent system $\scal$ (called
$\scal$-derivation) is a finite rooted tree with nodes labeled by
sequents, axioms at the top nodes, and where each node is connected
with the (immediate) successor nodes (if any) according to its
inference rules. A sequent $S$ is \emph{derivable} in the sequent
system $\scal$, notation $\deduc{\scal}{S}$, iff there is a derivation
of $S$ in $\scal$.  The system $\scal$ is usually omitted when it can
be inferred from the context.

\end{definition}

It is important to distinguish the two different notions associated to
the symbols $\vdash$ and $\DED$, namely: the former is used to build
sequents and the latter denotes derivability in a sequent system.

\begin{definition}[Height of derivation]\label{def:height}
The {\em height} of a derivation is the greatest number of successive
applications of rules in it, where an axiom has height 1.  The
expression
\[
\deducn{\scal}{n}{\Gamma \vdash \Delta}
\]
denotes that the sequent $\Gamma\SEQ\Delta$ is derivable in $\scal$
with a height of derivation {\em at most} $n$.  A property is said
{\em height-preserving} if such $n$ is an invariant. The annotated
sequent
\[
\Gamma\seqn{n}\Delta
\]
is a shorthand for $\deducn{\scal}{n}{\Gamma \vdash \Delta}$ when
$\scal$ is clear from the context. Moreover, when $\Gamma$ and
$\Delta$ are unimportant or irrelevant, this notation is further
simplified to $n:S$.

\end{definition}

In what follows, annotated sequents are freely used in inference
rules. For instance, if $s(\cdot)$ represents the successor function
on natural numbers, then
\[
\infer[r]{s(n):S}{n:S_1& \ldots & n:S_k}
\]
represents the annotated rule $r$ stating that
\begin{center}
if $\deducn{\scal}{n}{S_i}$ for each $i=1,\ldots,k$, then
$\deducn{\scal}{s(n)}{S}$
\end{center}

It may be useful to derive new rules from the ones initially proposed
in a system, since this may ease the reasoning when proving
properties. Such derived new rules are said to be {\em admissible} in
the system. {\em Invertibility}, on the other hand, is the
admissibility of ``upside-down'' rules, where the premises of a rule
are derived from the conclusion. Invertibility is one of the most
important properties in proof theory, since it is the core of proofs
of cut-elimination~\cite{gentzen35}, as well as the basis for
tailoring focused proof systems~\cite{andreoli92jlc,liang09tcs}.

\begin{definition}[Admissibility and Invertibility]\label{def:inv}
  Let $\scal$ be a sequent system. An inference  rule 
  $\infer{S}{S_1& \cdots&
    S_k}$ 
     is called:

\begin{itemize}
\item[i.] {\em admissible in $\scal$} if $S$ is derivable in
  $\scal$ whenever $S_1,\ldots, S_k$ are derivable in $\scal$.
\item[ii.]  {\em invertible} in $\scal$ if the rules
  $\frac{S}{S_1}, \ldots, \frac{S}{S_k}$ are admissible in
  $\scal$.
\end{itemize}
\end{definition}

The admissibility of structural rules is often required in the proof
of other properties. A \textit{structural rule} does not introduce
logical connectives, but instead changes the structure of the
sequent. Since sequents are built from multisets, such changes are
related to the cardinality of a formula or its presence/absence in a
context.

In the intuitionistic setting, the structural rules for {\em
  weakening} and {\em contraction}
\begin{equation}\label{eq:wc}
\vcenter{\infer[\mathsf{W}]{\Gamma,\Delta\vdash C}{\Gamma\vdash C}}
\qquad\text{and}\qquad
\vcenter{\infer[\mathsf{C}]{\Gamma,\Delta\vdash C}{\Gamma,\Delta,\Delta\vdash C}} 
\end{equation}
are height-preserving  admissible in $\Gtip$. A proof of the
admissibility of weakening can be built by induction on the height of
derivations (considering all possible rule applications) and it is
often independent of any other results.  For example, suppose that
$\deducn{\Gtip}{n}{\Gamma \vdash C}$ and consider, e.g., the case
where $C=A\cimp B$ and that the last rule applied in the proof of $\Gamma \vdash A\cimp B$ is $\cimp_{R}$
\[
\infer[\cimp_{R}]{\Gamma \seqn{s(i)} A\cimp B}{
\deduce{\Gamma, A \seqn{i} B}
{}}
\]
By inductive hypothesis, the sequent $\Gamma, \Delta,A \seqn{i} B$ is
derivable and then,
$$
\vcenter{\infer[\cimp_{R}]{\Gamma, \Delta \seqn{s(i)} A\cimp B}{
\deduce{\Gamma,  \Delta,A \seqn{i} B}
{}}} 
$$
The same exercise 
can be done for the other rules of the system, thus showing that $\mathsf{W}$ is height-preserving admissible: 
\begin{center}
$\deducn{\Gtip}{n}{\Gamma \vdash C}$
implies 
$\deducn{\Gtip}{n}{\Gamma,\Delta \vdash C}$
\end{center}

On the other hand, proving invertibility may require the admissibility
of weakening.  For example, for proving that $\cimp_R$ is
height-preserving invertible in $\Gtip$, one has to show that
$\deducn{\Gtip}{n}{\Gamma,F \vdash G}$ whenever
$\deducn{\Gtip}{n}{\Gamma \vdash F\cimp G}$. The proof follows  by
induction on the height of the derivation of $\Gamma \SEQ F\cimp
G$. Consider, e.g., the case when $\Gamma=\Gamma',A \cimp B$ and the
last rule applied is $\cimp_{L}$
\[
\infer[\cimp_L]{\Gamma' , A \cimp B \seqn{s(i)} F \cimp G}{
\deduce{\Gamma', A \cimp B \seqn{i} A}{}
&
\deduce{\Gamma', B \seqn{i}  F \cimp G}{}
}
\]
By inductive hypothesis on the right premise, $\Gamma', B , F \seqn{i}
G$ is derivable. Considering the left premise, since $\Gamma \seqn{i}
A$ is derivable, height-preserving admissibility of weakening implies
that $\Gamma,F \seqn{i} A$ is derivable and  the result follows:  
\[
\infer[\cimp_L]{\Gamma,F \seqn{s(i)} G}{
\deduce{\Gamma, F \seqn{i} A}{}
&
\deduce{\Gamma', B,F \seqn{i}  G}{}
}
\]

Note that not all introduction rules in $\Gtip$ are invertible: if
$p_1,p_2$ are different atomic propositions, then $p_i\vdash p_1\vee
p_2$ is derivable for $i=1,2$, but $p_i \vdash p_j$ is not for $i\not=
j$. Indeed, $\vee_{R_i}$ and $\cimp_{L}$ are the only non-invertible
rules in $\Gtip$.

Finally, admissibility of contraction ($\mathsf{C}$) often depends on
invertibility results. As an example, consider 
that $\deducn{\Gtip}{n}{\Gamma,F\vee G,F\vee G\vdash C}$ with 
last rule applied $\vee_{L}$
$$
\infer[\vee_{L}]{\Gamma,F\vee G,F\vee G\seqn{s(i)} C}{
\deduce{\Gamma,F\vee G,F\seqn{i} C}{}&
\deduce{\Gamma,F\vee G,G\seqn{i} C}{}
}
$$
The inductive hypothesis cannot be applied since the
premises do not have duplicated copies of formulas.  Since
$\vee_L$ is height-preserving  invertible, the derivability of
$\Gamma,F\vee G,F\seqn{i} C$ and $\Gamma,F\vee G, G\seqn{i} C$ implies
the derivability of $\Gamma,F,F\seqn{i} C$ and $\Gamma,G,G\seqn{i} C$.
By induction, $\Gamma,F\seqn{i} C$ and
$\Gamma,G\seqn{i} C$ are derivable and the result follows: 
$$
\infer[\vee_{L}]{\Gamma,F\vee G\seqn{s(i)} C}{
\deduce{\Gamma,F\seqn{i} C}{}&
\deduce{\Gamma,G\seqn{i} C}{}
}
$$

Invertibility and admissibility results will be largely used for
showing cut-elimination in Section~\ref{sec:cut}.


\section{Rewriting Logic Preliminaries}
\label{sec.prelim}

This section briefly explains order-sorted rewriting
logic~\cite{meseguer-rltcs-1992} and its main features as a logical
framework.  Maude~\cite{clavel-maudebook-2007} is a language and tool
supporting the formal specification and analysis of rewrite theories,
which are the specification units of rewriting logic.

An \emph{order-sorted signature} $\Sigma$ is a tuple $\Sigma
{=}(S,\leq,F)$ with a finite poset of sorts $(S,\leq)$ and a set of
function symbols $F$ typed with sorts in $S$, which can be
subsort-overloaded.  For $X = \{X_s\}_{s\in S}$ an $S$-indexed family
of disjoint variable sets with each $X_s$ countably infinite, the {\em
  set of terms of sort $s$} and the {\em set of ground terms of sort
  $s$} are denoted, respectively, by $T_\Sigma(X)_s$ and
$T_{\Sigma,s}$; similarly, $T_\Sigma(X)$ and $T_\Sigma$ denote the set
of terms and the set of ground terms.  A {\em substitution} is an
$S$-indexed mapping $\func{\theta}{X}{T_\Sigma(X)}$ that is different
from the identity only for a finite subset of $X$ and such that
$\theta(x) \in T_\Sigma(X)_s$ if $x \in X_s$, for any $x \in X$ and
$s\in S$.  A substitution $\theta$ is called {\em ground} iff
$\theta(x) \in T_\Sigma$ or $\theta(x) = x$ for any $x \in X$.  The
application of a substitution $\theta$ to a term $t$ is denoted by
$t\theta$.  Acquaintance with the usual notions of \textit{position}
$p$ in a term $t$, \textit{subterm} $t|_{p}$ at position $p$, and
\textit{term replacement} $t[u]_{p}$ at of $t$'s subterm at position
$p$ with term $u$ is assumed (see, e.g.,~\cite{dershowitz-jouannaud}).
The expression $u \preceq t$ (resp., $u \prec t$) denotes that term
$u$ is a subterm (resp., proper subterm) of term $t$.  Given a term $t
\in T_{\Sigma}(X)$, $\overline{t} \in T_{(S,\leq, F\cup
  C_{\overline{t}})}(X)$ is the ground term obtained from $t$ by
turning each variable $x \in \ovars{t}$ of sort $s\in S$ into the
fresh constant $\overline{x}$ of sort $s$ and where $C_{\overline{t}}
= \{\overline{x} \mid x \in \ovars{t} \}$.

A \emph{rewrite theory} is a tuple $\rcal = (\Sigma, E \uplus B, R)$
with: (i) $(\Sigma, E \uplus B)$ an order-sorted equational theory
with signature $\Sigma$, $E$ a set of (possibly conditional) equations
over $T_\Sigma(X)$, and $B$ a set of structural axioms -- disjoint
from the set of equations $E$ -- over $T_\Sigma(X)$ for which there is
a finitary matching algorithm (e.g., associativity, commutativity, and
identity, or combinations of them); and (ii) $R$ a finite set of
rewrite rules over $T_\Sigma(X)$ (possibly with equational
conditions). A rewrite theory $\rcal$ induces a rewrite relation
$\rews_{\rcal}$ on $T_{\Sigma}(X)$ defined for every $t,u \in
T_\Sigma(X)$ by $t \rews_\rcal u$ if and only if there is a
\textit{one-step} rewrite proof of $(\forall X) t \rews u$ in $\rcal$.
More precisely, $t \rews_\rcal u$ iff there is a rule
$(\ccrl{l}{r}{\cond}) \in R$, a term $t'$, a position $p$ in $t'$, and
a substitution $\func{\theta}{X}{T_\Sigma(X)}$ satisfying $t
=_{E\uplus B} t' = t'[l\theta]_{p}$, $u =_{E\uplus B}
t'[r\theta]_{p}$, and $\rcal \vdash \cond\theta$.

The tuple $\tcal_\rcal = (\tcal_{\Sigma/E},\stackrel{*}\rews_\rcal)$,
where $\stackrel{*}\rews_\rcal$ is the reflexive-transitive closure of
$\rews_\rcal$, is called the {\em initial reachability model of
  $\rcal$}~\cite{bruni-semantics-2006}.  An inductive property of a
rewrite theory $\rcal$ does not need to hold for any model of $\rcal$,
but just for $\tcal_{\rcal}$. Using a suitable inductive inference
system, for example, one based on a convenient notion of constructors
as proposed in \cite{rocha-sufflpar-2010}, the semantic entailment
$\models$ in $\tcal_\rcal$ can be under-approximated by an inductive
inference relation $\Vdash$ in $\rcal$, which is shown to be sound
with respect to $\models$ (i.e., for any property $\varphi$, if $\rcal
\Vdash \varphi$, then $\tcal_\rcal \models \varphi$). A
$\Sigma$-sentence $(\forall X) \, \varphi$ is called an
\textit{inductive consequence} of $\rcal$ iff $\rcal \Vdash (\forall
X) \; \varphi$, and this implies that $\tcal_\rcal \models \varphi$.

Appropriate requirements are needed to make an equational theory
$\rcal$ {\em executable} in Maude.  It is assumed that the equations
$E$ can be oriented into a set of (possibly conditional)
{sort-decreasing}, {operationally terminating}, and {confluent}
rewrite rules $\overrightarrow{E}$ modulo
$B$~\cite{clavel-maudebook-2007}. For a rewrite theory $\rcal$, the
rewrite relation $\rews_\rcal$ is undecidable in general, even if its
underlying equational theory is executable, unless conditions such as
coherence~\cite{viry-coherence-2002} are given (i.e., rewriting with
$\rews_{R/E \uplus B}$ can be decomposed into rewriting with
$\rews_{E/B}$ and $\rews_{R/B}$). The executability of a rewrite
theory $\rcal$ ultimately means that its mathematical and execution
semantics coincide.

In this paper, the rewriting logic specification of a sequent system
$\scal$ is a rewrite theory $\rcal_\scal = (\Sigma_\scal, E_\scal
\uplus B_\scal, R_\scal)$ where: $\Sigma_\scal$ is an order-sorted
signature describing the syntax of the logic $\scal$; $E_\scal$ is a
set of executable equations modulo the structural axioms $B_\scal$;
and $R_\scal$ is a set of executable rewrite rules modulo $B_\scal$
capturing those non-deterministic aspects of logical inference in
$\scal$ that require proof search.  The point is that although both
the computation rules $E_\scal$ and the deduction rules $R_\scal$ are
executed by rewriting modulo $B_\scal$, by the executability
assumptions on $\rcal_\scal$, the rewrite relation $\rews_{E_\scal /
  B_\scal}$ has a single outcome in the form of a canonical form and
thus can be executed blindly with a ``don't care'' non-deterministic
strategy. Furthermore, $B_\scal$ provides yet one more level of
computational automation in the form of $B_\scal$-matching and
$B_\scal$-unification algorithms. This interplay between axioms,
equations, and rewrite rules can ultimately make the executable
specification $\rcal_\scal$ very efficient with modest memory
requirements.

Finally, the expression $\textit{CSU}_B(t,u)$ denotes the
\textit{complete set of unifiers} of terms $t$ and $u$ modulo the
structural axioms $B$. Recall that for each substitution
$\func{\sigma}{X}{T_{\Sigma}(X)}$, there are substitutions $\theta \in
\textit{CSU}_{B}(t, u)$ and $\func{\gamma}{X}{T_{\Sigma}(X)}$ such
that $\sigma =_{B} \theta\gamma$. For a combination of free and
associative and/or commutative and/or identity axioms $B$, except for
symbols that are associative but not commutative, such a finitary
unification algorithm exists~\cite{hendrix-osunif-2012}. In the case
of $\rcal_\scal$, the structural axioms are identity, commutativity,
and associativity, which are the usual structural axioms for multisets
of formulas and sequents.


\section{A Rewriting View of Sequent Systems}
\label{sec.framework}

This section presents the machinery used for specifying a
propositional sequent system $\scal$ as a rewrite theory
$\rcal_\scal$.  Such a framework is equipped with sorts that represent
formulas, set of formulas, sequents, and lists of sequents.  The user
of the framework is expected to inhabit the sort for formulas in
$\rcal_\scal$ with the concrete syntax of the system $\scal$. This has
the immediate effect of fully inhabiting the remaining sorts of
$\rcal_\scal$.  As shown below, rewrite rules in $\rcal_\scal$
correspond to backwards inference rules of $\scal$, so that
proof-search in the former is successful whenever all leaves in a
proof-tree are instances of axioms. The system $\Gtip$
(Fig.~\ref{fig:G3ip}) is used throughout the section for illustrating
the proposed approach.

The notation of Maude~\cite{clavel-maudebook-2007} is adopted as an
alternative representation to the rewriting logic one introduced in
Section~\ref{sec.prelim}. This decision has the immediate effect of
producing an executable specification of sequent systems, while
providing a precise mathematical semantics of the given definitions.
Additional details about the implementation in Maude are given in
Section~\ref{sec.rew}.

As a reference for the development of this section, Maude declarations
are summarized next.

\begin{lstlisting}

mod M is ... endm               --- rewrite theory M
sort S .                        --- declaration of sort S
subsort S1 < S2 .               --- subsort relation
op f : S1,...,Sn -> T .         --- function symbol f of sort S1,...,Sn --> T
op c : -> T .                   --- constant c of sort T
rl  [rule-name] : l => r .      --- rewrite rule with name 'rule-name'
crl [rule-name] : l => r if c . --- conditional rewrite rule
\end{lstlisting}

An order-sorted signature $\Sigma_\scal$ defining the sorts for
building formulas (and multisets of formulas and sequents, which are
introduced later), is assumed:

\begin{lstlisting}

sort Prop .               --- Atomic propositions                                
sort Formula .            --- Formulas                                        
subsort Prop < Formula .  --- Atomic propositions are formulas
\end{lstlisting}
The object logic to be specified must provide suitable constructors
for these two sorts. For instance:

\begin{lstlisting}

mod G3i is 
 ...
 op p : Nat -> Prop .                       --- Atomic  Propositions
 op _/\_ : Formula Formula -> Formula .     --- Conjunction
 op False : -> Formula .                    --- False/bottom
 ...
endm 
\end{lstlisting}

\noindent
Atomic propositions take the form $p(n)$, where the natural number $n$
is the identifier of the proposition (e.g., the proposition $p_3$ is
written as \lstinline{p(3)}). Constructors for \lstinline{Prop} are
not allowed to have arguments of type \lstinline{Formula}.

Multisets of formulas are built in the usual way:

\begin{lstlisting}

sort MSFormula .              --- Multiset of Formulas
subsort Formula < MSFormula . --- A formula is a singleton multiset of formulas
op * : -> MSFormula .         --- Empty multiset
op _;_ : MSFormula MSFormula -> MSFormula [assoc comm id: * ] .  --- Union
\end{lstlisting}
  
\noindent
Given two multisets $M$ and $N$, the term \lstinline{M;N} denotes
$M\uplus N$. Note that the mixfix operator \lstinline|op _;_| (in
Maude, \lstinline{_} denotes the position of the parameters) is
declared with three structural axioms: associativity, commutativity,
and the empty multiset as its identity element.  Due to the
subsort relation \lstinline{Formula < MSFormula}, the sort
\lstinline{Formula} is the least sort of a formula $F$, while $F$ is
also a singleton containing $F$.

Sequents, as expected, are built as pairs of multisets of formulas. A
\emph{goal} is a list of sequents to be proved:

\begin{lstlisting}

sorts Sequent  Goal .                        --- Sequents and lists of sequents 
subsort Sequent < Goal .
op _|--_ : MSFormula MSFormula -> Sequent .  --- Sequents
--- Goals: List of sequents to be proved
op proved : -> Goal [ctor] .                 --- The empty list of sequents
op _|_ : Goal Goal -> Goal [frozen(2) id: proved] . 
\end{lstlisting}  

\noindent
The attribute \lstinline{frozen} used in the declaration of the
operator \lstinline{op _|_} (goals' concatenation) means that
inference steps can only be performed on the head of a non-empty list
of goals (i.e., the usual `head-tails' recursive structure). More
precisely, the attribute \lstinline{frozen(2)} defines a rewriting
strategy where the second parameter cannot be subject of rewriting.
In this way, only the first sequent \lstinline{S} in the list
\lstinline{S | G} can be reduced until it becomes \lstinline{proved}
(when possible), as will be explained shortly.

Inference rules in the rewrite theory $\rcal_\scal$ are specified as
rewrite rules that rewrite list of sequents.  There are two
options for expressing an inference rule as a rewrite rule. Namely,
they can be axiomatized as {\em backwards} inference (i.e., from
conclusion to premises) or as {\em forward} inference (i.e., from
premises to conclusion).  In this paper, as explained in
Section~\ref{sec.prelim}, the backwards inference approach is adopted,
so that a proof-search process advances by rewriting the target goal
of an inference rule to its premises, thus aiming at a bottom-up
proof-search procedure.  For instance, the initial rule, and the left
and right introduction rules for conjunction in the system $\Gtip$ are
specified as follows:

\begin{lstlisting}

var P : Prop .
var Gamma : MSFormula .
vars A B C : Formula .
--- Rules
rl [I] : P ; Gamma  |-- P => proved . 
rl [AndL] : A /\ B ; Gamma |-- C => A ; B ; Gamma |-- C .
rl [AndR] : Gamma |-- A /\ B => (Gamma |-- A) | (Gamma |-- B) .
\end{lstlisting}

Variables in rules are implicitly universally quantified. The type of
variables is specified with the syntax \lstinline{var x : S }.  Hence,
the initial rule must be read as
\[(\forall P:\texttt{Prop}, \Gamma:\texttt{MSFormula})((P, \Gamma \vdash P) \; \rews \; \proved)
\]
The constant \lstinline{proved} denotes the empty list of goals or,
equivalently, the empty collection of (pending) proof
obligations. Rules with more than one premise, such as
\lstinline{AndR}, are specified using \lstinline{op: _|_}. Modulo the
axioms $B_\scal$, a term \lstinline{proved | G} is structurally equal to
\lstinline{G}, thus making the goal \lstinline{G} automatically active
for proof-search under the rewrite strategy declared for goals.

As illustrated with the running example, the syntax of the object
logic in $\rcal_\scal$, as well as its inference rules, is
straightforward. This is usually called 
in rewriting
logic 
the {\em
  $\epsilon$-representational distance} ~\cite{meseguer-twenty-2012}, where the system being specified
mathematically as a rewrite theory and its codification in Maude as a
system module are basically the same.  This is certainly an appealing
feature that can widen the adoption of the framework proposed here for
implementing and analyzing sequent systems.

In face of the $\epsilon$-representational distance, the adequacy of
 $\rcal_\scal$ with respect to $\scal$ follows from the
soundness of rewriting logic itself.

\begin{proposition}[Adequacy]\label{th:adq}
  Let $\scal$ be a sequent system, $\rcal_\scal$ the resulting rewrite
  theory encoding the syntax and inference rules of $\scal$, $S$ a
  sequent in $\scal$, and $t_S$ its representation in
  $\rcal_\scal$. Then,
  \[\deduc{\scal}{S} \qquad \text{iff} \qquad \deducr{\rcal_\scal}{t_S \stackrel{*}{\rews} \proved}.\]
\end{proposition}

Observe that the proof of a sequent $S$ in $\scal$ may follow
different strategies depending on the order the subgoals are proved.
Such strategies are irrelevant because all the branches in the
derivation of a proof must be closed (i.e., ending with an instance of
an axiom).  Hence, if $S$ is provable, it can be proved using the
strategy enforced by \lstinline{op _|_} in $\rcal_\scal$: always
trying to first solve the left-most subgoal of the pending goals.

Let $r$ be a sequent rule in $\scal$ and $R$ the corresponding rewrite
rule in $\rcal_\scal$.  Modulo the associativity and commutativity of
multisets of formulas, it is easy to show that, for any sequent $S$
and its representation $t_S$, $t_S \stackrel{1}{\rews}_{R_\scal} t_S'$
iff $t_S'$ is the representation of the premises (where $\proved$
means no premises) obtained when $r$ is applied (on the same active
formula) in $S$.  Hence, rewrite steps are in one-to-one
correspondence with proof search (bottom-up) steps in
$\scal$-derivations.

When using the proposed framework, the resulting rewrite theory
becomes a proof search procedure. For instance, Maude's command

\lstinline{search p(1) /\ p(2) |-- p(2) /\ p(1) =>* proved . }

\noindent
answers the question of whether the sequent $p_1 \wedge p_2 \vdash p_2
\wedge p_1$ is provable in $\Gtip$.

For the proof analyses later developed in this paper, operations to
the rewrite theory $\rcal_\scal$ are added. A new constructor for
annotated sequents (see Definition~\ref{def:height}) and also a copy
of the inference rules dealing with annotations are included:
\begin{lstlisting}

op _:_ : Nat Sequent -> Sequent .  --- Annotated sequents
var k : Nat .
--- Automatically generated annotated sequent rules
rl [I]    : s(k) : P ; Gamma  |-- P => proved. 
rl [AndL] : s(k) : A /\ B ; Gamma |-- C => k : A ; B ; Gamma |-- C .
rl [AndR] : s(k) : Gamma |-- A /\ B => 
                   ( k : Gamma |-- A) |  ( k : Gamma |-- B) .

\end{lstlisting}

The function application \lstinline{s(.)} denotes the successor
function on \lstinline{Nat}.  Note that axioms are annotated with an
arbitrary height $k\geq 1$. In rules with two premises, both of them
are marked with the same label. This is, without loss of generality,
since it is always possible to obtain larger annotations from shorter
ones (see Theorem~\ref{th.height}).

In the rest of the paper, given a sequent system $\scal$,
$\rcal_\scal$ will denote the resulting rewrite theory that encodes
the syntax, inference, and annotated inference rules of $\scal$. By
abusing the notation, when a sequent $S\in \scal$ is used in the
context of the rewrite theory $\rcal_\scal$ (e.g., in $S
\rews_{\rcal_\scal} S'$), such $S$ must be understood as the
corresponding term $t_S \in \Sigma_{\rcal_\scal}(X)$ representing $S$
in $\rcal_\scal$. Similarly, ``a sequent rule $r$'' in the context of
the theory $\rcal_\scal$ is to be understood as ``the representation
of $r$ in $\rcal_\scal$''.  The expression $\rcal_1 \cup \rcal_2$
denotes the extension of the theory $\rcal_1$ by adding the inference
rules of $\rcal_2$ (and {\em vice versa}); in this case, the sequents
in the resulting theory $\rcal_1 \cup \rcal_2$ are terms in the
signature $\Sigma_{\rcal_1} \cup \Sigma_{\rcal_2}$.  If $S$ is a
sequent, the expression $\rcal \cup \{S\}$ denotes the extension
resulting from $\rcal$ by adding the sequent $S$ as an axiom,
understood as a zero-premise rule (i.e., $\rcal$ is extended with the
rule \lstinline{rl [ax] S => proved .}).  Moreover, given a rewrite
theory $\rcal$ and a sequent $S$, the notation $\deducr{\rcal}{S}$
means $\deducr{\rcal} S \stackrel{*}{\rews} \proved$, i.e., there is a
derivation of $S$ in the system specified by $\rcal$. Similarly, for
annotated sequents, $\deducr{\rcal}{n:S}$ means $\deducr{\rcal} (n:S)
\stackrel{*}{\rews} \proved$.

\begin{theorem}\label{th.height}
 Let $\scal$ be a sequent system, $S$ a sequent, and $k\geq 1$. Then,
\[
 \deducr{\rcal_\scal}{}{k:S} \qquad\text{implies}\qquad
 \deducr{\rcal_\scal}{s(k):S}.
\] 
\end{theorem}
\begin{proof}
  By induction on $k$. If $k=1$, then $S$ is necessarily an instance
  of an axiom and $s(1):S \rews_{\rcal_\scal} \proved$ using the same
  axiom. The case $k>1$ is immediate by applying induction on the
  premises (with shorter derivations).
\end{proof}


\section{Proving Admissibility and Invertibility}
\label{sec.check}

This section presents rewrite- and narrowing-based techniques for
proving admissibility and invertibility of inference rules of a
sequent system specified as a rewrite theory $\rcal_\scal$ (see
Section~\ref{sec.framework}).
They are presented as meta-theorems about sequent systems with the
help of rewrite-related scaffolding, such as terms and substitutions,
and they provide sufficient conditions for obtaining the desired
properties.
The system $\Gtip$ is, as in previous sections, used as running
example (Section~\ref{sec.case} presents the complete set of case
studies).

The procedures proposed here heavily depend on {\em unification},
hence a brief discussion on the subject is presented next.  First,
note that, if no additional axioms are added to the symbols in the
syntax of the object logic $\scal$, the existence of a unification
algorithm for terms in the sort \lstinline{Sequent} is guaranteed.
As a matter of example, consider the unification of the conclusions of
the right and left rules for conjunction in $\Gtip$. Unification
problems take the form
\lstinline{t1 =? t1' /\ ... /\ tn =? tn'}
and Maude's command \lstinline{variant unify} computes the set of
unifiers modulo the declared axioms in the theory:
  
\begin{lstlisting}
  
variant unify suc(n) : A /\ B ; Gamma |-- C =?  suc(n') : Gamma' |-- A' /\ B' .
                
Unifier #1
n -->  %3:FNat                       n' --> %3:FNat
A -->  %1:Formula                    A' --> %4:Formula
B -->  %2:Formula                    B' --> %5:Formula
C -->  %4:Formula /\ %5:Formula
Gamma --> %6:MSFormula            
Gamma' --> %6:MSFormula ; %1:Formula /\ %2:Formula

No more unifiers.                
  \end{lstlisting}

\noindent
Observe that the variables of the second rule are renamed to avoid
clash of names.  The term \lstinline|%1:Formula| denotes a fresh
variable of sort \lstinline{Formula}.  Let $t$ and $t'$ be the two
terms in the above unification problem and $\theta$ the (unique)
substitution computed.  Consider the least signature
$\Sigma'_{\rcal_\scal}$ that contains $\Sigma_{\rcal_\scal}$ as well
as a fresh constant \lstinline{
\lstinline{
\lstinline{
\lstinline{
inferred from the context, the constant \lstinline{
is written as \lstinline{
carry their typing information. In this example,
$\Sigma_{\rcal_\scal}'$ extends $\Sigma_{\rcal_\scal}$
with constants
\lstinline{op 
\lstinline{op 
The ground term $\overline{t\theta}$ in the signature
$\Sigma'_{\rcal_\scal}$ is ~~
\lstinline{s(
As expected, on this (ground) sequent it is possible to apply both
\lstinline{AndL} and \lstinline{AndR}.

\subsection{Admissibility of Rules}\label{sec:adm-cond}

This section introduces sufficient conditions to prove theorems of the
form
\[
\mbox{if } \deducn{\scal}{n}{\seq{\Gamma'}{\Delta'}} 
\mbox{ , then  }
\deducn{\scal}{n}{\seq{\Gamma}{\Delta}}
\]
i.e., height preserving admissibility of the rule~\footnote{The
admissibility problems considered in this section correspond to
one-premise rules only. Observe that a general approach for proving
admissibility of rules with more than one premise does not
exist. Indeed, cut-elimination itself is an admissibility problem,
which will be addressed in Section~\ref{sec:cut}.}
\[
\infer[r]{\Gamma\SEQ \Delta}{\Gamma' \SEQ \Delta'}
\] 

Admissibility is often proved by induction on the height of the
derivation $\pi$ of the sequent $\Gamma' \SEQ \Delta'$ in combination
with case analysis on the last rule applied in $\pi$.  It turns out
that this analysis may depend on other results. For example, as
illustrated in Section~\ref{sec.proof}, proving admissibility of
contraction often depends on invertibility results.  Hence, any
general definition of admissibility of rules in the rewriting setting
has to internalize such reasoning.  This will be formalized by {\em
  closing} the leaves in $\pi$ w.r.t. theorems of the form ``if $S_1$
is provable, so is $S_2$''. Such theorems will be encoded as rewrite
rules of the form $r : S_1 \rews S_2$.  More precisely, let $t_s$ be a
ground term denoting a sequent and $r: t \rews t'$ be a rule. The set
\[\closure{t_s}{r} = \{t_s\} \cup \{t'\theta \mid \theta \in \textit{CSU}(t, t_s) \}
\]
is the least set containing $t_s$ and the resulting premises after the
instantiation of $r$ in $t_s$.  Let $T_s$ be a set of ground terms and
$R= \{r_1,...r_m\}$ be a set of theorems of the form $r_k : t^k \to
t^k_1$.
Then,
\[\closure{T_s}{R} = T_s \cup 
\bigcup\limits_{k\in 1..m} \{t^k_1\theta \mid t \in T_s , \theta \in
\textit{CSU}(t^k, t)\}.
\]

Proofs of admissibility need to consider all the cases of the last
rule applied in a proof $\pi$. Definition~\ref{def.check.admis}
identifies rules that are height-preserving admissible {\em relative}
to one of the sequent rules of the system.

\begin{definition}[Local admissibility]\label{def.check.admis}
Let $\scal$ be a sequent system, $\mathcal{I}$ be a (possibly empty)
set of rules, and $r_t \in \scal$ and $r_s$ be rules given by
\[
\infer[r_t]{s(k): T}{k:T_1 ~\cdots~ k:T_n}
\qquad 
\infer[r_s]{S}{S_1}
\]
The rule \emph{$r_s$ is height-preserving admissible relative to $r_t$
in $\scal$} under the assumptions $\mathcal{I}$ iff assuming that
$i:S_1$ is provable then, for each $\theta \in
\textit{CSU}(i:S_1,s(k):T)$,
  \[\deducr{ \rcal_\scal \cup \closure{\{\overline{(k:T_j)\theta} \mid j \in 1 ..  n \}}{\mathcal{I}}
    \cup  \{ (\forall \vec{x}) (\overline{k\theta} : S_1 \to \overline{k\theta} : S )\} }{\overline{(i:S)
      \theta}}, \]
where $vars(i:S_1) \cap vars(s(k):T) = \emptyset$ and $\vec{x} =
vars(S)\cup vars(S_1)$.
\end{definition}

For proving admissibility of the rule $r_s$, the goal is to prove
that, if $S_1$ is derivable with height $i$, then $S$ is also
derivable with at most the same height. The proof follows by induction
on $i$ (the height of a derivation $\pi$ of $S_1$). Suppose that the
last rule applied in $\pi$ is $r_t$.  This is only possible if $S_1$
and $T$ ``are the same'', up to substitutions.  Hence, the idea is
that each unifier $\theta$ of $i:S_1$ and $s(k): T$ covers the cases
where the rule $r_t$ can be applied on the sequent $S_1$.  For
computing this unifier, it is assumed that the rules do not share
variables (the implementation takes care of renaming the variables if
needed). A proof obligation is generated for each unifier. Consider,
for instance, the proof obligation of the ground sequent
$\overline{(i:S)\theta}$ for a given $\theta \in \textit{CSU}(i:S_1,
s(k):T)$. This means that, as hypothesis, the derivation below is
valid
\begin{equation}\label{eq:der1}
  \infer[r_t]{(i:S_1)\theta}{(k:T_1)\theta & \cdots & (k:T_n)\theta}
\end{equation}
Hence, all the premises in this derivation are assumed derivable.
This is the purpose of extending the theory with the following set of
ground sequents (which are interpreted as rules of the form $t \to
\proved$):
\begin{equation}
\label{eq:ind1}
\left\{\overline{(k:T_j)\theta}
\mid j \in 1 ..  n \right\}
\end{equation}

The proof of admissibility theorems may require auxiliary
lemmas. Assuming the theorems $\mathcal{I}$, the sequents resulting
after an application of $r\in \mathcal{I}$ to the sequents in Equation
\ref{eq:ind1} can be also assumed to be provable (notation
$\closure{\cdot}{\mathcal{I}}$).  The typical instantiation of
$\mathcal{I}$ in admissibility  analysis will be the already proved invertibility
lemmas.

If $\theta \in \textit{CSU}(i:S_1, s(k):T)$, then $\theta$ should
map $k$ to a fresh variable, say
\lstinline{
and $i$ to
\lstinline{s(
Hence, the (ground) goal
  $\overline{(i:S) \theta}$ to be proved takes the form
\lstinline{s(
$\overline{S \theta}$ where
  \lstinline{
is the freshly generated constant in the extended signature.  By
induction, it can be assumed that the theorem (i.e., $S_1$ implies
$S$) is valid for shorter derivations, i.e., derivations of height at
most
\lstinline{
\begin{equation}
\label{ind:2}
\{ (\forall \vec{x}) (\overline{k\theta} : S_1 \to \overline{k\theta} : S )\}
\end{equation}
where the height of the derivation is ``frozen'' to be the constant
$\overline{k\theta}$.  This allows for applying $r_s$ only on sequent
of height $\overline{k\theta}$. In particular, induction can be used on all the premises of the rule $r_t$ in Equation
\eqref{eq:der1}.

If it is possible to show that the ground sequent
$\overline{(i:S)\theta}$ is derivable from the extended rewrite
theory, then the admissibility result will work for the particular case when $r_t$
is the last applied rule in the derivation $\pi$ of $S_1$. Since a
complete set of unifiers is finite for terms of the sort
\lstinline{Sequent}, then there are finitely many proof obligations to
discharge in order to check if a rule is admissible relative to a rule
in a sequent system.  Observe that the set $\textit{CSU}(i:S,s(k):T)$
may be empty. In this case, the set of proof obligations is empty and
the property vacuously holds.

The base cases in this proof correspond to axiom rules.  Consider for
instance the case where $r_t$ is the initial rule in the system
$\Gtip$. Since there are no premises, the set in Equation
\eqref{eq:ind1} is empty and hence there are no ground sequents of the
form
\lstinline{
as hypothesis.  Consider the hypothetical case that
\lstinline{s(
is rewritten to a term of the form
\lstinline{
by using another rule of the system (different from the initial rule
$r_t$).  Note that it is not possible to use the rule in Equation
\eqref{ind:2} to reduce the goal to
\lstinline{proved}:
an application of this rule 
produces inevitably yet another sequent of the form
\lstinline{
Thus the only hope to finishing the proof is to apply the initial rule
on the sequent
\lstinline{s(
and the rule in Equation \eqref{ind:2} is never used in such a proof.

The notion of admissibility relative to a rule is the key step in the
inductive argument, when establishing sufficient conditions for the
admissibility of a rule in a sequent system.

\begin{theorem}\label{thm.check.admis}
  Let $\scal$ be a sequent system, $\mathcal{I}$ be a (possibly empty)
  set of rules, and $\infer[r_s]{{\Gamma}\SEQ{\Delta}}{{\Gamma' }\SEQ{
      \Delta'}}$ be an inference rule.  If $r_s$ is admissible
  relative to each $r_t$ in $\scal$ under the assumption
  $\mathcal{I}$, then $r_s$ is height-preserving admissible in $\scal$
  under the assumption $\mathcal{I}$, i.e.,
  $\deducn{\scal}{n}{\seq{\Gamma'}{\Delta'}}$ implies
  $\deducn{\scal}{n}{\seq{\Gamma}{\Delta}}$ assuming the theorems
  $\mathcal{I}$.
\end{theorem}

\begin{proof}
Assume that $\Gamma' \seqn{n} \Delta'$ is derivable in the system $\mathcal{S}$.  The proof
proceeds by induction on $n$  with case
analysis on the last rule applied. Assume that the last applied rule
is $r_t$.  By hypothesis (using Definition~\ref{def.check.admis}), it
can be concluded that $\Gamma \seqn{n} \Delta$ is derivable and the result follows.
\end{proof}

It is important to highlight the rationale behind
Definition~\ref{def.check.admis}, which is similar to the results in
the forthcoming sections.  The proof search procedures try to show
that $\overline{(i:S)\theta}$ is provable by reducing it to
\code{provable} (Proposition \ref{th:adq}). Since rules in
$\rcal_\scal$ are encoded in a backward fashion (rewriting the
conclusion into the premises), the procedure attempts to build a
bottom-up derivation of that sequent.  The set of assumptions
(Equation \eqref{eq:ind1}) is added as axioms and it is closed under
the rules in $\mathcal{I}$.  The closure reflects a forward reasoning:
if the (ground) sequent $S$ is provable and theorem $r\in\mathcal{I}$
applies on it, then the right-hand side of $r$ can be also assumed to
be provable.  During the search procedure, a goal is immediately
discharged if it belongs to that set of assumptions.  Finally, the
induction principle is encoded by specializing the theorem to be
proved, and allowing applications of it to ground sequents with
shorter height annotations (Equation~\eqref{ind:2}).

The proof of admissibility of weakening for $\Gtip$ and some other
systems studied in Section \ref{sec.case} does not rely on any result
(and hence, $\mathcal{I}$ is empty).  Also, the proof of Theorem
\ref{th.height}, mechanized by stating it as the admissibility of the
annotated rule
\begin{equation}\label{eq:H}
\infer[\mathsf{H}]{s(n):S}{n:S}
\end{equation}
is also proved with $\mathcal{I}= \emptyset$ in all the systems in
Section \ref{sec.case}.  The proof of contraction requires some
invertibility lemmas and they must be added to $\mathcal{I}$.  The use
of $\mathcal{I}$ thus presents a convenient and modular approach
because properties can be proved incrementally.

\subsection{Invertibility of Rules}\label{sec:inv-cond}

This section gives a general definition for proving height-preserving
invertibility of rules.  Observe that such analysis is done
premise-wise. The case of rules with several premises is performed for
each one of them separately. For instance, consider the rule $\cimp_L$
of $\Gtip$, and let $\cimp_{L1}$ and $\cimp_{L2}$ be the rules
\[
\infer[\cimp_{L1}]{\Gamma, F\cimp G \vdash F}{
\deduce{\Gamma, F\cimp G \vdash C}{}
}
\qquad
\infer[\cimp_{L2}]{\Gamma,  G \vdash C}{
\deduce{\Gamma, F\cimp G \vdash C}{}
}
\]
It can be shown that $\cimp_{L1}$ is not admissible while $\cimp_{L2}$
is. Hence, $\cimp_L$ is invertible only in its right premise (see
Definition \ref{def:inv}).

Some invertibility results depend on, e.g., the admissibility of
weakening.  Hence, for modularity, the invertibility analysis is
parametric under a set of rules $\mathcal{H}$ 
of admissible
rules of the system (that can be used during the proof search
procedure).


\begin{definition}[Local invertibility]\label{def.check.invert}
  Let $\scal$ be a sequent system and $\mathcal{H}$ be a (possibly
  empty) set of rules.  Consider the following annotated inference
  rules:
\[
\infer[r_s]{s(k):S}{k:S_1
    ~\cdots~ k:S_m}
\qquad
\infer[r_t]{s(m):T}{m:T_1 ~\cdots~ m:T_n}
\]
Under the assumption $\mathcal{H}$, the premise $l\in 1 .. m$ of the
rule \emph{$r_s$ is height-preserving invertible relative to the rule
$r_t$} iff for each $\theta \in \textit{CSU}(s(k):S,s(m):T)$:\\
\noindent\resizebox{\textwidth}{!}{
$
\deducr{\mathcal{H} \cup \rcal_\scal \cup \left\{\overline{(m:T_j)\theta} \mid j \in 1.. n \right\} \cup \bigcup\limits_{j \in 1.. n}\{\overline{(m:S_l)\gamma} \mid \gamma \in \textit{CSU}(k:S,\overline{(m:T_j)\theta})\}
}{ \overline{(k:S_l) \theta}}
$}\\
where the variables in $S$ and $T$ are assumed disjoint.  Under the
assumption $\mathcal{H}$, the rule $r_s$ is height-preserving
invertible relative to $r_t$ if all its premises are.
\end{definition}

In order to show the invertibility of a rule $r_s$, the goal is to
check that derivability is not lost when moving from the conclusion
$S$ to the premises $S_1, \cdots, S_m$.  Each premise $S_l$ entails a
different proof obligation.  Let $\infer[r_{sl}]{S}{S_l}$ be the
``sliced'' version of rule $r_s$ when the case of the premise $l\in
1..m$ is considered.
The proof is by induction on the derivation $\pi$ of
$S$.  Suppose that the last rule applied in $\pi$ is $r_t$. 
Observe that this is only possible if 
$S$ and $T$ unify.
Let 
$\theta \in \textit{CSU}(s(k):S,s(m):T)$ and assume that 
$\theta(k)=$\lstinline{
(and hence $\theta(m)=$\lstinline{
The resulting derivation $\pi$ takes the form
\begin{equation}
\label{eq:inv}
\infer[r_t]{s(\%1):\overline{S\theta}}{
\%1:\overline{T_1\theta} \dots \%1:\overline{T_n\theta}
}
\end{equation}
The premises of this derivation are assumed to be provable and the
theory is extended with the axioms $\overline{(m:T_j)\theta}$.  Since
those premises have shorter derivations, induction applies on them.
More precisely, given that the ground sequent
$\overline{(m:T_j)\theta}$ is provable and the rule $r_{sl}$ can be
applied on it, the resulting sequent after the application of $r_{sl}$
is also provable with the same heigh of derivation
$\overline{m\theta}$:

\begin{equation}\label{eq-inv-ih}
\begin{aligned}[c]
\mbox{ if~~ }
\end{aligned}
\begin{aligned}[c]
\infer[r_{sl}]{\overline{T_j \theta}}{T_jS_l }
\end{aligned}
\begin{aligned}[c]
\mbox{~~ then } 
\mbox{the sequent ${T_j{S_l}}$ is provable with height $\%1$}
\end{aligned}
\end{equation}
This inductive reasoning is captured by the expression 

\[
\bigcup\limits_{j \in 1.. n}\{\overline{(m:S_l)\gamma} \mid \gamma \in \textit{CSU}(k:S,\overline{(m:T_j)\theta})\}
\]

The unifier $\gamma$ checks whether it is possible to apply $r_{sl}$
on the premise $T_j$. If this is the case, the resulting premise
$S_l\gamma$ is added as an axiom.
If, from the extended theory, it is possible to prove derivable the
premise $S_l$ with height $\overline{k\theta}$, then the invertibility result will
work for the particular case when $r_t$ was the last applied rule in
the derivation $\pi$ of $S$.  If the set $ \textit{CSU}(s(k)S,s(m):T)$
is empty, it means that the rules $r_t$ and $r_s$ cannot be applied on
the same sequent and the property vacuously holds for that particular
case of $r_t$.  For instance, in $\Gtip$, the proof of invertibility
of $\wedge_R$ does not need to consider the case of invertibility
relative to $\vee_R$ since it is not possible to have, at the same
time, a conjunction and a disjunction on the succedent of the
sequent. In multiple-conclusion systems as e.g, $\Gtcp$ (see Section
\ref{case:gtcp}), this proof obligation is not vacuously discarded.

Theorem~\ref{thm.check.invert} presents sufficient conditions for
checking the invertibility of a rule $r_s$ in a sequent system. Note
that if $r_s$ is a rule without premises (i.e., an axiom), the result
vacuously holds. Hence, only the case for $m\geq 1$ premises is
considered.  The proof of the next theorem is similar to the one given for
Theorem~\ref{thm.check.admis}.

\begin{theorem}\label{thm.check.invert}
  Let $\scal$ be a sequent system, $\mathcal{H}$ be a set of rules,
  and $r_s$ be an inference rule with $m \geq 1$ premises in $\scal$.
  Let $l \in 1..m$.  If the premise $l$ of $r_s$ is height-preserving
  invertible relative to each $r_t$ in $\scal$ under the assumption
  $\mathcal{H}$, then the premise $l$ in $r_s$ is height-preserving
  invertible in $\scal$ under the assumption $\mathcal{H}$.
\end{theorem}

Most of the proofs of invertibility require Theorem~\ref{th.height}
and, in some cases, admissibility of weakening.  Hence, the assumption
$\mathcal{H}$ must contain those theorems for each case.  The
dependencies among the different proofs will be stated in
Section~\ref{sec.case} for each of the systems under analysis.


\section{Proving Cut-Elimination and Identity Expansion}\label{sec:cut}

One of the most fundamental properties of a proof system is
\emph{analyticity}, expressing that in a proof of a given formula $F$,
only subformulas of $F$ need to appear. In sequent calculus, this
property is often proved by showing that the \emph{cut-rule} is
admissible.  Roughly, the cut-rule introduces an auxiliary lemma $A$
in the reasoning ``if $A$ follows from $C$ and $B$ follows from $A$,
then $B$ follows from $C$''.  The admissibility of the cut rule (known
as the cut-elimination property) states that adding intermediary
lemmas does not change the set of theorems of the logical system. That
is, the lemma $A$ is not needed in the proofs of the system. This
implies that, if $B$ is provable from the hypothesis $C$, then there
exists a direct (cut-free) proof of this fact.

The cut-rule may take different forms depending on the logical system
at hand. For concreteness, consider the following cut-rule for the
system $\Gtip$:

\begin{equation}\label{eq-cut-g3ip}
\infer[Cut]{\Gamma \vdash B}{
\deduce{\Gamma \vdash A}{}
&
\deduce{\Gamma, A \vdash B}{}
}
\end{equation}

This rule has an {\em additive} flavor: the context $\Gamma$ is shared by
the premises.  Later, different cut-rules will be considered,
including multiplicative-like cut-rules (splitting the context
$\Gamma$ in the premises) and also cut-rules for one-sided sequent
systems.

Gentzen's style proof of admissibility of the
cut-rule~\cite{gentzen35} generally follows by proving that top-most
applications of cut can be eliminated.  This is usually done via a
nested induction on the {\em complexity} of the cut-lemma $A$ and
sub-induction on the {\em cut-height}, i.e., the sum of the heights of
the derivations of the premises.  In the following, the rationale
behind this cut-elimination procedure will be formalized by means of
rewrite-based conditions. The next section discusses how these
conditions become a semi-automatic procedure for checking the
cut-elimination property for different logical systems.
 
The analyses in Section \ref{sec.check} showed how to inductively
reason on the height of derivations inside the rewriting logic
framework.  But what about the induction on the complexity of
formulas?  In general, it is possible to inductively reason about
terms built from algebraic specifications.  The unification of a
sequent against the conclusion of an inference rule $r_s\in \scal$
uniquely determines the active formula $F$ introduced by $r_s$. Since
terms in the sort \lstinline{Formula} are inductively generated from
the constructors of that sort, special attention can be given to the
sub-terms (if any) of $F$ since those are the \textit{only} candidates 
that are useful to build the needed inductive hypothesis in
the cut-elimination proof. Such sub-terms are called {\em auxiliary
  formulas}.

Gentzen's procedure consists of reducing topmost cut-applications to
the atomic case, then showing that these cuts can be eliminated. The
reduction step to be performed depends on the status of the
cut-formula in the premises: {\em principal} on both or {\em
  non-principal} on at least one premise.  This procedure is formalized next.

\paragraph{Principal Cases.} If the cut-formula is principal (see
Definition~\ref{def:seqsys}) in both premises of the cut-rule, then
induction on the {\em complexity} of the cut-formula is applied. For
instance, consider the case when the cut-formula is $A\wedge B$:

\noindent\resizebox{\textwidth}{!}{
$
\begin{aligned}
\infer[Cut]{\Gamma \SEQ C}{
  \infer[\wedge_R]{\Gamma \seqn{s(n)} A \wedge B}{
   \deduce{\Gamma \seqn{n} A}{}
   &
   \deduce{\Gamma \seqn{n} B}{}
  }
  &
  \infer[\wedge_L]{\Gamma, A\wedge B \seqn{s(m)} C}{
   \deduce{\Gamma, A,  B \seqn{m} C}{}
  }
}
\end{aligned}
\begin{aligned}
\qquad\leadsto\qquad
\end{aligned}
\begin{aligned}
\infer[Cut]{\Gamma \SEQ C}{
 \deduce{\Gamma \SEQ A}{}
 &
 \infer[Cut]{\Gamma, A \SEQ C}{
  \infer[W]{\Gamma,A \SEQ B}{
   \deduce{\Gamma \SEQ B}{}
  }
  &
  \deduce{\Gamma,A, B \SEQ C}{}
 }
}
\end{aligned}
$}\\

\noindent
Both applications of the cut-rule in the right-hand derivation are on
smaller formulas and then induction applies. Note that this kind of
reduction does not necessarily preserve the height of the
derivation. Hence, no height-annotation can be used on the resulting
premises.  Note also that weakening is needed in order to match the derivation leaves.

If the cut-formula $A\wedge B$ is frozen (making $A,B$ constants of
type \lstinline{Formula}), then the inductive reasoning is formalized
with the following rules:
\begin{center}
\begin{lstlisting}
rl [cutF] : Gamma |-- C => (Gamma |-- cA) | (Gamma, cA |- C)
rl [cutF] : Gamma |-- C => (Gamma |-- cB) | (Gamma, cB |- C)
\end{lstlisting}
\end{center}
Here $cA, cB$ are constants representing the sub-terms of the ground
term $\overline{A \wedge B}$. The other terms are variables.  More
generally, if the cut formula is a term of the form $f(t_1,...,t_n)$,
each $t_i$ of sort \lstinline{Formula} gives rises to a different
rule.  In the case of constants (e.g., \lstinline{False}) and atomic
propositions (no sub-terms of sort \lstinline{Formula}), the set of
generated rules is empty.

\paragraph{Non-principal Cases.}  The non-principal cases in the
proof of cut-elimination require permuting up the application of the
cut-rule w.r.t. the application of an inference rule, thus reducing
the {\em cut-height}. As an example, assume that the left premise of
the cut-rule is the conclusion of an application of the rule
$\wedge_L$. Hence, it must be the case that the antecedent of the
sequent contains a conjunct $F\wedge G$. The reduction is as
follows:

\noindent\resizebox{\textwidth}{!}{
$
\begin{aligned}
\infer[Cut]{\Gamma, F\wedge G \SEQ B}{
  \infer[\wedge_L]{\Gamma, F\wedge G \seqn{s(n)} A}{
   \deduce{\Gamma, F,G \seqn{n} A}{}
  }
  &
  \deduce{\Gamma, F\wedge G,A \seqn{s(m)} B}{}
}
\end{aligned}
\begin{aligned}
\qquad\leadsto\qquad
\end{aligned}
\begin{aligned}
\infer[\wedge_L]{\Gamma, F\wedge G \SEQ B}{
 \infer[Cut]{\Gamma, F, G \SEQ B}{
  \deduce{\Gamma, F,G \seqn{n} A}{}
  &
  \deduce{\Gamma, F, G,A \seqn{s(m)} B}{}
 }
}
\end{aligned}
$}

\noindent
Permuting up cuts results in an application of the cut-rule on shorter
derivations. The top-rightmost sequent in the right-hand side
derivation is deduced via the height-preserving invertibility of the
$\wedge_L$ rule and the fact that $\Gamma, F\wedge G,A \seqn{s(m)} B$
is provable.  Similar reductions are possible when the cut formula $A$
is not principal in the right premise of the cut-rule.

If the height of the premises in the above derivation is frozen, it is
possible to extend the theory $\rcal_\scal$ with two new rules that
exactly mimic the behavior of replacing a non-principal cut with a
smaller one.  Those rules are:

\begin{lstlisting}

rl [CutH] : Gamma |-- G  => (cn : Gamma |-- cA) | (s(cm) : Gamma, cA |-- G) .
rl [CutH] : Gamma |-- G  => (s(cn) : Gamma |-- cA) | (cm : Gamma, cA |-- G) .
\end{lstlisting}

\noindent
where $cn$ (resp., $cm$) is the frozen constant resulting from $n$
(resp., $m$) and $cA$ the ground term representing the cut-formula.
The first rule reflects the principle behind the above reduction where
the height of the left premise of the cut-rule is reduced. The second
rule reflects the case where the cut-formula is not principal in the
right premise. Note that the first rule cannot be applied directly on
the sequent $\Gamma, F\wedge G \SEQ B$: the left premise $\Gamma, F
\wedge G \SEQ A$ is provable with height $s(n)$ but, not necessarily,
with height $n$.  Similarly for the second rule and the right premise
of the cut-rule.

\subsection{Cut Elimination}
\label{sec.cut.necond}

The scenario is ready for establishing the necessary conditions for
cut-elimination. The specification of the additive cut-rule for the
system $\Gtip$ can be written as

\begin{lstlisting}

rl [cut] : Gamma |-- C => (Gamma |-- A) | (Gamma, A |-- C) [nonexec] .
\end{lstlisting}

This rewrite rule has an extra variable ($A$) in the right-hand side
and cannot be used for execution (unless a strategy is
provided). Hence, the attribute \lstinline{nonexec} identifies this
rule as no executable. In the following, \lcut\ (resp., \rcut) is used
to denote the term \lstinline{Gamma |-- A} (resp.,
\lstinline{Gamma, A |-- C})
whose set of variables is \lstinline|{Gamma, A}| (resp.,
\lstinline|{Gamma, A, C}|). Moreover, \hcut\ denotes the
head/conclusion of the cut-rule, i.e., the term
\lstinline{Gamma |-- C}.

As already illustrated, admissibility results and invertibility lemmas
are usually needed in order to complete the proof of
cut-elimination. Such auxiliary results are specified in the analysis,
respectively, as $\mathcal{H}$ and $\mathcal{I}$. As explained in
Sections \ref{sec:adm-cond} and \ref{sec:inv-cond}, a rule in
$\mathcal{H}$ encodes an admissible rule of the system that can be
used, in a backward fashion, during the proof-search
procedure. Moreover, a rule in $\mathcal{I}$ is used to close the
assumptions under the invertibility results.

Given a logical system $\scal$, every possible derivation of the
premises of the cut-rule should be considered.  This means that there
is a proof obligation for each $r_s, r_t \in \scal$ s.t. $r_s$ (resp.,
$r_t) $ is applied on the left (resp., right) premise (i.e., when
$r_s$ matches \lcut\ and $r_t$ marches \rcut). More precisely:

\begin{definition}[Local cut-elimination]\label{def.cond.cut}
   Let $\scal$ be a sequent system, and $\mathcal{H}$ and
   $\mathcal{I}$ a set of rules. Let
   \[\infer[r_s]{s(n): S}{n:S_1 ~\cdots~
     n:S_m}
     \quad
     \infer[r_t]{s(k): T}{k:T_1 ~\cdots~ k:T_n}
     \]
be inference rules in
   $\scal$. Under the assumptions $\mathcal{H}$ and $\mathcal{I}$, the cut rule is admissible relative to $r_s$ and $r_t$ iff for each  $\theta \in \textit{CSU}(S,\lcut)$ and $\gamma \in \textit{CSU}(T,\rcut\theta)$:
   \[
   \begin{array}{rll}
   \mathcal{H} \cup
	\rcal_\scal 
 \cup \textit{ind-F} \cup \textit{ind-H}  ~\cup
 \\
\closure{\{\overline{(n:S_j)\gamma}, \overline{S_j\gamma} \mid j \in 1  ..  m \}
   \cup \{\overline{(n:T_j)\gamma}, \overline{T_j\gamma} \mid j \in 1  ..  n \}}{\mathcal{I}}
 & \DEDR & 
 \overline{\hcut\gamma}
 \end{array}
   \]
where the variables in $S$ and $T$ are assumed disjoint and
   \[
   \begin{array}{lll}
   \textit{ind-F} &=&  \left\{ (\hcut \to \lcut ~|~ \rcut ) [t/A] ~\mid~ t \prec \overline{A\gamma}  \mbox{~ and $t$ has sort \texttt{Formula}}\right\}  \\
   \textit{ind-H} &=& \{ \hcut \to (\overline{n\gamma}:\lcut ~|~ \overline{s(n)\gamma}:\rcut)[ \overline{A\gamma}/A] \} \cup \\
   &&\{ \hcut \to (\overline{s(n)\gamma}:\lcut ~|~ \overline{n\gamma}:\rcut)[ \overline{A\gamma}/A] \} 
   \end{array}
   \]
\end{definition}

The rules in $\scal$ are extended with axioms corresponding to the
sequents resulting after the application of $r_s$ and $r_t$ on the
premises of the cut-rule.  Note that those premises are added with and
without the height annotations, which 
can be used in
applications of inductive hypothesis with shorter derivations and
(non-height preserving) applications of the cut-rule
over simpler formulas, respectively. Note also that the set of axioms is closed
under applications of the rules in $\mathcal{I}$.

The set of rules in \textit{ind-F} specifies a valid application of
the cut-rule with sub-terms of the cut-formula ($t \prec
\overline{A\gamma}$). As usual, this rule is assumed to be universally
quantified on the remaining variables ($t$ is a ground term).  On the
other hand, the set of rules in \textit{ind-H} specifies a valid
application of the cut-rule with shorter derivations. As discussed
previously, two cases need to be considered: when the left and right
premises are shorter. In both cases, the height of the derivation is
fixed ($\overline{n\gamma}$) as well as the cut-formula ($[
  \overline{A\gamma}/A]$).

Regarding the base-cases of the induction, if the cut-formula is a
constant or an atomic proposition, the set $\textit{ind-F}$ is
empty. If the rule $r_s$ is an axiom, then the set
$\left\{\overline{n:S_j\gamma}, \overline{S_j\gamma} \mid j \in 1 ..
m \right\}$ is also empty. In this case, an attempt of proving
$\overline{\hcut\gamma}$ by starting with the first rule in
$\textit{ind-H}$ leads to a goal of the form
$\overline{n\gamma}:\texttt{lcut}$ where no inference rule $r\in
\scal$ can be applied: $\overline{n\gamma}$ is a constant of the form
$\%1$ and it does not unify with the annotation $s(n)$ in the
conclusion of $r$.  Hence, if $r_s$ is an axiom, a proof of
$\overline{\hcut\gamma}$ cannot start with $\textit{ind-H}$. A similar
analysis applies for $r_t$.

The cut-elimination proof needs to consider all the possible matchings
of the rules of the system and the premises of the cut-rule.

\begin{theorem}\label{thm.cut.elim}
  Let $\scal$ be a sequent system, and $\mathcal{H}$ and $\mathcal{I}$
  be set of rules.  If for each $r_s$ and $r_t\in \scal$ the cut-rule
  is admissible relative to $r_s$ and $r_t$ under the assumptions
  $\mathcal{H}$ and $\mathcal{I}$, then the cut-rule is admissible in
  $\scal$ (relative to $\mathcal{H}$ and $\mathcal{I}$).
\end{theorem}

\begin{proof}
  Consider the annotated cut-rule below (a similar analysis applies
  for the other cut-rules introduced in Section \ref{sec.case}):
\[
\infer[Cut]{\Gamma \vdash B}{
\deduce{\Gamma \seqn{s(n)} A}{}
&
\deduce{\Gamma, A \seqn{s(m)} B}{}
}
\]
Assume that there exists a derivation of the premises starting,
respectively with the rules $r_s$ and $r_t$.  By the hypothesis and
Definition \ref{def.cond.cut}, there is a valid derivation of the
sequent in the conclusion and the result follows.
\end{proof}

\subsection{Identity Expansion}
\label{sec.cut.idexp}

The identity axiom states that any atomic formula is a consequence of
itself.  It has a dual flavor w.r.t. the cut-rule, in the sense that,
while in the cut-rule a formula is eliminated, in the identity axiom
an atomic formula is introduced. In $\Gtip$, $I$ represents the
general schema for the identity axiom (see Fig. \ref{fig:G3ip}), where
$p$ is an atomic formula.

In Section~\ref{sec.cut.necond}, the cut-rule was proved admissible
relative to some assumptions. The correspondent dual property
w.r.t. the identity axiom is the {\em identity expansion}: assuming
$I$, is {\em any} well founded formula a consequence of itself? Or
equivalently: Is the general identity axiom -- not restricted to atoms
-- admissible? For example, in $\Gtip$, proving identity expansion
requires proving the admissibility of the axiom $I_A$ for an arbitrary
formula $A$:
\[
\infer[I_A]{\Gamma, A \vdash A}{}
\] 
Observe that the cut/identity duality is also reflected during
(bottom-up) proof-search: while applications of the cut-rule should be
avoided because arbitrary formulas need to be produced ``out of thin
air'', applications of general identity axioms are most welcome, since
they may make the proof smaller and proof-search simpler.

The proof of identity expansion proceeds by induction on the formula
$A$. Consider a constructor $C=f(\vec{t})$ of type
$S_1,...,S_n \to \texttt{Formula}$ and let $\overline{C}$ be
the ground term $f(\overline{\vec{t}})$ where each $t_i \in \vec{t}$
is replaced by a fresh constant $\vec{t_i}$ of sort
$S_i$. Hence, the goal is to prove $\overline{C} \vdash
\overline{C}$ assuming that $\overline{{t_i}} \vdash \overline{{t_i}}$
is provable for each $t_i\in \vec{t}$ of sort \lstinline{Formula}. 

\begin{theorem}\label{thm.cut.idexp}
Let $\scal$ be a propositional sequent system and $\rcal_\scal$, with
signature ${(S,\leq,F)}$, be the resulting rewrite theory encoding
$\scal$.  Identity expansion holds in $\scal$ iff for each symbol $f
\in F$ of type $S_1,...,S_n \to \texttt{Formula}$ ($n\geq
0$),

 \[
\rcal_\scal  \cup \{ t' \vdash t' \mid t' \prec \overline{f(\vec{t})} \mbox{ and $t'$ has sort \texttt{Formula}} \}
  \DEDR  
 (\overline{f(\vec{t}) \vdash f(\vec{t})})
   \]
   
\end{theorem}

In one-sided systems, the goal is to show that $\vdash A, A^\perp$ is
provable for any $A$, where $A^\perp$ denotes the {\em dual} of $A$
(see Section \ref{ex:ll} for the definition of duality).
Theorem~\ref{thm.cut.idexp} can be easily adapted to the one-sided case.


\section{Reflective Implementation}\label{sec.rew}

This section details the design principles behind the
$\toolname$\footnote{\url{https://carlosolarte.github.io/L-framework/}},
a tool implementing the narrowing procedures described in
Sections~\ref{sec.check} and \ref{sec:cut}.  The $\toolname$ receives
as input the object logic sequent system (OL) to be analyzed, as a
rewrite theory $\rcal_\scal$, and the specification of the properties
of interest. Then, it outputs \LaTeX\ files with the results of the
analyses of, e.g., the proof reductions needed to establish
cut-elimination.  The specification of the OL follows as in
Section~\ref{sec.framework} (details can be found in Section
\ref{sec:ss-maude}).  As explained in Section \ref{sec:reflection},
the implementation of the algorithms heavily relies on the reflective
capabilities of rewriting logic.  Moreover, the specification of the
properties for each kind of analysis follows a similar pattern where a
suitable functional theory needs to be instantiated (Section
\ref{sec:gen-impl}).  The subsequent sections offer further details
about the implementation of each kind of analysis: admissibility of
rules (Section \ref{sec:adm-analysis}), invertibility of rules
(Section \ref{sec-impl-inv}), cut elimination (Section
\ref{imp:cut-elim}), and identity expansion (Section
\ref{sec-impl-id-exp}).

For readers interested in the details of the implementation, pointers
to the Maude files and definitions are given in this section. However,
readers interested only in the results of the analyses can safely skip
this section; several examples and instances of the definitions
presented here can be found in Section~\ref{sec.case}.

\subsection{Sequent System Specification}\label{sec:ss-maude}

The starting point for a sequent system specification is the
definition of its syntax and inference rules.  The file
\lstinline{syntax.maude} contains the functional module (i.e., an
equational theory and no rewriting rules) \lstinline{SYNTAX-BASE}.
Such a module defines the sorts \lstinline{Prop} and
\lstinline{Formula}, as well as the subsort relation
\lstinline{Prop < Formula}.
No constructors for these sorts are given since those depend on the OL
and hence must be provided.  The sort \lstinline{MSFormula}, for
multiset of formulas, is pre-defined here with the constructors
presented in Section \ref{sec.framework}: \lstinline{op *} denoting
the empty multiset and \lstinline{op _;_ } for multiset union.  Some
auxiliary functions, needed to produce \LaTeX\ outputs, are declared
in this module. In particular, the OL may define a mapping to replace
symbols of the syntax with \LaTeX\ macros, e.g.,
\begin{lstlisting}

eq TEXReplacement = ('|-- |-> '\vdash), ('/\ |-> '\wedge) , ('\/ |-> '\vee) ...
\end{lstlisting}

The \lstinline{sequent.maude} file defines the functional module
\lstinline{SEQUENT} with the sort \lstinline{Sequent} and the
following constructors:

\begin{lstlisting}

op  |--_ :           MSFormula -> Sequent . --- one sided
op _|--_ : MSFormula MSFormula -> Sequent . --- two sided
\end{lstlisting}

This syntax should suffice for most of the sequent-based inference
systems. As shown in Section~\ref{sec.case}, it is also possible to
provide more constructors to deal, for instance, with dyadic systems (i.e.,
one-sided sequents with two separated contexts).

Internally, the tool annotates sequents with inductive measures. For
this purpose, the constructor
\begin{lstlisting}

op _:_ : INat Sequent -> Sequent .          --- height annotations
\end{lstlisting}
is added to the specification. The sort \lstinline{INat} (file
\lstinline{nat-inf.maude}) extends the natural numbers expressed in
Peano-like notation as $s^n(z)$ with a constant \lstinline{inf}
denoting ``unknown''. This constant is used, e.g., in the
cut-elimination procedure where structural cuts do not
preserve/decrease the height of the derivation and, therefore, the
resulting sequent does not have any (known) measure.

OLs are also allowed to define equations to complete the definition of
the mapping \lstinline|TEXReplacementSeq| that replaces the name of
the rules with suitable \LaTeX\ symbols:

\begin{lstlisting}

eq TEXReplacementSeq = ('AndL |-> '\wedge_L), ('AndR |-> '\wedge_R) ...
\end{lstlisting}

The sort \lstinline{Goal}, the subsort relation
\lstinline{Sequent < Goal}, the constructors \lstinline{op _|_}, and
\lstinline{proved} (for building list of sequents) are also defined in
the module \lstinline{SEQUENT}.

The \lstinline{sequent.maude} file also specifies the module
\lstinline{SEQUENT-SOLVING} with auxiliary procedures for building
derivation trees and outputting \LaTeX\ code. This module uses
reflection heavily (more on this in the next section) in order to
deal, in a general and uniform way, with the representation of any
sequent regardless of its specific syntax. Moreover, Maude's module
\lstinline{LEXICAL} is used for converting between strings and lists
of Maude's quoted identifiers (terms of the form \lstinline{'identifier} with sort \lstinline{Qid}). As shown below, 
Qids materialize the meta-representation of
any term.

\subsection{Reflection and the Core System}
\label{sec:reflection}

A reflective logic is a logic in which important aspects of its
meta-theory can be represented consistently at the object level.  In a
nutshell, a reflective logic is a logic that can be faithfully
interpreted in itself. In this way, the object-level representation
can correctly simulate the relevant deductive aspects of its
meta-theory. Maude's language design and implementation make
systematic use of the fact that rewriting logic is reflective, making
the meta-theory of rewriting logic accessible to the user as a
programming module~\cite{DBLP:journals/tcs/ClavelDELMMQ02}.

For the purpose of this paper, the focus is on two meta-theoretic
notions, namely, those of theory/module and the deductive entailment
relation $\Vdash$. Formally, there is an \textit{universal} rewrite
theory $\ucal$ in which any finitely represented rewrite theory
$\rcal$ can be represented as a term $\metal{\rcal}$, including
$\ucal$ itself, any terms $t,u$ in $\rcal$ as terms
$\metal{t},\metal{u}$, respectively, and a pair $(\rcal, t)$
as a term $\langle \metal{\rcal}, \metal{t}~\rangle$ in such a
way that the equivalence 
$ \rcal \Vdash t \rews u \quad \Leftrightarrow \quad \ucal \Vdash \langle
\metal{\rcal}, \metal{t}~\rangle \rews \langle
\metal{\rcal}, \metal{u}~\rangle$ holds. 
Since $\ucal$ is representable in itself, a ``reflective tower'' can
be achieved with an arbitrary number of levels of reflection.
%

In general, simulating a single step of rewriting at one level
involves many rewriting steps one level up. Therefore, in naive
implementations, each step up the reflective tower comes at
considerable computational cost. In Maude, key functionalities of the
universal theory $\ucal$ have been efficiently implemented in the
functional module \lstinline{META-LEVEL}, providing ways to perform
reflective computations.

Additional utilities for manipulating modules and terms in the
$\toolname$ are implemented in the \lstinline{meta-level-ext.maude}
file. For instance, all the operations on theories described in
Section~\ref{sec.check} are contained there (some of them are detailed
below).

The module \lstinline{META-LEVEL} implements the so called \emph{descent}
functions that manipulate (meta-) terms.  The function
\lstinline{op upTerm : Universal -> Term .}  returns the (meta)
representation $\metal{t}$ of a term $t$. For example, constants are
represented as \lstinline{Id.Type} (e.g., \lstinline{'proved.Goal}),
variables as \lstinline{Id:Type} (e.g., \lstinline{'F:Formula}),
and functions as
\lstinline{Id[Params]} (e.g., the term
\lstinline{'_;_[Gamma:MSFormula,Delta:MSFormula]} represents the
multiset of formulas \lstinline{Gamma ; Delta}).

From a well-formed (meta-) term $\metal{t}$ it is possible to recover
the term $t$ (one level down). Note, however, that not all meta-terms
have a suitable representation in the theory in the level below, for
instance, \lstinline{'_|_[S:Sequent, F:Formula]} is not the image of any
valid sequent or formula in the module specifying the system
$\Gtip$.
The function \lstinline{op downTerm : Term Universal -> Universal}
takes as parameter the (meta) representation of a term $\metal{t}$ and
a term $t'$. It returns the canonical form of $t$ if it is a term
having the same sort of $t'$; otherwise, it returns $t'$.  Usually,
$t'$ is an error term used to denote that the descent translation was
not possible. For instance, if $\metal{t}$ is expected to be the
meta-representation of a formula, then $t'$ can be the constant
\lstinline{op error : -> Formula .}

At the meta-level, modules in Maude (i.e., rewrite theories) are
represented as terms with sort \lstinline{Module}.  The function
\lstinline{upModule} can be used to obtain such a term. All the
components of a module (sorts, functions, equations and rules) have a
suitable sort and representation in the meta-level, thus making them
first-class citizens. For instance, most of the analyses require to
extend the sequent theory with new rules. The function below adds to
the module \lstinline{M} a set of rules \lstinline{RS}:
\begin{lstlisting}

op newModuleRls : Module RuleSet -> Module .
eq newModuleRls(M, RS) 
  = (mod 'NEW-MOD is
      getImports(M)    --- Sorts, imports, equations, etc remain the same
      sorts getSorts(M) .
      getSubsorts(M) getOps(M) getMbs(M) getEqs(M)
      (getRls(M) RS)   --- Union of the rules of M and RS
     endm ) .
\end{lstlisting}

Rules at the meta-level are terms of sort \lstinline{Rule} built from the
constructors

\begin{lstlisting}

op rl_=>_[_]. : Term Term AttrSet -> Rule .
op crl_=>_if_[_]. : Term Term Condition AttrSet -> Rule . --- conditional rule

\end{lstlisting}

The parameter of sort \lstinline{AttrSet} is a set of attributes of the
rule, including, e.g., the label used as identification
(\lstinline{[label('AndR)]}).  This representation opens the
possibility of manipulating rules in a simple way.  For instance,
consider a sequent rule $r = (S \to S')$, for some conclusion $S$ and
premise $S'$, to be proved height-preserving admissible. Consider also
that the current goal is to show that a given ground sequent is
provable with height at most \lstinline{s(h)}. A call to the function below
with parameters \lstinline{h} and $r$ produces a restricted version of $r$
where it can be applied only on sequents annotated with (the constant
height) \lstinline{h}:

\begin{lstlisting}

op inductive-rule : GroundTerm Rule -> Rule .
eq inductive-rule(gt, rl '_:_['n:FNat , S] => '_:_['n:FNat , S'] [ AS ].) =
                      rl '_:_[gt ,      S] => '_:_[gt ,      S'] [ AS ].
\end{lstlisting}

The module \lstinline{META-LEVEL} offers functions to perform Maude's
operations at the meta-level.  For instance, the function
\lstinline{metaRewrite} allows for rewriting the meta-representation
of a term in a given module and \lstinline{metaVariantDisjointUnify}
to solve unification problems.  Moreover, given a theory $\rcal$ and
two terms $t,u$ in that theory, \lstinline{metaSearchPath} allows for
checking the entailment $\rcal \Vdash t \rews u$ by testing whether
$\ucal \Vdash \langle \metal{\rcal}, \metal{t}\rangle \rews \langle
\metal{\rcal}, \metal{u}\rangle$.  For instance, in the following
assignment:

\begin{lstlisting}

Ans := metaSearchPath(M, SGoal, upTerm(proved), nil, '*, bound-spec, 0) .
\end{lstlisting}

\noindent
the term \lstinline{M} is the meta-representation of a sequent system;
\lstinline{SGoal} is a term representing the goal/sequent to be proved;
\lstinline{nil} specifies that there are no additional conditions to
be satisfied by the solution; \lstinline{'*} means that the reduction
may take 0 or more steps; \lstinline{bound-spec} indicates the maximum
depth of the search; and the final $0$ is the solution number. The
term \lstinline{Ans} is a term of type \lstinline{Trace?}. It can be
\lstinline{failure} or a list of trace steps showing how to perform
each of the reductions $t \rews_r t'$ where the rule $r$ is applied to
$t$ leading to $t'$.  It is worth noticing that \lstinline{metaSearchPath}
implements a breadth-first search strategy. Hence, if \lstinline{SGoal} can
be rewritten into \texttt{proved} with at most \lstinline{bound-spec}
steps, then \lstinline{metaSearchPath} will eventually find such proof (if
the executability conditions for the rewrite theory are met
\cite{clavel-maudebook-2007}).  Moreover, since \lstinline{Ans} is a proof
term evidencing how to prove the rewriting
\lstinline{SGoal}$\stackrel{*}\rews$\lstinline{proved}, it can be used to rebuild the
needed derivation of \lstinline{SGoal} in the sequent system at hand.

\subsection{The General Approach for Implementing the Algorithms}\label{sec:gen-impl}

All the analyses implemented in \toolname\ are instances of the same
template:
\begin{enumerate}
 \item A module interface (called \emph{theory} in Maude) specifies
   the input for the analysis. This theory includes parameters such
   as: the name of the module implementing the OL; the specification,
   as a rewrite rule, of the theorem to be proved; the extra
   hypotheses (set of rewrite rules) corresponding to the already
   proved theorem; the bound for the search depth; etc.
 \item A module implementing the decision procedures proposed in
   Sections \ref{sec.check} and \ref{sec:cut}. All algorithms follow
   the same principles:

   \begin{enumerate}
   \item A function \lstinline{generate-cases} uses unification to
     generate all the proof obligations to be discharged. In each kind
     of analysis, there are suitable sorts and constructors to
     represent the proof obligations. Normally, the cases include the
     terms denoting the premises (that can be assumed to be provable)
     as well as the goal to be proved.
   \item Auxiliary definitions to extend the theory with new axioms
     and the right inductive hypothesis. Take for instance the
     function \lstinline{inductive-rule} explained above.

 \item A function \lstinline{holds?}  that receives as parameter one
   of the proof obligations, uses the functions in (b) to extend the
   theory and calls to \lstinline{metaSearchPath} to check if the goal
   can be proved. 
\end{enumerate}

 \item An extra module providing facilities to produce the \LaTeX\ output.
\end{enumerate}

The following sections give some details about the components (1) and
(2) for each kind of analysis.  Common facilities for all the analyses
are implemented in the \lstinline{theorem-proving-base.maude} file,
including: converting the term $t$ into the ground term $\overline{t}$
(i.e., replacing variables with fresh constants) and generating axioms
(rules of the form \lstinline{rl [ax] T => proved .}) from the
assumptions of the theorem.

\subsection{Admissibility Analysis} \label{sec:adm-analysis}

The core procedures for automating the proof of admissibility
theorems, following the definitions in Section~\ref{sec:adm-cond}, are
specified in \lstinline{admissibility.maude}. The input point is the
definition of the \emph{functional theory} below.
\begin{lstlisting}

fth ADMISSIBILITY-SPEC is
 pr META-LEVEL .
 op th-name    : -> String .  --- Name of the theorem
 op mod-name   : -> Qid .     --- Module with the OL specification
 op file-name  : -> String .  --- Output file
 op rule-spec  : -> Rule .    --- Rule to be proved admissible (goal => premise)
 op bound-spec : -> Nat .     --- Depth-search 
 --- Already proved admissibility lemmas (set of rules)
 op already-proved-theorems : -> RuleSet .
 --- Identifiers of the height preserving invertible rules
 op inv-rules  : -> QidList .
 --- Mutual inductive rule (for mutual inductive proofs).     
 op mutual-ind : GroundTerm -> RuleSet .
endfth

\end{lstlisting}

Theories in Maude are module interfaces used for declaring
parameterized modules. Such interfaces define the syntactic and
semantic properties to be satisfied by the actual parameter modules
used in an instantiation~\cite{DBLP:journals/tcs/ClavelDELMMQ02}.

The name of the theorem is specified with the string
\lstinline{th-name} (e.g., ``Admissibility of weakening.'').  The
identifier of the module \lstinline{mod-name} (e.g.,
\lstinline{'G3ip}) is used to obtain the meta-representation of the OL
module defining the syntax and inference rules of the sequent system.
The field \lstinline{file-name} specifies the output (\LaTeX) file.

The theorem to be proved is specified via a rewriting rule (a term of
sort \lstinline{Rule}). As an example, the height-preserving
admissibility of weakening for   $\Gtip$ is:
\begin{lstlisting}

eq rule-spec =
  rl '_:_['n:FNat,  '_|--_['_;_['Gamma:MSFormula, 'F:Formula], 'G:Formula]] =>
     '_:_['n:FNat,  '_|--_['Gamma:MSFormula,'G:Formula]] 
     [ label('W) ]. ) .
\end{lstlisting}

\noindent This term is the meta-representation of the rule

\begin{lstlisting}

rl [W] n : Gamma, F |-- G => n : Gamma |-- G .
\end{lstlisting}

Note that the premise and the conclusion have the same height,
specifying that if the premise is provable with height at most $n$ so
is the conclusion.  Since the entailment relation is undecidable in
general, all the analyses are performed up to a given search depth
(field \lstinline{bound-spec}).  Hence the procedures are sound (in
the sense of the theorems in Sections \ref{sec.check} and
\ref{sec:cut}), but not complete.

As already explained, for modularity, the analyses can depend on
external lemmas.  Those auxiliary results are specified as
\lstinline{already-proved-theorem} (backward reasoning) and
\lstinline{inv-rules} (forward reasoning).  Finally, some theorems
require mutual induction. For instance, admissibility of contraction
for the system of classical logic (Section \ref{case:gtcp}) requires a
mutual induction on the the right and left rule for contraction.  The
field \lstinline{mutual-ind} specifies the other theorems that can be
applied on shorter derivations.  For that, if the parameter of type
\lstinline{GroundTerm} is of the form \lstinline|'s[h.Nat]|, the
mutual-theorem is instantiated with sequents of (constant) height
\lstinline{h}.


The main procedures implementing the analysis of admissibility of
rules can be found in the parametric module
\lstinline|ADMISSIBILITY-ALG{SPEC :: | \lstinline|ADMISSIBILITY-SPEC}|.
Consider, for instance, the task of proving the admissibility of a
rule \lstinline{rs}. Such a rule is specified as the parameter
\lstinline{rule-spec} in \lstinline{SPEC}.  For each rule \lstinline{rt} of the
system, proof obligations are generated by the module's
functionality. Following Definition \ref{def.check.admis}, this is
done by unifying the premise/body of \lstinline{rs} with the
conclusion/head of \lstinline{rt}:
\begin{lstlisting}

U :=  metaVariantDisjointUnify(module, 
	 getBody(rule-spec, module) =? getHead(rt, module), empty, 0, N) .
\end{lstlisting}

\noindent
Here, \lstinline{N} is a natural number used to enumerate the
unifiers. By identifying the unifier \lstinline{U}, it is possible to
obtain the resulting premises when \lstinline{rt} is applied on the
body/premise of \lstinline{rs} (see Equation \eqref{eq:der1}).  For that,
the descent function \lstinline{metaXapply} enables the application of a
rule to a term according to a given substitution. Computing the ground
term that substitutes the variables by fresh constants is a routine
exercise by following the inductive definition of the sort
\lstinline{Term}.

The cases for admissibility take the form 
\lstinline{adm-case(Q, M, GTC, GTP, GG)}, where:
\lstinline{Q} is the identifier of the rule \lstinline{rt} in Definition
\ref{def.check.admis}; \lstinline{M} is the module implementing the OL,
\lstinline{GTC} and \lstinline{GTP} correspond to the conclusion and the
premises in Equation \eqref{eq:der1}, and \lstinline{GG} is the goal to be
proved ($\overline{(i:S) \theta}$ in Definition
\ref{def.check.admis}).

A useful mnemonic that applies from now on: \lstinline{G} is for
\emph{ground}, \lstinline{T} for \emph{term}, \lstinline{C} for
\emph{conclusion}, \lstinline{P} for \emph{premise(s)}, and the last
\lstinline{G} in \lstinline{GG} for \emph{goal}.
 
The module \lstinline{M} above is extended as follows:

\begin{lstlisting}

M' := newModuleRls(M, 
  inductive(getHeight(GG))        --- theorem instantiated on shorter derivations 
  mutual-ind(getHeight(GG))
  premises(GTP)                   --- all premises in GTP as axioms
  inv-premises(M, inv-rules, GTP) --- closing w.r.t invertible rules
  already-proved-theorems ) .
\end{lstlisting}

The function \lstinline{inductive} takes as parameter the annotated height
of the current goal \lstinline{GG}. If it is of the form \lstinline{s(h)},
\lstinline{rule-spec} is instantiated with the height \lstinline{h} (as
explained in the Section~\ref{sec:reflection}). Otherwise,
\lstinline{inductive}  and  \lstinline{mutual-ind}  
 return \lstinline{none} (an empty list of rules). The  function  call 
\lstinline{premises(GTP)} 
generates, for each sequent \lstinline{S} in the set of premises \lstinline{GTP}, 
a new axiom rule (\lstinline{S => proved});  \lstinline{inv-premises} generates further axioms  by applying the invertible rules \lstinline{inv-rules} in the
premises \lstinline{GTP}; and \lstinline{already-proved-theorems} adds already
proved theorems (specified as rules).

From the extended theory \lstinline{M'}, \lstinline{metaSearchPath} is used to
check whether \lstinline{GG} can reach \lstinline{proved} and determine the status of the current proof obligation. 

\subsection{Invertibility of Rules}\label{sec-impl-inv}

The core procedures are in \lstinline{invertibilty.maude}.  The input
to the invertibility analysis is specified as a realization of the
functional theory \lstinline{INV-SPEC}:
\begin{lstlisting}

fth INV-SPEC is
    pr META-LEVEL .
    op th-name : -> String .    --- Name of the theorem
    op mod-name : -> Qid .      --- Module with the OL specification
    op bound-spec : -> Nat .    --- Bound of the search procedure
    op file-name : -> String .  --- File name to write the output
    op already-proved-theorems : -> RuleSet . --- Admissible rules
endfth


\end{lstlisting}

There is no need to specify the theorem to be proved, because it
suffices to take each of the rules of the input module and ``flip'' it
in order to obtain the invertibility lemma to be proved (see
Definition \ref{def:inv}).  This procedure outputs a \LaTeX\ file with
the invertibility status of each rule in the system. In the case of
rules with two premises, each premise is analyzed separately (see
Def. \ref{def.check.invert}). This allows for proving, e.g.,
that the rule $\cimp_L$ in the system $\Gtip$ is invertible only in
its right premise (see Section \ref{ex:g3i}).

Following the same recipe for admissibility, the proof obligations for
invertibility are generated by solving unification problems. Assume
that the rule being analyzed is \lstinline{Q} and it is to be tested
invertible with respect to the rule \lstinline{Q'}.  If \lstinline{Q} is a
two-premise rule, then it is split into two different rules to be
analyzed separately.  Assume that the resulting sliced rule with at
most one premise is \lstinline{R}.  The \lstinline{N}-th unifier between the
heads/conclusions of the rules is

\begin{lstlisting}

T  := getHead(R)
T' := getHead(Q', module)
U  := metaVariantDisjointUnify(module, getHead(T) =? getHead(T'), empty, 0, N)

\end{lstlisting}

\noindent
The case to be analyzed is then:
\begin{lstlisting}

inv-case(R, Q',
  apply(T', U), --- the sequent where R and Q' can be applied
  applyRule(apply(T, U), Q', module, U), --- applying Q' on the resulting seq.
  applyRule(apply(T, U), R,  module, U)) --- applying R  on the resulting seq.
\end{lstlisting}
The first two parameters identify the case. The third parameter
corresponds to the sequent where both rules can be applied, i.e., the
conclusion in Equation \eqref{eq:inv}. The fourth parameter
corresponds to the premises after the application of \lstinline{Q'}, i.e.,
the premises in Equation \eqref{eq:inv}. The last parameter is the
goal being proved, i.e., the premise resulting after the application
of \lstinline{R} on the same sequent ($\overline{(k:S_l) \theta}$ in
Definition \ref{def.check.invert}).  Call \lstinline{GTC}, \lstinline{GTP}, and
\lstinline{GG} to the last three parameters of the case once their
variables are replaced with fresh constants.

The inductive reasoning consists in applying, when possible, the rule
\lstinline{R} on the sequents in \lstinline{GTP}.  By induction, such
application of \lstinline{R} on a sequent $S$ in \lstinline{GTP} must preserve
the height annotation of $S$.  Hence, a modified version of \lstinline{R}
is needed:
\begin{lstlisting}

op inductive-rule : Rule -> Rule .
eq inductive-rule( ( rl '_:_['s[T],T'] => T'' [ AS ]. ) ) 
      =            ( rl '_:_[   T, T'] => T'' [ label('IND) ]. ) .
\end{lstlisting}

\noindent
Note the use of the height \lstinline{T} (instead of \lstinline{s[T]}) in the
resulting rule.  Call \lstinline{RI} the rule resulting from the
application of this function to the rule \lstinline{R} being analyzed.

Computing the set of sequents that can be assumed as axioms by
induction becomes now a simple task. It suffices to compute one step
of rewriting for each sequent \lstinline{GT} in the set \lstinline{GTP} as
follows:
\begin{lstlisting}

 metaSearch(MRI, GT, 'G:Goal, nil, '+, 1, k) .
\end{lstlisting}

\noindent
The term \lstinline{MRI} inherits all the functional description of the
module specifying the OL, but it contains only one single rule, namely,
\lstinline{RI}.  The (ground) sequent \lstinline{GT} is rewritten to any
possible list of sequents (variable \lstinline{'G:Goal}) in this theory in
exactly one step (\lstinline{'+} means one or more steps and the bound in
the 6th parameter forces it to be exactly one). The last parameter is
used to enumerate all the possible solutions.

Finally, the OL theory \lstinline{M} can be extended before attempting a
proof of the goal \lstinline{GG} of the current case:
\begin{lstlisting}

M' := newModuleRls(M,  premises(GTP)        --- axioms from GTP
		       inductive(M, GTP, R) --- inductive reasoning
		       already-proved-theorems)

\end{lstlisting}

\subsection{Cut Elimination}\label{imp:cut-elim}

There are different cut-rules depending on the shape of the sequent
system at hand (e.g., one-sided, two-sided, dyadic) and also on the
structural rules allowed.  The file \lstinline{cut-elimination-base.maude}
defines common facilities for all the cut-elimination procedures.
Impressive enough, minor extensions to this module have been required
to implement cut-elimination procedures for the systems in
Section~\ref{sec.case}.

The common interface is the following functional theory:

\begin{lstlisting}

fth CUT-SPEC is
    pr META-LEVEL .
    op th-name      : -> String .  --- Name of the theorem
    op mod-name     : -> Qid .     --- Module with the OL specification
    op bound-spec   : -> Nat .     --- Bound of the search procedure
    op file-name    : -> String .  --- File name to write the output
    op already-proved-theorems : -> RuleSet . --- Admissible rules
    op inv-rules    : -> QidList .            --- Invertible rules
endfth
\end{lstlisting}

\noindent
The parametric module \lstinline|CUT-BASE{SPEC :: CUT-SPEC}| contains
the common definitions.  An operator for the cut rule is defined
\lstinline|op cut-rule : -> Rule . |, but no equation for it is
provided. Each OL is responsible for completing this definition. For
instance, in $\Gtip$, the cut-rule shares the context in the
antecedent of the sequent between the two premises. Hence, the file
\lstinline{cut-add-scon.maude} (additive cut for single conclusion
systems) extends \lstinline|CUT-BASE| with the following equation:
\begin{lstlisting}

eq cut-rule =
 (rl '_:_['inf.INat, '_|--_['Gamma:MSFormula, 'F:Formula]] => 
 '_|_[
   '_:_['h1$$:FNat,'_|--_['_;_['Gamma:MSFormula, 'FCut$$:Formula],'F:Formula]],
   '_:_['h2$$:FNat,'_|--_[     'Gamma:MSFormula             ,'FCut$$:Formula]]]
  [ label('\Cut) ]. ) .

\end{lstlisting}
The conclusion of the rule is annotated with \lstinline{inf} (the constant
of type \lstinline{INat} denoting ``don't know'').  This specification is
nothing else that the meta-representation of the rule in Equation
\eqref{eq-cut-g3ip}. In all the analyses, the cut-formula is expected
to be named as \lstinline|FCut$$| and the height of the two premises
as \lstinline|h1$$| and \lstinline|h2$$|.

The module \lstinline|CUT-BASE| offers mechanisms to generate, in a
uniform way, the proof obligations for the cut-elimination procedures
considered here.  This is done in two steps. First, one of the rules
of the system identified as \lstinline{Q1} is matched against the first
premise of the cut-rule (\lstinline{lcut} below):

\begin{lstlisting}

U :=  metaVariantDisjointUnify(module, lcut =? getHead(Q1, module), empty, 0, N)
\end{lstlisting}

\noindent
The unifier \lstinline{U} must map the variables of \lstinline{lcut} to some
fresh variables. Hence, before unifying a second rule \lstinline{Q2}
against the second premise of the cut-rule, denoted as \lstinline{rcut}, the
substitution \lstinline{U} must be applied on \lstinline{rcut} (see Definition
\ref{def.cond.cut}):

\begin{lstlisting}

U' :=  metaVariantDisjointUnify(module, apply(rcut, U) =? 
					getHead(Q2, module), empty, Nvars, M) .

\end{lstlisting}

A proof obligation takes the following form:
\begin{lstlisting}

cut-case(   
	  cut-sub-case(TC, TP),   --- left premise  (conclusion and premises)
	  cut-sub-case(TC', TP'), --- right premise (conclusion and premises)
	  apply(getHead(cut-rule), (U ; U')),  --- goal to be proved
	  apply('FCut$$:Formula, (U ; U'))     --- the cut formula
	 )
\end{lstlisting}
The notation \lstinline{U ; U'} is used for composition of substitutions.

Induction on the height of the derivation is also defined in
\lstinline{CUT-BASE} in a general way and independently of the
cut-rule.
\begin{lstlisting}

op induct-height : GroundTerm GroundTerm GroundTerm -> RuleSet .  
eq induct-height('suc[gh], 'suc[gh'], GF) =
 	( rl getHead(cut-rule) =>
	  '_|_[   --- Fixing heights (gh and suc[gh']) and the cut-formula (GF)
	    apply(lcut, ('h1$$:FNat <- gh ; 'FCut$$:Formula <- GF)),
	    apply(rcut, ('h2$$:FNat <- 'suc[gh'] ; 'FCut$$:Formula <- GF))]
	  [ label('\hCut) ]. )
	( rl getHead(cut-rule) =>
	  '_|_[   --- Fixing heights (suc[gh] and gh') and the cut-formula (GF)
	    apply(lcut, ('h1$$:FNat <- 'suc[gh] ; 'FCut$$:Formula <- GF)),
	    apply(rcut, ('h2$$:FNat <- gh' ; 'FCut$$:Formula <- GF))]
	  [ label('\hCut) ]. ) .
 
\end{lstlisting}	

\noindent
The syntax \lstinline{x <- t} is used to denote the substitution $[t/x]$.  This specification implements $\textit{ind-H}$ in
Definition~\ref{def.cond.cut}.  The first two parameters are the
height of the left and the right premises above the cut. The last
parameter is the cut-formula considered in the current goal.

Induction on the structure of the formula is partially defined in
\lstinline|CUT-BASE|, but additional work is needed in the specific
OL.  General definitions include, e.g.,
\begin{lstlisting}

op induct-struct : GroundTerm -> RuleSet .
eq induct-struct(Q[LGT]) = $induct-struct(LGT) .  
eq induct-struct(GT)     = none [owise] .
\end{lstlisting}

The first equation applies to constructors for formulas (e.g.,
\lstinline|'_/\_[A, B]|, representing $A\wedge B$) and then, induction applies on all the
sub-terms in the list of ground terms \lstinline{LGT} (if they are of sort
\lstinline{Formula}). If the parameter is a constant, e.g., 
\lstinline{'False.Formula}, only the second equation applies and no
inductive rule is generated.  The keyword \lstinline{owise} (a shorthand
for \lstinline{otherwise}) means ``use this equation if all the others
defining the corresponding function symbol fail''. This is syntactic
sugar to simplify the specification and it can be encoded in plain
conditional equational theories~\cite{clavel-maudebook-2007}.

The function \lstinline|$induct-struct| calls to
\lstinline{induct-struct-formula} for each term in \lstinline{LGT}. The
definition of \lstinline{induct-struct-formula} is specific for each
OL. Here the one defined in \lstinline{cut-add-scon.maude}:

\begin{lstlisting}

op induct-struct-formula : GroundTerm -> RuleSet . 
eq induct-struct-formula(GTA) 
    =  if getType(GTA) == 'Formula           --- only sub-terms of sort Formula
       then ( rl  getHead(cut-rule) =>       --- head unchanged
	  '_|_[  --- premises with height 'inf and the cut-formula fixed to GTA
	    apply(lcut, ('h1$$:FNat <- 'inf.INat ; 'FCut$$:Formula <- GTA)),
	    apply(rcut, ('h2$$:FNat <- 'inf.INat ; 'FCut$$:Formula <- GTA)) ]
	  [ label('\sCut) ]. )
        else none --- no rule if GTA is not a formula
        fi .
\end{lstlisting}

\noindent
The parameter \lstinline{GTA} is a ground term denoting a proper
sub-formula of the cut-formula. The cut-rule is instantiated as
follows: the head of the rule is unchanged, the height of the premises
is \lstinline{'inf.INat}, and the cut-formula is fixed to be (the ground
term) \lstinline{GTA}. The change in the height of the premises is due to
the fact that cuts on smaller formulas do not preserve necessarily the
height of the derivation.

The above definition seems to be general enough for any cut-rule and,
in theory, it would be possible to define it once and for all the
systems.  However, the generated rule is problematic from the point of
view of proof search. Note that both premises are annotated with
\lstinline{'inf.INat}. Hence, even if the rule does not need to ``guess''
the cut-formula (since it is fixed to a ground term \lstinline{A}), it is
always possible to rewrite the goal $\Gamma \vdash G$ into $\Gamma, A
\vdash G$ and later into $\Gamma, A, A \vdash G$, etc. For this
reason, in some systems, extra conditions are needed to restrict the
application of this rule and reduce the search space (more about this
in Section~\ref{sec.case}). Of course, a bad choice of such conditions
may render the analysis inconclusive. A natural rule of thumb that
worked in most of the cases was to restrict the application of this
rule when the sub-formula $A$ is not already in the context $\Gamma$.
 
Finally, the main search procedure must be also tailored for each
OL. There is a template that, with little modifications, can be used
in all  systems reported here. In particular, OLs may define
different conditions for the application of the rules with the aim of
reducing the search space. The main definition in
\lstinline{cut-add-scon.maude} is:
\begin{lstlisting}
 
ceq holds$?(cut-case( ... ))
 = output ...
 if      --- Premises with implicit weakening
   RS := premises-W(GTC) premises-W(GTC') premises-W(GTP) premises-W(GTP')
	  --- invertibility lemmas on the premises
	  inv-premises(M, inv-rules, GTP) inv-premises(M, inv-rules,  GTP')
	  already-proved-theorems --- adding  admissibility  lemmas 
	  --- Induction on the height of the derivation
	  induct-height(getHeight(gtc), getHeight(gtc'), GF)
	  --- Induction on the formula only in principal cases
	  if numFormulas(GG) <= 1 then induct-struct(GF) else none fi
 \end{lstlisting}
In the code above, \lstinline{RS} is the set of rules used to extend the
theory of the OL before calling the search procedure.  This extension
is similar to the ones presented for admissibility and
invertibility. There are, however, two new ingredients: (a) the way
the axioms are generated and (b) the additional condition deciding
whether the rule for structural induction is added or not to solve the
current goal \lstinline{GG}:.

\begin{description}

\item[(a)] \lstinline{premises-W(S)} converts into an axiom the sequent
  \lstinline{S}. Unlike \lstinline{premises(S)} used in the previous sections,
  the resulting rule in \lstinline{premises-W(S)} internalizes weakening:
  if $S$ is the (ground) sequent $\Gamma \vdash F$, then $\Delta
  \vdash F$ is provable for any $\Delta \supseteq \Gamma$.  This
  avoids the need for adding the rule $W$ into
  \lstinline{already-proved-theorems}, thus reducing the search space.
  This simplification cannot be used, of course, in substructural
  logics, such as linear logic (Section \ref{ex:ll}).

\item[(b)] As already explained, the rule for structural induction is
  problematic from the point of view of proof search. In some OLs it
  is possible to control its use.  For example, in $\Gtip$, the
  application of $\textit{ind-F}$ can be restricted to solve only the
  principal cases (and the non-principal cases will run faster).  Such
  cases can be identified by counting the number of constants of type
  \lstinline{Formula} in the current goal \lstinline{GG} (see the last line in
  the code above). In other systems, however, this simplification does
  not work as in the case of the system $\mLJ$ (Section \ref{sec:mlj})
  where structural induction is also needed in some of the
  non-principal cases.
\end{description}
 
Wrapping up, configuring the cut-elimination procedure requires
adjusting and tuning some parameters. As shown in Section
\ref{sec.case}, some logics that share the same structural properties
may reuse a common infrastructure.  For instance, the cut-rule and
 procedures defined in file \lstinline{cut-add-mcon.maude} (two-sided
and multi-conclusion systems where weakening and contraction are
allowed) can be used to prove cut-elimination for the system $\Gtcp$
(Sec. \ref{case:gtcp}), $\mLJ$ (Sec.  \ref{sec:mlj}), and some
systems for modal logics (Sec.  \ref{case:modal}).  However,
cut-elimination is a non-trivial property and hence, full automation
is impossible for `the' general case.  In each of these systems, the
user must determine the invertibility lemmas that will be considered
during the search procedure. This is done by simply modifying the
input parameter \lstinline{inv-rules} in \lstinline{CUT-SPEC}.

\subsection{Identity Expansion}\label{sec-impl-id-exp}

This analysis uses the following functional theory as interface (\lstinline{id-expand.maude}): 
\begin{lstlisting}

fth ID-EXP-SPEC is
 op mod-name : -> Qid .
 op file-name : -> String .
 op bound-spec : -> Nat .
 op goal : GroundTerm -> GroundTerm . --- e.g.,   F |-- F  or  |- F, dual(F) 
 op already-proved-theorems : -> RuleSet .
 op types-formula : -> TypeList . --- types different from 'Formula to be analyzed
endfth
\end{lstlisting}

Given a
ground term $F$ denoting a formula, the call \lstinline{goal(F)} returns
the sequent to be proved. This definition allows to consider
id-expansion theorems for one-sided and two-sided systems. The last
field is used to define other sorts that need to be considered in the
analysis. For instance, the specification of modal logics includes the
sort \lstinline{BFormula < Formula} for boxed formulas (see Section
\ref{case:modal}). By adding \lstinline{BFormula} in \lstinline{types-formula},
the case $\Box F \vdash \Box F$ is generated.

\section{Case Studies}
\label{sec.case}

This section explains how structural meta-properties of several
propositional sequent systems can be specified and proved with the
approach presented in this paper. The site hosting the
\toolname\footnote{\url{https://carlosolarte.github.io/L-framework/}} includes all the logical systems described here, the
proof-search implementation of the strategies, and the PDFs generated
by the proof-search algorithms.  The chosen methodology for proving
the (meta-)theorems in this section is modular: first, it attempts to
build a proof without any external lemma; second, when needed, it
analyzes the failing cases and adds already proved theorems for
completing the proof. This methodology allows for analyzing
interdependencies between the different results inside various logical
systems.  Since there are no similar tools to compare the performance
of the \toolname, the main benchmark pursued is to show that it is
flexible enough to deal with several proof-theoretic properties with
little effort in defining the object logic and the properties of
interest. In all the cases, but those reported in Section \ref{ex:ll}
(where the proof of cut-elimination requires induction on two
different rules), implementing the analyses amounts only to
instantiate the interfaces/theories described in Section
~\ref{sec.rew}.

In order to show the feasibility of the \toolname, time-related
observations are reported. In all the cases, the depth bound
(\lstinline{bound-spec}) is set to 15 and the experiments are
performed on a MacBook Pro, 4 cores 2.3 GHz, and 8GB of RAM, running
Maude 3.0. On average, the admissibility analyses are completed in
about 5 seconds.  Once all the auxiliary lemmas are added, the time
needed to complete the proof of cut-elimination depends on whether
explicit structural rules are used (about 30 seconds) or not (about 15
seconds). 

\subsection{System $\Gtip$ for Propositional Intuitionistic Logic}\label{ex:g3i}

The system $\Gtip$ and its specification as a rewrite theory are
presented in Section \ref{sec.check}. The proof of meta-theoretical
properties is next described in detail.

\paragraph{Weakening.}
Theorem \ref{th.height} (see rule $\mathsf{H}$ in Equation
\eqref{eq:H}) and height-preserving admissibility of weakening can be
proved without auxiliary lemmas.

\begin{theorem}[Weakening and Weak-height]\label{g3ip-w}\label{g3ip-wh}
  If $\deducn{\Gtip}{n}{\Gamma \vdash C}$, then
  $\deducn{\Gtip}{s(n)}{\Gamma \vdash C}$.  Moreover, the rule
  $\mathsf{W}$ (Equation \eqref{eq:wc}) is height-preserving
  admissible in $\Gtip$.
\end{theorem}

\begin{proof}
  See the specification of the theorems in
  \lstinline{g3i/prop-W.maude} and \lstinline{g3i/prop-H.maude}. The
  resulting proofs are in \lstinline{g3i/g3i.pdf}. Both properties are
  proved in less than 1 second without any additional lemma.  \qed
\end{proof}

\paragraph{Invertibility.}
All the rules in $\Gtip$ are invertible, but $\vee_{R_i}$ and
$\cimp_L$. However, the tool initially fails (in 7.6s) to prove the
invertibility of $\cimp_R$, $\wedge_R$, $\wedge_L$, $\vee_L$, and
$\top_L$.

Consider the rule $\cimp_R$. Recall that the invertibility of a rule
is proven by testing the local invertibilities relative to all
possible rules (see Definition~\ref{def.check.invert}).  Hence, the
tool proves, e.g., that $\cimp_R$ is invertible w.r.t $\wedge_L$ (the
symbol $\bullet$ denotes successor):
\[
\reduction{\infer[\wedge_L]{\suc \wspace \HEIGHT{3} \wspace : \CTX{6}, \FORMULA{4} \wedge \FORMULA{5} \vdash \FORMULA{1} \iimp \FORMULA{2}}{\deduce{\HEIGHT{3} : \CTX{6}, \FORMULA{4}, \FORMULA{5} \vdash \FORMULA{1} \iimp \FORMULA{2}}{}}}
{\infer[\wedge_L]{\suc \wspace \HEIGHT{3} \wspace : \CTX{6}, \FORMULA{1}, \FORMULA{4} \wedge \FORMULA{5} \vdash \FORMULA{2}}{\infer[\AXHI]{\HEIGHT{3} : \CTX{6}, \FORMULA{1}, \FORMULA{4}, \FORMULA{5} \vdash \FORMULA{2}}{}}}\]
Note the application of the inductive hypothesis on the shorter
derivation of height $h_3$. All the other cases of local invertibility
of $\cimp_R$ are similar, but the cases w.r.t. $\cimp_R$ and $\cimp_L$
fail.  As shown below, when the admissible rules $\mathsf{H}$ and
$\mathsf{W}$ (Theorem \ref{g3ip-w}) are added to the set of
\lstinline{already-proved-theorems}, the tool completes the missing
cases.  The failure w.r.t. $\cimp_R$ is:

\reductionFail{\infer[\cimp_R]{\suc \wspace \HEIGHT{1} \wspace : \CTX{2} \vdash \FORMULA{3} \iimp \FORMULA{4}}{\deduce{\HEIGHT{1} : \CTX{2}, \FORMULA{3} \vdash \FORMULA{4}}{}}}{\textcolor{red}{\infer[\failTK]{\suc \wspace \HEIGHT{1} \wspace : \CTX{2}, \FORMULA{3} \vdash \FORMULA{4}}{}}}

\noindent
This is the dummy case where the same rule is applied on the same
formula and the proof transformation should be trivial.  In fact, an
application of $\mathsf{H}$ completes the case:
\[
\reduction{\infer[\cimp_R]{\suc \wspace \HEIGHT{1} \wspace : \CTX{2} \vdash \FORMULA{3} \iimp \FORMULA{4}}{\deduce{\HEIGHT{1} : \CTX{2}, \FORMULA{3} \vdash \FORMULA{4}}{}}}
{\infer[\heightRule]{\suc \wspace \HEIGHT{1} \wspace : \CTX{2}, \FORMULA{3} \vdash \FORMULA{4}}{\infer[\Assumption]{\HEIGHT{1} : \CTX{2}, \FORMULA{3} \vdash \FORMULA{4}}{}}}\]
Regarding the failure of the invertibility of $\cimp_R$ w.r.t $\cimp_L$ 

\reductionFail{\infer[\iimp_L]{\suc \wspace \HEIGHT{3} \wspace : \CTX{6}, \FORMULA{4} \iimp \FORMULA{5} \vdash \FORMULA{1} \iimp \FORMULA{2}}{\deduce{\HEIGHT{3} : \CTX{6}, \FORMULA{4} \iimp \FORMULA{5} \vdash \FORMULA{4}}{} & \deduce{\HEIGHT{3} : \CTX{6}, \FORMULA{5} \vdash \FORMULA{1} \iimp \FORMULA{2}}{}}}{\textcolor{red}{\infer[\failTK]{\suc \wspace \HEIGHT{3} \wspace : \CTX{6}, \FORMULA{1}, \FORMULA{4} \iimp \FORMULA{5} \vdash \FORMULA{2}}{}}}

\noindent
If $\cimp_L$ is applied on the sequent $\Delta_6,F_1, F_4\cimp F_5
\seqn{s(h_3)} F_2$, two premises are obtained: $\Delta_6, F_1, F_4
\cimp F_5 \seqn{h_3} F_4$ and $\Delta_6, F_1, F_5 \seqn{h_3} F_2$. The
second premise is provable by induction. The proof of the first
premise requires weakening on $F_1$:
\[
\reduction{\infer[\iimp_L]{\suc \wspace \HEIGHT{3} \wspace : \CTX{6}, \FORMULA{4} \iimp \FORMULA{5} \vdash \FORMULA{1} \iimp \FORMULA{2}}{\deduce{\HEIGHT{3} : \CTX{6}, \FORMULA{4} \iimp \FORMULA{5} \vdash \FORMULA{4}}{} & \deduce{\HEIGHT{3} : \CTX{6}, \FORMULA{5} \vdash \FORMULA{1} \iimp \FORMULA{2}}{}}}
{\infer[\iimp_L]{\suc \wspace \HEIGHT{3} \wspace : \CTX{6}, \FORMULA{1}, \FORMULA{4} \iimp \FORMULA{5} \vdash \FORMULA{2}}{\infer[W]{\HEIGHT{3} : \CTX{6}, \FORMULA{1}, \FORMULA{4} \iimp \FORMULA{5} \vdash \FORMULA{4}}{\infer[\Assumption]{\HEIGHT{3} : \CTX{6}, \FORMULA{4} \iimp \FORMULA{5} \vdash \FORMULA{4}}{}} & \infer[\AXHI]{\HEIGHT{3} : \CTX{6}, \FORMULA{1}, \FORMULA{5} \vdash \FORMULA{2}}{}}}\]

\noindent
Once $\mathsf{W}$ is added, the proofs of invertibility of $\wedge_R$,
$\wedge_L$, $\vee_L$ and $\top_L$ are also completed.

Some cases are vacuously discharged. For instance, there are no proof
obligations for the invertibility of $\cimp_R$ w.r.t. $\wedge_R$ (it
is impossible to unify the conclusion of these rules). Moreover, the
cases of the axioms $I$, $\top_R$, and $\bot_L$ are trivial since the
only proof obligation is to reduce \lstinline{proved} into
\lstinline{proved}:

\[
\reduction{\infer[\iimp_L]{\suc \wspace \HEIGHT{1} \wspace : \CTX{4}, \FORMULA{2} \iimp \FORMULA{3} \vdash \top}{\deduce{\HEIGHT{1} : \CTX{4}, \FORMULA{2} \iimp \FORMULA{3} \vdash \FORMULA{2}}{} & \deduce{\HEIGHT{1} : \CTX{4}, \FORMULA{3} \vdash \top}{}}}
{\trivial}\]

The rules $\vee_{R_1}$, $\vee_{R_2}$, and the left premise of
$\cimp_L$ are clearly not invertible.  The following failures provide
good evidences that these cases do not succeed:

$\small
\begin{array}{c}
\reduction{\infer[\vee_2]{\suc \wspace \HEIGHT{1} \wspace : \CTX{2} \vdash \FORMULA{3} \vee \FORMULA{4}}{\deduce{\HEIGHT{1} : \CTX{2} \vdash \FORMULA{4}}{}}}{\textcolor{red}{\infer[\failTK]{\suc \wspace \HEIGHT{1} \wspace : \CTX{2} \vdash \FORMULA{3}}{}}}\qquad\mbox{ and }\qquad
\reduction{\infer[\top_R]{\wspace \HEIGHT3 \wspace \suc : \CTX4, \FORMULA1 \iimp \FORMULA2 \vdash \top}{}}{\textcolor{red}{\infer[\failTK]{\wspace \HEIGHT3 \wspace \suc : \CTX4, \FORMULA1 \iimp \FORMULA2 \vdash \FORMULA1}{}}}
\end{array}
$

\begin{theorem}[Invertibility]\label{g3ip-inv}
  All the rules, but $\vee_{R_i}$ and $\cimp_L$, are height-preserving
  invertible in $\Gtip$. Moreover, the right premise of $\cimp_L$ is
  height-preserving invertible.
\end{theorem}

\begin{proof}
  The invertibility of $\wedge_R$, $\wedge_L$, $\vee_L$, and $\top_L$
  depends on the admissibility of $\mathsf{W}$ (Theorem
  \ref{g3ip-wh}).  The invertibility of $\cimp_R$ and the
  invertibility of the right premise of $\cimp_L$ require Theorem
  \ref{g3ip-wh} ($\mathsf{W}$ and $\mathsf{H}$). The specification of
  the property is in \lstinline{prop-inv.maude}.  The analysis is
  completed in 7.7 seconds.  \qed
\end{proof}

\paragraph{Contraction.}
When attempting a proof of admissibility of contraction, the local
admissibility cases w.r.t $\cimp_L$, $\wedge_L$, $\vee_L$, and
$\top_L$ fail. Here is the failing case for $\vee_L$:

\reductionFail{\infer[\vee_L]{\suc \wspace \HEIGHT{1} \wspace : \CTX{5}, \FORMULA{2} \vee \FORMULA{3}, \FORMULA{2} \vee \FORMULA{3} \vdash \FORMULA{4}}{\deduce{\HEIGHT{1} : \CTX{5}, \FORMULA{2}, \FORMULA{2} \vee \FORMULA{3} \vdash \FORMULA{4}}{} & \deduce{\HEIGHT{1} : \CTX{5}, \FORMULA{3}, \FORMULA{2} \vee \FORMULA{3} \vdash \FORMULA{4}}{}}}{\textcolor{red}{\infer[\failTK]{\suc \wspace \HEIGHT{1} \wspace : \CTX{5}, \FORMULA{2} \vee \FORMULA{3} \vdash \FORMULA{4}}{}}}

\noindent
Note that the inductive hypothesis cannot be used neither on the left
nor on the right premise.  After adding \lstinline{'OrL} to the set of
already-proved invertible rules (field \lstinline{inv-rules}),
the \toolname\ completes this case as follows:

\[
\reduction{\infer[\vee_L]{\suc \wspace \HEIGHT{1} \wspace : \CTX{5}, \FORMULA{2} \vee \FORMULA{3}, \FORMULA{2} \vee \FORMULA{3} \vdash \FORMULA{4}}{\deduce{\HEIGHT{1} : \CTX{5}, \FORMULA{2}, \FORMULA{2} \vee \FORMULA{3} \vdash \FORMULA{4}}{} & \deduce{\HEIGHT{1} : \CTX{5}, \FORMULA{3}, \FORMULA{2} \vee \FORMULA{3} \vdash \FORMULA{4}}{}}}
{\infer[\vee_L]{\suc \wspace \HEIGHT{1} \wspace : \CTX{5}, \FORMULA{2} \vee \FORMULA{3} \vdash \FORMULA{4}}{\infer[\IH]{\HEIGHT{1} : \CTX{5}, \FORMULA{2} \vdash \FORMULA{4}}{\infer[\AssumptionInv]{\HEIGHT{1} : \CTX{5}, \FORMULA{2}, \FORMULA{2} \vdash \FORMULA{4}}{}} & \infer[\IH]{\HEIGHT{1} : \CTX{5}, \FORMULA{3} \vdash \FORMULA{4}}{\infer[\AssumptionInv]{\HEIGHT{1} : \CTX{5}, \FORMULA{3}, \FORMULA{3} \vdash \FORMULA{4}}{}}}}\]

\[
\reduction{\infer[\vee_L]{\suc \wspace \HEIGHT{1} \wspace :(\CTX{5}, \FORMULA{2} \vee \FORMULA{3}), \FORMULA{6}, \FORMULA{6} \vdash \FORMULA{4}}{\deduce{\HEIGHT{1} : \CTX{5}, \FORMULA{2}, \FORMULA{6}, \FORMULA{6} \vdash \FORMULA{4}}{} & \deduce{\HEIGHT{1} : \CTX{5}, \FORMULA{3}, \FORMULA{6}, \FORMULA{6} \vdash \FORMULA{4}}{}}}
{\infer[\vee_L]{\suc \wspace \HEIGHT{1} \wspace : \CTX{5}, \FORMULA{6}, \FORMULA{2} \vee \FORMULA{3} \vdash \FORMULA{4}}{\infer[\IH]{\HEIGHT{1} : \CTX{5}, \FORMULA{2}, \FORMULA{6} \vdash \FORMULA{4}}{\infer[\Assumption]{\HEIGHT{1} : \CTX{5}, \FORMULA{2}, \FORMULA{6}, \FORMULA{6} \vdash \FORMULA{4}}{}} & \infer[\IH]{\HEIGHT{1} : \CTX{5}, \FORMULA{3}, \FORMULA{6} \vdash \FORMULA{4}}{\infer[\Assumption]{\HEIGHT{1} : \CTX{5}, \FORMULA{3}, \FORMULA{6}, \FORMULA{6} \vdash \FORMULA{4}}{}}}}\]

\noindent
Due to unification, there are indeed two cases: one in which the
disjunctive formula is contracted and one of the copies is principal,
and another case with the disjunction not being contracted (instead,
another formula $F_6$ is). The second case follows without using the
invertibility lemma.

\begin{theorem}[Contraction]
  The contraction rule $\mathsf{C}$ (Equation \eqref{eq:wc}) is
  height-preserving admissible in $\Gtip$.
\end{theorem}

\begin{proof}
  The cases $\vee_L$, $\wedge_L$, and $\top_L$ require the
  invertibility of the respective rules (Theorem \ref{g3ip-inv}). The
  case $\cimp_L$ requires invertibility of the right premise of this
  rule (specified in \lstinline{inv-rules} as
  \lstinline{'impL$$1}). The proof takes 1.3 seconds.  \qed
\end{proof}

\paragraph{Cut-Elimination. } 
Controlling the structural rules is one of the key points to reduce
the search space in sequent systems.  The system $\Gtip$ embeds
weakening in its initial rule ($\Gamma$ can be an arbitrary context)
and contraction in rules with two premises (all rules are additive and
contraction is explicit in the left premise of $\cimp_L$).  Thus,
$\mathsf{W}$ and $\mathsf{C}$ are admissible in $\Gtip$, and the proof
search procedure does not need to guess how many times a formula must
be contracted or whether some of them need to be weakened.  The
cut-rule considered for $\Gtip$ is additive (see Equation
\eqref{eq-cut-g3ip}), hence also carrying an implicit contraction.

The cut-elimination procedure for $\Gtip$ implements two strategies
for reducing the search space (see Section \ref{imp:cut-elim}):

\begin{enumerate} 
 \item If \lstinline{GTP} is the resulting set of sequents that can be
   assumed to be provable, then such sequents are added as axioms that
   internalize weakening. For instance, if ${\Gamma \vdash F}$ is a
   (ground) sequent in \lstinline{GTP}, then, $\Delta \vdash F$ is assumed
   to be provable whenever $\Delta \supseteq \Gamma$.  This avoids the
   need for adding the rule $\mathsf{W}$ to the set of already proved
   theorems, thus reducing the non-determinism during proof-search:
   weakening is ``delayed'' until the leaves of the derivation are
   reached.

 \item The inductive hypothesis on the structure of the formula is
   only used in the principal cases.  Hence, less alternatives are
   explored when proving the non-principal cases.
\end{enumerate}

The strategy (1) considerably affects the performance in both failing
and successful attempts as explained below. The strategy (2) saves few
seconds when all the needed auxiliary lemmas are added and the proof
succeeds. In failing attempts, this strategy has an important impact.

Due to (1), $\mathsf{W}$ and $\mathsf{C}$ are not  added to
\lstinline{already-proved-theorems} and, for the moment, consider also
that none of the invertibility lemmas is added.  This experiment leads
to proofs for the trivial cases ( $\top_R$, $\bot_L$, $I$ and
$\top_L$) and fails for the other rules in almost 15
seconds\footnote{In the same experiment, if the strategy (2) is not
considered, the case $\cimp_R$ does not finish after 15 min.  }.

\paragraph{Non-principal Cases.}
Usually the non-principal cases are easily solved by permuting down
the application of a rule and reducing the height of the cut. Some of
this cases are already proved in this first iteration. For instance,
the case ($\wedge_L$, $\vee_1$) --- $\wedge_L$ applied on the left
premise and $\vee_1$ on the right premise --- is solved as follows:

\cutEnv{\cutReduction{\infer[\Cut]{\noheight : \CTX{11}, \FORMULA{6} \wedge \FORMULA{7} \vdash \FORMULA{9} \vee \FORMULA{10}}{\infer[\wedge_L]{\suc \wspace \HEIGHT{1} \wspace : \CTX{11}, \FORMULA{6} \wedge \FORMULA{7} \vdash \FORMULA{12}}{\deduce{\HEIGHT{1} : \CTX{11}, \FORMULA{6}, \FORMULA{7} \vdash \FORMULA{12}}{}} & \infer[\vee_1]{\suc \wspace \HEIGHT{8} \wspace :(\CTX{11}, \FORMULA{6} \wedge \FORMULA{7}), \FORMULA{12} \vdash \FORMULA{9} \vee \FORMULA{10}}{\deduce{\HEIGHT{8} : \CTX{11}, \FORMULA{12}, \FORMULA{6} \wedge \FORMULA{7} \vdash \FORMULA{9}}{}}}}
{\infer[\vee_1]{\noheight : \CTX{11}, \FORMULA{6} \wedge \FORMULA{7} \vdash \FORMULA{9} \vee \FORMULA{10}}{\infer[\hCut]{\noheight : \CTX{11}, \FORMULA{6} \wedge \FORMULA{7} \vdash \FORMULA{9}}{\infer[\AssumptionW]{\suc \wspace \HEIGHT{1} \wspace : \CTX{11}, \FORMULA{6} \wedge \FORMULA{7} \vdash \FORMULA{12}}{} & \infer[\AssumptionW]{\HEIGHT{8} : \CTX{11}, \FORMULA{12}, \FORMULA{6} \wedge \FORMULA{7} \vdash \FORMULA{9}}{}}}}}\\

\noindent
In the right derivation, $\hCut$ is an application of the cut-rule
with shorter derivations. Moreover, $\AssumptionW$ finishes the proof
due to the sequents assumed to be provable (on the left derivation),
possibly applying $\mathsf{W}$. In this particular case, weakening is
not needed.  Some other similar cases, however, fail. Take for
instance the case ($\wedge_L,\wedge_L$):

\reductionFail{\infer[\Cut]{\noheight : \CTX{12}, \FORMULA{6} \wedge \FORMULA{7} \vdash \FORMULA{11}}{\infer[\wedge_L]{\suc \wspace \HEIGHT{1} \wspace : \CTX{12}, \FORMULA{6} \wedge \FORMULA{7} \vdash \FORMULA{9} \wedge \FORMULA{10}}{\deduce{\HEIGHT{1} : \CTX{12}, \FORMULA{6}, \FORMULA{7} \vdash \FORMULA{9} \wedge \FORMULA{10}}{}} & \infer[\wedge_L]{\suc \wspace \HEIGHT{8} \wspace :(\CTX{12}, \FORMULA{6} \wedge \FORMULA{7}), \FORMULA{9} \wedge \FORMULA{10} \vdash \FORMULA{11}}{\deduce{\HEIGHT{8} : \CTX{12}, \FORMULA{9}, \FORMULA{10}, \FORMULA{6} \wedge \FORMULA{7} \vdash \FORMULA{11}}{}}}}{\textcolor{red}{\infer[\failTK]{\noheight : \CTX{12}, \FORMULA{6} \wedge \FORMULA{7} \vdash \FORMULA{11}}{}}}

\noindent
What is missing here is the invertibility of $\wedge_L$ on the
assumption $\Delta_{12}, F_9, F_{10}, F_6\wedge F_7 \vdash_{h_8}
F_{11}$. If this invertibility lemma is added, the tool completes the
case:

\cutEnv{\cutReduction{\ }
{\infer[\wedge_L]{\noheight : \CTX{12}, \FORMULA{6} \wedge \FORMULA{7} \vdash \FORMULA{11}}{\infer[\hCut]{\noheight : \CTX{12}, \FORMULA{6}, \FORMULA{7} \vdash \FORMULA{11}}{\infer[\AssumptionW]{\HEIGHT{1} : \CTX{12}, \FORMULA{6}, \FORMULA{7} \vdash \FORMULA{9} \wedge \FORMULA{10}}{} & \infer[\wedge_L]{\suc \wspace \HEIGHT{8} \wspace : \CTX{12}, \FORMULA{6}, \FORMULA{7}, \FORMULA{9} \wedge \FORMULA{10} \vdash \FORMULA{11}}{\infer[\AssumptionInv]{\HEIGHT{8} : \CTX{12}, \FORMULA{10}, \FORMULA{6}, \FORMULA{7}, \FORMULA{9} \vdash \FORMULA{11}}{}}}}}}

Inspecting similar failing cases suggests the need for including also
the invertibility of the rules $\wedge_R$ and $\vee_L$, and also the
invertibility of the right premise of $\cimp_L$. This solves some
missing cases but still, the cases for $\cimp_L$, $\wedge_L$ and
$\vee_L$ are not complete. One of the failures for ($\cimp_L,
\cimp_L$) is the following:

\reductionFail{\infer[\Cut]{\noheight : \CTX{10}, \FORMULA{7} \iimp \FORMULA{8} \vdash \FORMULA{9}}{\infer[\iimp_L]{\suc \wspace \HEIGHT{1} \wspace : \CTX{10}, \FORMULA{7} \iimp \FORMULA{8} \vdash \FORMULA{11}}{\deduce{\HEIGHT{1} : \CTX{10}, \FORMULA{7} \iimp \FORMULA{8} \vdash \FORMULA{7}}{} & \deduce{\HEIGHT{1} : \CTX{10}, \FORMULA{8} \vdash \FORMULA{11}}{}} & \infer[\iimp_L]{\suc \wspace \HEIGHT{6} \wspace :(\CTX{10}, \FORMULA{7} \iimp \FORMULA{8}), \FORMULA{11} \vdash \FORMULA{9}}{\deduce{\HEIGHT{6} : \CTX{10}, \FORMULA{11}, \FORMULA{7} \iimp \FORMULA{8} \vdash \FORMULA{7}}{} & \deduce{\HEIGHT{6} : \CTX{10}, \FORMULA{8}, \FORMULA{11} \vdash \FORMULA{9}}{}}}}{\textcolor{red}{\infer[\failTK]{\noheight : \CTX{10}, \FORMULA{7} \iimp \FORMULA{8} \vdash \FORMULA{9}}{}}}

The cut-formula is $F_{11}$ and it is not principal in any of the
premises.  Once $\cimp_L$ is applied on the goal $\Delta_{10},F_7
\cimp F_8 \vdash F_9$, the resulting left premise is already proved
(see the left-most sequent in the left derivation). The right premise
can be proved with $\mathsf{H}$ (Theorem \ref{g3ip-wh}) if it is added to 
\lstinline{already-proved-theorems}:

\cutEnv{\cutReduction{\ }
{\infer[\iimp_L]{\noheight : \CTX{10}, \FORMULA{7} \iimp \FORMULA{8} \vdash \FORMULA{9}}{\infer[\AssumptionW]{\noheight : \CTX{10}, \FORMULA{7} \iimp \FORMULA{8} \vdash \FORMULA{7}}{} & \infer[\hCut]{\noheight : \CTX{10}, \FORMULA{8} \vdash \FORMULA{9}}{\infer[\AssumptionW]{\HEIGHT{1} : \CTX{10}, \FORMULA{8} \vdash \FORMULA{11}}{} & \infer[H]{\suc \wspace \HEIGHT{6} \wspace : \CTX{10}, \FORMULA{11}, \FORMULA{8} \vdash \FORMULA{9}}{\infer[\AssumptionW]{\HEIGHT{6} : \CTX{10}, \FORMULA{11}, \FORMULA{8} \vdash \FORMULA{9}}{}}}}}}

\paragraph{Principal Cases.} 
Note that in all the above derivations, due to unification, there is
always more than one constant of sort \lstinline{Formula} in the goal:
besides the formula in the succedent of the sequent, there are
formulas in the antecedent that are needed for the application of a
left rule in the left premise of the cut. Due to the strategy (2),
cuts on smaller formulas are not considered during the proof search
for these cases.  The situation is different in the principal
cases. Consider for instance the case $(\cimp_R, \cimp_L)$: \\

\noindent\resizebox{.7\textwidth}{!}{
$
\begin{array}{c}
\infer[\Cut]{\noheight : \CTX{9} \vdash \FORMULA{8}}{\infer[\iimp_R]{\suc \wspace \HEIGHT{1} \wspace : \CTX{9} \vdash \FORMULA{6} \iimp \FORMULA{7}}{\deduce{\HEIGHT{1} : \CTX{9}, \FORMULA{6} \vdash \FORMULA{7}}{}} & \infer[\iimp_L]{\suc \wspace \HEIGHT{5} \wspace : \CTX{9}, \FORMULA{6} \iimp \FORMULA{7} \vdash \FORMULA{8}}{\deduce{\HEIGHT{5} : \CTX{9}, \FORMULA{6} \iimp \FORMULA{7} \vdash \FORMULA{6}}{} & \deduce{\HEIGHT{5} : \CTX{9}, \FORMULA{7} \vdash \FORMULA{8}}{}}} \to\\\\
{\infer[\sCut]{\noheight : \CTX{9} \vdash \FORMULA{8}}{\infer[\hCut]{\noheight : \CTX{9} \vdash \FORMULA{6}}{\infer[\AssumptionW]{\suc \wspace \HEIGHT{1} \wspace : \CTX{9} \vdash \FORMULA{6} \iimp \FORMULA{7}}{} & \infer[\AssumptionW]{\HEIGHT{5} : \CTX{9}, \FORMULA{6} \iimp \FORMULA{7} \vdash \FORMULA{6}}{}} & \infer[\sCut]{\noheight : \CTX{9}, \FORMULA{6} \vdash \FORMULA{8}}{\infer[\AssumptionW]{\noheight : \CTX{9}, \FORMULA{6} \vdash \FORMULA{7}}{} & \infer[\AssumptionW]{\noheight : \CTX{9}, \FORMULA{6}, \FORMULA{7} \vdash \FORMULA{8}}{}}}}
\end{array}
$
}

\noindent
Rule $\sCut$ corresponds to a cut on a sub-formula. Note that the
antecedent of the the goal is just a constant of sort \lstinline{MSFormula}
($\Delta_9$).

\begin{theorem}[Cut elimination]\label{th-cut-g3ip}
  The cut-rule in Equation \eqref{eq-cut-g3ip} is admissible in
  $\Gtip$.
\end{theorem}

\begin{proof}
  See the specification and dependencies in
  \lstinline{g3i/prop-cut.maude}. The proof requires Theorem
  \ref{g3ip-wh}.  All the cases --except for $\top_R$, $\top_L$,
  $\bot_L$, and $I$-- require Theorem \ref{g3ip-inv}.  The proof is
  completed in 13.8 sec.  \qed
\end{proof}

\paragraph{Multiplicative Cut in $\Gtip$.}  Observe that, by adopting
the additive version of the cut-rule, several common proof search
problems are avoided, e.g., loops created by the uncontrolled use of
contraction. But what if the following {\em multiplicative} cut is
considered in $\Gtip$ instead, where the context is split between the
two premises?

\begin{equation}\label{eq-cut-lin-g3ip}
\infer[Cut]{\Gamma,\Delta \vdash B}{
\deduce{\Gamma \vdash A}{}
&
\deduce{\Delta, A \vdash B}{}
}
\end{equation}

\noindent
The search tree now is considerably bigger since all the alternatives
on how to split the context $\Gamma, \Delta$ need to be considered
(see \lstinline{cut-mul-scon.maude}). This is an interesting question,
and the discussion presented next will serve to pave the way to the
analysis of sequent systems with linear contexts (see
Section~\ref{ex:ll}).

Note that, if only the rule $\mathsf{H}$ (Theorem \ref{g3ip-wh}) is
added, then all the cases but the principal cases for $\iimp$ and
$\wedge$ succeed in 26 seconds.  The failure on $(\wedge_R,
\wedge_L)$ is:

\reductionFail{\infer[\Cut]{\noheight : \CTX{2}, \CTX{9} \vdash \FORMULA{8}}{\infer[\wedge_R]{\suc \wspace \HEIGHT{1} \wspace : \CTX{2} \vdash \FORMULA{6} \wedge \FORMULA{7}}{\deduce{\HEIGHT{1} : \CTX{2} \vdash \FORMULA{6}}{} & \deduce{\HEIGHT{1} : \CTX{2} \vdash \FORMULA{7}}{}} & \infer[\wedge_L]{\suc \wspace \HEIGHT{5} \wspace : \CTX{9}, \FORMULA{6} \wedge \FORMULA{7} \vdash \FORMULA{8}}{\deduce{\HEIGHT{5} : \CTX{9}, \FORMULA{6}, \FORMULA{7} \vdash \FORMULA{8}}{}}}}{\textcolor{red}{\infer[\failTK]{\noheight : \CTX{2}, \CTX{9} \vdash \FORMULA{8}}{}}}

\noindent
This case is solved by cutting with $F_6$ and $F_7$. However, since
the cut-rule is multiplicative, the contexts $\Delta_2$ and $\Delta_9$
need to be contracted first. Adding contraction, on terms of sort
\lstinline{MSFormula} makes infeasible the proof search procedure: any
subset of the antecedent context can be chosen for contraction and
such rule can be applied on any sequent/goal. Instead, a more
controlled version of contraction can be added: contract the whole
context only if there are no duplicated elements in it. This more
restricted rule cannot be applied twice on the same goal, thus
reducing the number of alternatives leading to the following proof
transformation:

\resizebox{.65\textwidth}{!}{
$
\begin{array}{c}
{\infer[\Cut]{\noheight : \CTX{2}, \CTX{9} \vdash \FORMULA{8}}{\infer[\iimp_R]{\suc \wspace \HEIGHT{1} \wspace : \CTX{2} \vdash \FORMULA{6} \iimp \FORMULA{7}}{\deduce{\HEIGHT{1} : \CTX{2}, \FORMULA{6} \vdash \FORMULA{7}}{}} & \infer[\iimp_L]{\suc \wspace \HEIGHT{5} \wspace : \CTX{9}, \FORMULA{6} \iimp \FORMULA{7} \vdash \FORMULA{8}}{\deduce{\HEIGHT{5} : \CTX{9}, \FORMULA{6} \iimp \FORMULA{7} \vdash \FORMULA{6}}{} & \deduce{\HEIGHT{5} : \CTX{9}, \FORMULA{7} \vdash \FORMULA{8}}{}}}} \qquad \to
\\\\
{\infer[C]{\noheight : \CTX{2}, \CTX{9} \vdash \FORMULA{8}}{\infer[\sCut]{\noheight : \CTX{2}, \CTX{2}, \CTX{9}, \CTX{9} \vdash \FORMULA{8}}{\infer[\hCut]{\noheight : \CTX{2}, \CTX{9} \vdash \FORMULA{6}}{\infer[\AssumptionW]{\suc \wspace \HEIGHT{1} \wspace : \CTX{2} \vdash \FORMULA{6} \iimp \FORMULA{7}}{} & \infer[\AssumptionW]{\HEIGHT{5} : \CTX{9}, \FORMULA{6} \iimp \FORMULA{7} \vdash \FORMULA{6}}{}} & \infer[\sCut]{\noheight : \CTX{2}, \CTX{9}, \FORMULA{6} \vdash \FORMULA{8}}{\infer[\AssumptionW]{\noheight : \CTX{2}, \FORMULA{6} \vdash \FORMULA{7}}{} & \infer[\AssumptionW]{\noheight : \CTX{9}, \FORMULA{7} \vdash \FORMULA{8}}{}}}}}
\end{array}
$
}

\begin{theorem}[Multiplicative Cut-elim. ]
  The rule in Eq. \eqref{eq-cut-lin-g3ip} is admissible in $\Gtip$.
\end{theorem}

\begin{proof}
  The specification is in \lstinline{prop-cut-mul.maude}. See the
  contraction rule used in the definition of
  \lstinline{already-proved-theorems}. No invertibility lemma is
  needed for this proof.  The analysis is completed in 25.6 seconds.
  \qed
\end{proof}

If $\mathsf{W}$ is not embedded in the initial axioms and $\mathsf{C}$
is allowed on arbitrary contexts, there is little hope to conclude
these proofs in reasonable time.

\paragraph{Identity Expansion.}
Finally, the dual property of cut-elimination, identity expansion, is
 easily proved in the \toolname.

\begin{theorem}[Identity-Expansion]
  If $F$ is a formula, then $F \vdash F$ is provable in $\Gtip$.
\end{theorem}

\begin{proof}
  See \lstinline{prop-ID.maude}. The rule $\mathsf{W}$ needs to be
  added to the set of already proved theorems. It is used, e.g., in
  the following case:
$
 \small{\infer[\iimp_R]{\noheight : \FORMULA{0} \iimp \FORMULA{1} \vdash \FORMULA{0} \iimp \FORMULA{1}}{\infer[\iimp_L]{\noheight : \FORMULA{0}, \FORMULA{0} \iimp \FORMULA{1} \vdash \FORMULA{1}}{\infer[W]{\noheight : \FORMULA{0}, \FORMULA{0} \iimp \FORMULA{1} \vdash \FORMULA{0}}{\infer[\IH]{\noheight : \FORMULA{0} \vdash \FORMULA{0}}{}} & \infer[W]{\noheight : \FORMULA{0}, \FORMULA{1} \vdash \FORMULA{1}}{\infer[\IH]{\noheight : \FORMULA{1} \vdash \FORMULA{1}}{}}}}
 }
 $\qed
\end{proof}

\subsection{Multi-conclusion Propositional  Intuitionistic Logic ($\mLJ$)}\label{sec:mlj}

Maehara's $\mLJ$~\cite{maehara54nmj} is a multiple conclusion system
for intuitionistic logic.  The rules in $\mLJ$ have the exact same
shape as in $\Gtip$, except for the right rules for disjunction and
implication (see Fig.~\ref{fig:mLJ}). The disjunction right rule in
$\mLJ$ matches the correspondent rule in classical logic where the
disjunction is interpreted as the comma in the succedent of
sequents. The right implication, on the other hand, forces all
formulas in the succedent of the premise to be weakened. This
guarantees that, when the $\cimp_R$ rule is applied on $A\cimp B$, the
formula $B$ should be proved assuming {\em only} the pre-existent
antecedent context extended with the formula $A$. This creates an
interdependency between $A$ and $B$.

\paragraph{Weakening.}
The proof of admissibility of weakening in $\mLJ$ is similar to
$\Gtip$, only noting that, in the former, weakening is also
height-preserving admissible in the succedent of sequents.

\begin{theorem}[Weakening and Weak-height]\label{th:weak-mlj}
  If $\deducn{\mLJ}{n}{\Gamma \vdash \Delta}$, then
  $\deducn{\mLJ}{s(n)}{\Gamma \vdash \Delta}$
  ($\mathsf{H}$). Moreover, the following rules are height-preserving
  admissible in $\mLJ$:
\[
\infer[\mathsf{W}_L]{\Gamma, F \vdash \Delta}{\Gamma \vdash \Delta}
\qquad
\infer[\mathsf{W}_R]{\Gamma \vdash \Delta, F}{\Gamma \vdash \Delta}
\]
\end{theorem}

\begin{proof}
  The three properties are specified in
  \lstinline{mLJ/prop-WH.maude}. No auxiliary lemmas are needed. This
  theorem is proved in less than 3 seconds.  \qed
\end{proof}

\begin{figure}
$\small
\infer[{\cimp_L}]{\Gamma, A\cimp B \vdash \Delta}{\Gamma, A\cimp B
 \vdash A, \Delta \quad \Gamma, B \vdash
\Delta }
\qquad
\infer[{\cimp_R}]{\Gamma \vdash A \cimp B, \Delta}{\Gamma, A
\vdash B}
\qquad
\infer[{\lor_R}]{\Gamma \vdash A \lor B, \Delta}{\Gamma \vdash A,B, \Delta }
$
\caption{The multi-conclusion intuitionistic sequent system $\mLJ$.}\label{fig:mLJ}
\end{figure}

\paragraph{Invertibility.}
All the rules in $\mLJ$ are invertible, with the exception of
$\cimp_R$.

\begin{theorem}[Inv. ]
  All the rules but $\cimp_R$ are height-preserving invertible in
  $\mLJ$.
\end{theorem}

\begin{proof}
  See the specification in
  \lstinline{mLJ/prop-inv.maude}. $\mathsf{H}$ (Theorem
  \ref{th:weak-mlj}) is needed for all the cases but $\top_R$,
  $\bot_L$ and $I$. This proof takes 15.1 seconds.  \qed
\end{proof}

\paragraph{Contraction.}
Contraction is also admissible, on both sides of
sequents.

\begin{theorem}[Contraction]
  The rules
\[
\infer[\mathsf{C}_L]{\Gamma, F \vdash \Delta}{\Gamma,F,F \vdash
  \Delta} \qquad \infer[\mathsf{C}_R]{\Gamma \vdash F, \Delta}{\Gamma
  \vdash F, F, \Delta}
\]
  are both height-preserving admissible in $\mLJ$.
\end{theorem}

\begin{proof}
  The specification is in \lstinline{mLJ/prop-C.maude}. The proof of
  admissibility of $\mathsf{C}_L$ (resp. $\mathsf{C}_R$) requires the
  invertibility of $\top_L$, $\vee_L$, $\wedge_L$, and $\iimp_L$
  (resp., $\bot_R$, $\vee_R$ and $\wedge_R$).  \qed
\end{proof}

\paragraph{Cut-elimination and Identity Expansion.}
In the cut-elimination procedure for $\Gtip$, the application of the
rule for structural induction (\lstinline{sCut}) is restricted to goals
with at most one term of sort \lstinline{Formula} (corresponding to
the principal cases). That simplification is not possible in $\mLJ$:
with that restriction, the cases for $\wedge_R$ and $\vee_R$ fail
w.r.t. $\cimp_R$.  Here is the case for $(\vee_R, \cimp_R)$:
 
\reductionFail{\infer[\Cut]{\noheight : \CTX{12} \vdash \CTX{11}, \FORMULA{9} \iimp \FORMULA{10}}{\infer[\vee_R]{\suc \wspace \HEIGHT{1} \wspace : \CTX{12} \vdash(\CTX{11}, \FORMULA{9} \iimp \FORMULA{10}), \FORMULA{6} \vee \FORMULA{7}}{\deduce{\HEIGHT{1} : \CTX{12} \vdash(\CTX{11}, \FORMULA{9} \iimp \FORMULA{10}), \FORMULA{6}, \FORMULA{7}}{}} & \infer[\iimp_R]{\suc \wspace \HEIGHT{8} \wspace : \CTX{12}, \FORMULA{6} \vee \FORMULA{7} \vdash \CTX{11}, \FORMULA{9} \iimp \FORMULA{10}}{\deduce{\HEIGHT{8} : \CTX{12}, \FORMULA{9}, \FORMULA{6} \vee \FORMULA{7} \vdash \FORMULA{10}}{}}}}{\textcolor{red}{\infer[\failTK]{\noheight : \CTX{12} \vdash \CTX{11}, \FORMULA{9} \iimp \FORMULA{10}}{}}}

\noindent
Note that the cut formula $F_6\vee F_7$ is principal on the left
premise, but it is not on the right premise. This case cannot be solved
by reducing the height of the cut by first applying $\iimp_R$ since
this would remove the context $\Delta_{11}$ (needed in the left premise). 

The cut-elimination procedure for $\mLJ$ is based on the module
defined in the file \lstinline{cut-add-mcon.maude} (additive
multiple-conclusion) where \lstinline{sCut} is added in cases where the
goal sequent has at most two terms of sort \lstinline{Formula}.  Hence,
\lstinline{sCut} is allowed also in non-principal cases.  This solves the
previous case and the search procedure finds the following proof
transformation: \\

\resizebox{0.95\textwidth}{!}{$
{\to\quad
{\infer[\sCut]{\noheight : \CTX{12} \vdash \CTX{11}, \FORMULA{9} \iimp \FORMULA{10}}{\infer[\sCut]{\noheight : \CTX{12} \vdash \CTX{11}, \FORMULA{6}, \FORMULA{9} \iimp \FORMULA{10}}{\infer[\AssumptionW]{\noheight : \CTX{12} \vdash \CTX{11}, \FORMULA{6}, \FORMULA{7}, \FORMULA{9} \iimp \FORMULA{10}}{} & \infer[\iimp_R]{\noheight : \CTX{12}, \FORMULA{7} \vdash \CTX{11}, \FORMULA{6}, \FORMULA{9} \iimp \FORMULA{10}}{\infer[\AssumptionInv]{\noheight : \CTX{12}, \FORMULA{7}, \FORMULA{9} \vdash \FORMULA{10}}{}}} & \infer[\iimp_R]{\noheight : \CTX{12}, \FORMULA{6} \vdash \CTX{11}, \FORMULA{9} \iimp \FORMULA{10}}{\infer[\AssumptionInv]{\noheight : \CTX{12}, \FORMULA{6}, \FORMULA{9} \vdash \FORMULA{10}}{}}}}}
$}

\begin{theorem}[Cut-elimination and ID-expansion]\label{th:cut-mlj}
  The following cut rule
\[
\infer[Cut]{\Gamma \vdash \Delta}{
\deduce{\Gamma \vdash \Delta, A}{}
&
\deduce{\Gamma, A \vdash \Delta}{}
}
\]
is admissible in $\mLJ$. Moreover, for any $F$, the sequent $F \vdash
F$ is provable.
\end{theorem}

\begin{proof}
  For cut-elimination (\lstinline{prop-cut.maude}), $\mathsf{H}$ as
  well as the invertibility of the rules $\wedge_L$, $\wedge_R$,
  $\vee_L$, $\vee_R$, and $\iimp_L$ are needed. For identity-expansion
  (\lstinline{prop-ID.maude}), the admissibility of $\mathsf{W}_R$ and
  $\mathsf{W}_L$ (Theorem \ref{th:weak-mlj}) is needed.  The proof
  takes 35.2 seconds.  \qed
\end{proof}

\subsection{System $\Gtcp$ for Propositional Classical Logic }\label{case:gtcp}

$\Gtcp$~\cite{troelstra96bpt} is a well known two-sided sequent system
for classical logic, where the structural rules are implicit and all
the rules are invertible. The rules are similar to those of $\mLJ$
with the exception of $\iimp_R$:
\[
\infer[\iimp_R]{\Gamma \vdash \Delta, F\iimp G}{\Gamma, F \vdash \Delta, G}
\]

\paragraph{Weakening.}
Admissibility of weakening of $\Gtcp$ follows that same lines as in
$\mLJ$.

\begin{theorem}[Weakening and Weak-height]\label{th:weak-g3cp}
  If $\deducn{\Gtcp}{n}{\Gamma \vdash \Delta}$, then
  $\deducn{\Gtcp}{s(n)}{\Gamma \vdash \Delta}$ ($\mathsf{H}$).  The
  rules $\mathsf{W}_L$ and $\mathsf{W}_R$ (see Theorem
  \ref{th:weak-mlj}) are height-preserving admissible in $\Gtcp$.
\end{theorem}

\begin{proof}
  The three properties are specified in
  \lstinline{g3c/prop-WH.maude}. No auxiliary lemmas are needed.
  \qed
\end{proof}

\paragraph{Invertibility.}
All the rules in $\Gtcp$ are invertible.
 
\begin{theorem}[Invertibility]
  All the rules in $\Gtcp$ are height-preserving invertible.
\end{theorem}

\begin{proof}
  See the specification in \lstinline{g3c/prop-inv.maude}.
  $\mathsf{H}$ (Theorem \ref{th:weak-g3cp}) is needed. This proof
  takes 15.1 seconds.  \qed
\end{proof}

\paragraph{Contraction.}
An attempt of proving admissibility of contraction on the left side of the sequent  fails
due to $\cimp_L$ (and on the right due to $\cimp_R$).  Here the
failing case for $\mathsf{C}_L$:

\reductionFail{\infer[\iimp_L]{\suc \wspace \HEIGHT{1} \wspace : \CTX{5}, \FORMULA{2} \iimp \FORMULA{3}, \FORMULA{2} \iimp \FORMULA{3} \vdash \CTX{4}}{\deduce{\HEIGHT{1} : \CTX{5}, \FORMULA{2} \iimp \FORMULA{3} \vdash \CTX{4}, \FORMULA{2}}{} & \deduce{\HEIGHT{1} : \CTX{5}, \FORMULA{3}, \FORMULA{2} \iimp \FORMULA{3} \vdash \CTX{4}}{}}}{\textcolor{red}{\infer[\failTK]{\suc \wspace \HEIGHT{1} \wspace : \CTX{5}, \FORMULA{2} \iimp \FORMULA{3} \vdash \CTX{4}}{}}}

\noindent
Due to invertibility of $\cimp_L$, the sequent $\Delta_5 \vdash
\Delta_4, F_2, F_2$ is provable. However, induction does not apply on
this sequent since $F_2$ is on the right side of the sequent.
 
Hence, the proof of admissibility of contraction is by mutual
induction on $\mathsf{C}_L$ and $\mathsf{C}_R$. For instance, the case
of $\cimp_L$ is solved by applying $\mathsf{C}_R$ on a shorter
derivation:

\[
\reduction{\infer[\iimp_L]{\suc \wspace \HEIGHT{1} \wspace : \CTX{5}, \FORMULA{2} \iimp \FORMULA{3}, \FORMULA{2} \iimp \FORMULA{3} \vdash \CTX{4}}{\deduce{\HEIGHT{1} : \CTX{5}, \FORMULA{2} \iimp \FORMULA{3} \vdash \CTX{4}, \FORMULA{2}}{} & \deduce{\HEIGHT{1} : \CTX{5}, \FORMULA{3}, \FORMULA{2} \iimp \FORMULA{3} \vdash \CTX{4}}{}}}
{\infer[\iimp_L]{\suc \wspace \HEIGHT{1} \wspace : \CTX{5}, \FORMULA{2} \iimp \FORMULA{3} \vdash \CTX{4}}{\infer[\IHMutual]{\HEIGHT{1} : \CTX{5} \vdash \CTX{4}, \FORMULA{2}}{\infer[\AssumptionInv]{\HEIGHT{1} : \CTX{5} \vdash \CTX{4}, \FORMULA{2}, \FORMULA{2}}{}} & \infer[\IH]{\HEIGHT{1} : \CTX{5}, \FORMULA{3} \vdash \CTX{4}}{\infer[\AssumptionInv]{\HEIGHT{1} : \CTX{5}, \FORMULA{3}, \FORMULA{3} \vdash \CTX{4}}{}}}}\]

\[
\reduction{\infer[\iimp_L]{\suc \wspace \HEIGHT{1} \wspace :(\CTX{5}, \FORMULA{2} \iimp \FORMULA{3}), \FORMULA{6}, \FORMULA{6} \vdash \CTX{4}}{\deduce{\HEIGHT{1} : \CTX{5}, \FORMULA{6}, \FORMULA{6} \vdash \CTX{4}, \FORMULA{2}}{} & \deduce{\HEIGHT{1} : \CTX{5}, \FORMULA{3}, \FORMULA{6}, \FORMULA{6} \vdash \CTX{4}}{}}}
{\infer[\iimp_L]{\suc \wspace \HEIGHT{1} \wspace : \CTX{5}, \FORMULA{6}, \FORMULA{2} \iimp \FORMULA{3} \vdash \CTX{4}}{\infer[\IH]{\HEIGHT{1} : \CTX{5}, \FORMULA{6} \vdash \CTX{4}, \FORMULA{2}}{\infer[\Assumption]{\HEIGHT{1} : \CTX{5}, \FORMULA{6}, \FORMULA{6} \vdash \CTX{4}, \FORMULA{2}}{}} & \infer[\IH]{\HEIGHT{1} : \CTX{5}, \FORMULA{3}, \FORMULA{6} \vdash \CTX{4}}{\infer[\Assumption]{\HEIGHT{1} : \CTX{5}, \FORMULA{3}, \FORMULA{6}, \FORMULA{6} \vdash \CTX{4}}{}}}}\]
 In the second proof transformation,    contraction is applied on a formula different from the implication and mutual induction is not needed.

\begin{theorem}[Contraction]
  The rules $\mathsf{C}_L$ and $\mathsf{C}_R$ are height-preserving
  admissible in $\Gtcp$.
\end{theorem}

\begin{proof}
  The specification is in \lstinline{prop-C.maude}. The proof of
  admissibility of $\mathsf{C}_L$ (resp., $\mathsf{C}_R$) requires the
  invertibility of $\top_L$, $\vee_L$, $\wedge_L$, and $\iimp_L$
  (resp., $\bot_R$, $\vee_R$, $\wedge_R$, and $\iimp_R$). For
  $\mathsf{C}_L$, the specification includes the following definition
  for mutual induction:
\begin{lstlisting}

eq mutual-ind('suc[GT]) = 
  rl '_:_[GT, '_|--_['Gm:MSFormula, '_;_['F:Formula, 'Dt:MSFormula]]] =>
     '_:_[GT, '_|--_['Gm:MSFormula, '_;_['_;_['F:Formula,'F:Formula],
      		     'Dt:MSFormula]]]   [ label('\IHMutual) ]. ) .
eq mutual-ind (GT) = none [owise] .
\end{lstlisting}    

\noindent
This means that $\mathsf{C}_R$ can be applied on shorter derivations of height
\lstinline{GT} (due to the pattern ``\lstinline{'suc[GT]}'' in the
definition of the equation).  Similarly, the proof of $\mathsf{C}_R$
requires mutual induction on $\mathsf{C}_L$ (specifically for the case
$\iimp_R$).  \qed
\end{proof}

\paragraph{Cut-elimination and Identity Expansion.}
The proofs of cut-elimination and identity expansion are also similar
to the cases of $\mLJ$.

\begin{theorem}[Cut-elimination and ID-expansion]
  The cut-rule in Theorem \ref{th:cut-mlj} is admissible in
  $\Gtcp$. Moreover, for any $F$, the sequent $F \vdash F$ is
  provable.
\end{theorem}

\begin{proof}
  For cut-elimination (\lstinline{prop-cut.maude}), $\mathsf{H}$ as
  well as the invertibility of the rules $\iimp_L$, $\iimp_R$,
  $\wedge_L$, $\wedge_R$, $\vee_L$, and $\vee_R$ are needed. For
  identity-expansion (\lstinline{prop-ID.maude}), the admissibility of
  $\mathsf{W}_R$ and $\mathsf{W}_L$ is needed.  The proof of
  cut-elimination takes 22.5 seconds and id-expansion less than one
  second.  \qed
\end{proof}

\subsection{Propositional linear logic}\label{ex:ll}

Linear logic ($\LLM$) \cite{girard87tcs} is a resource-conscious
logic. Formulas are consumed when used during proofs, unless they are
marked with the exponential $\quest$ (whose dual is $\bang$), in which
case they can be weakened and contracted.  Besides the exponentials,
propositional $\LLM$ connectives include the additive conjunction
$\with$ and disjunction $\oplus$, their multiplicative versions
$\otimes$ and $\lpar$, and the unities $\one,\zero,\top,\bot$. These
connectives form actually pairs of {\em dual} operators:
\[
\begin{array}{lc@{\quad}lc@{\quad}lc@{\quad}lc@{\quad}l}
(A\with B)^\perp \equiv A^\perp\oplus B^\perp & & (A\otimes B)^\perp \equiv A^\perp\lpar B^\perp & & (\bang A)^\perp \equiv \quest A^\perp & &
\zero^\perp \equiv \top & &\one^\perp \equiv \bot
\end{array}
\]
where $A^\perp$ denotes the negation of the formula $A$. All negations
in $\LLM$ can be pushed inwards and restricted to the atomic scope.

\begin{figure}[t]
\resizebox{\textwidth}{!}{
$ 
\begin{array}{c}
\infer[I]{\vdash p^{\bot}, p}{}
    \qquad
    \infer[1]{\vdash 1}{}
    \qquad
    \infer[\bot]{\vdash \Gamma, \bot}{\vdash \Gamma}
    \qquad
        \infer[\top]{\vdash \Gamma, \top}{}
    \\
    \infer[\otimes]{\vdash \Gamma_1, \Gamma_2, A \otimes B}{
      \vdash \Gamma_1, A & \vdash \Gamma_2, B}
    \qquad
    \infer[\lpar]{\vdash \Gamma, A \lpar B}{
      \vdash \Gamma, A, B}
    \qquad
        \infer[\with]{\vdash \Gamma, A \with B}{
      \vdash \Gamma, A & \vdash \Gamma, B}
 \qquad
    \infer[\oplus_1]{\vdash \Gamma, A \oplus B}{\vdash \Gamma, A}
    \quad
    \infer[\oplus_2]{\vdash \Gamma, A \oplus B}{\vdash \Gamma, B}

\end{array}
$
}
  \caption{One-sided multiplicative-additive linear logic ($\MALL$).}
  \label{fig:MALL}
\end{figure}

First, consider the fragment of $\LLM$ without the exponentials, that is, 
(classical) propositional multiplicative-additive linear logic
($\MALL$). The one-sided proof system for $\MALL$ is depicted in
Fig.~\ref{fig:MALL}.  As expected, the structural rules for weakening
and contraction are not admissible in this system.

\paragraph{Invertibility.}
The well known invertibility results for $\MALL$ are easily proved in
the \toolname.

\begin{theorem}[Weak-height and Invertibility]\label{th:wH-mall}
  If $\deducn{\MALL}{n}{ \vdash \Gamma}$, then $\deducn{\MALL}{s(n)}{
    \vdash \Gamma}$ ($\mathsf{H}$). Moreover, all the rules but
  $\otimes$ and $\oplus_i, i\in\{1,2\}$, are height-preserving
  invertible.
\end{theorem}

\begin{proof}
  See \lstinline{MALL/prop-H.maude} and
\lstinline{MALL/prop-inv.maude}. 
Due to the splitting of the context, clearly $\otimes$ is not
invertible. Here two (failing) cases showing that the left premise of
this rule is not invertible. The first one corresponds to
invertibility w.r.t. $\top$ and the second one w.r.t.  $\otimes$:

\noindent$
\begin{array}{c}
\reduction{\infer[\top]{\suc \wspace \HEIGHT{3} \wspace : \vdash \top, \CTX{4}, \CTX{5}, \FORMULA{\one} \tensor \FORMULA{2}}{}}{\textcolor{red}{\infer[\failTK]{\suc \wspace \HEIGHT{3} \wspace : \vdash \FORMULA{\one}, \CTX{4}}{}}}
\qquad
\reduction{\infer[\tensor]{\suc \wspace \HEIGHT{3} \wspace : \vdash(\CTX{6}, \CTX{7}, \FORMULA{\one} \tensor \FORMULA{2}),(\CTX{8}, \CTX{9}), \FORMULA{4} \tensor \FORMULA{5}}{\deduce{\HEIGHT{3} : \vdash \FORMULA{4}, \CTX{6}, \CTX{7}, \FORMULA{\one} \tensor \FORMULA{2}}{} & \deduce{\HEIGHT{3} : \vdash \FORMULA{5}, \CTX{8}, \CTX{9}}{}}}{\textcolor{red}{\infer[\failTK]{\suc \wspace \HEIGHT{3} \wspace : \vdash \FORMULA{\one}, \CTX{6}, \CTX{8}, \FORMULA{4} \tensor \FORMULA{5}}{}}}
\end{array}
$
\qed
\end{proof}

\paragraph{Cut-elimination and Identity Expansion.} 
Since the system under consideration is one-sided, the cut-rule has a
one-sided presentation. The theory in \lstinline{cut-lin-osided.maude}
(linear cut, one sided) specifies the operation
\lstinline{op dual :  Formula -> Formula .}
and the OL must define equations for it reflecting the De Morgan
dualities of the connectives. Here some examples for the $\LLM$
connectives:
 \begin{lstlisting}

    eq dual(p(i)) = perp(i) . eq dual(perp(i)) = p(i) . eq dual(1) = bot . 
    eq dual(F ** G) = dual(F) $ dual(G) . eq dual(F $ G) = dual(F) ** dual(G) .
 \end{lstlisting}

\begin{theorem}[Cut-elim. and identity-expansion]\label{th-cut-mall}
The rule 
$
\infer[Cut]{\vdash \Gamma, \Delta}{ 
 \deduce{\vdash \Gamma, F}{}
 &
 \deduce{\vdash \Delta, F^\perp}{}
}
$
is admissible in $\MALL$.  Moreover, for any formula $F$, the sequent
$\vdash F, F^\perp$ is provable.
\end{theorem}

\begin{proof}
  The notation $F^\perp$ corresponds to \lstinline{dual(F)}. The proof
  of cut-elimination (specification in
  \lstinline{MALL/prop-cut.maude}) relies only on the admissibility of
  $\mathsf{H}$.  Here the principal case $(\tensor, \lpar)$:

\cutEnv{\cutReduction{\infer[\Cut]{\noheight : \vdash(\CTX{4}, \CTX{5}), \CTX{9}}{\infer[\tensor]{\suc \wspace \HEIGHT{1} \wspace : \vdash \FORMULA{6} \tensor \FORMULA{7}, \CTX{4}, \CTX{5}}{\deduce{\HEIGHT{1} : \vdash \FORMULA{6}, \CTX{4}}{} & \deduce{\HEIGHT{1} : \vdash \FORMULA{7}, \CTX{5}}{}} & \infer[\lpar]{\suc \wspace \HEIGHT{8} \wspace : \vdash dual(\FORMULA{6} \tensor \FORMULA{7}), \CTX{9}}{\deduce{\HEIGHT{8} : \vdash \CTX{9}, dual(\FORMULA{6}), dual(\FORMULA{7})}{}}}}
{\infer[\sCut]{\noheight : \vdash \CTX{4}, \CTX{5}, \CTX{9}}{\infer[\Assumption]{\noheight : \vdash \CTX{4}, \FORMULA{6}}{} & \infer[\sCut]{\noheight : \vdash \CTX{5}, \CTX{9}, dual(\FORMULA{6})}{\infer[\Assumption]{\noheight : \vdash \CTX{5}, \FORMULA{7}}{} & \infer[\Assumption]{\noheight : \vdash \CTX{9}, dual(\FORMULA{6}), dual(\FORMULA{7})}{}}}}}
\qed
\end{proof}

\begin{figure}
$
\begin{array}{c}
        \infer[\bang]{\vdash \quest A_1,\ldots,\quest A_n, \bang A}
                 {\vdash \quest A_1,\ldots,\quest A_n, A}
\qquad
    \infer[\quest]{\vdash \Gamma, \quest A}{\vdash \Gamma, A}
   \qquad
    \infer[?_W]{\vdash \Gamma, \quest A}{\vdash \Gamma}
    \quad
    \infer[?_C]{\vdash \Gamma, \quest A}{\vdash \Gamma, \quest A, \quest A}

\end{array}
$
  \caption{Exponential rules for $\LLM$.}
  \label{fig:ll}
\end{figure}

\paragraph{Exponentials.}
Consider the one-sided system for linear logic obtained by adding the
exponential $?$ and its dual $!$. The system $\LLM$ results from the
inclusion of the inference in Fig. \ref{fig:ll} to those in
Fig. \ref{fig:MALL}.  Note the explicit rules for weakening and
contraction on formulas marked with $?$.

The specification of the rule $!$ (called \emph{promotion}) requires a
new sort to guarantee that the context only contains formulas marked
with $?$ (see \lstinline{LL.maude}):
\begin{lstlisting}

sorts ?Formula ?MSFormula . --- Formulas and multisets of formulas marked with ?
op ?_ : Formula -> ?Formula .    --- a ?-Formula is built with ?
subsort ?Formula < ?MSFormula .  --- A single ?-Formula is also a ?-multiset
subsort ?MSFormula < MSFormula . --- Multisets of ?-formulas are also multisets
--- In Rule !,  the context must contain only ?-Formulas
var CLq : ?MSFormula .   rl [bang]  : |-- CLq ; ! F => |-- CLq ; F . 


\end{lstlisting}

\begin{theorem}[Weak-height, Weak-! and inv.]\label{th-adm-wb}
  If $\deducn{\LLM}{n}{ \vdash \Gamma}$, then $\deducn{\LLM}{s(n)}{
    \vdash \Gamma}$ ($\mathsf{H}$).  The rule
  $\infer[\mathsf{W}!]{\vdash \Gamma, F}{\vdash \Gamma, !F} $ is
  height-preserving admissible.  Moreover, none of the rules in
  Fig. \ref{fig:ll} is height-preserving invertible.
\end{theorem}

\begin{proof}
  See \lstinline{LL/prop-H.maude}, \lstinline{LL/prop-WB.maude} and
  \lstinline{LL/prop-inv.maude}.  The invertibility results and the
  admissibility of $\mathsf{W}!$ depend on $\mathsf{H}$. Note that the
  rule $?_C$ is invertible: it is always possible to apply it to later
  use $?_W$ on the same contracted formula. However, such procedure
  does not preserve the height of the derivation. \qed
\end{proof}

The cut elimination procedure for this system is certainly more
involved. An attempt of proving this result fails for $!$ and $?_C$:

\reductionFail{\infer[\Cut]{\noheight : \vdash \CTXQ{3}, \CTX{6}}{\infer[!]{\suc \wspace \HEIGHT{1} \wspace : \vdash ! \FORMULA{4}, \CTXQ{3}}{\deduce{\HEIGHT{1} : \vdash \FORMULA{4}, \CTXQ{3}}{}} & \infer[?_C]{\suc \wspace \HEIGHT{5} \wspace : \vdash dual(! \FORMULA{4}), \CTX{6}}{\deduce{\HEIGHT{5} : \vdash \CTX{6}, ? dual(\FORMULA{4}), ? dual(\FORMULA{4})}{}}}}{\textcolor{red}{\infer[\failTK]{\noheight : \vdash \CTXQ{3}, \CTX{6}}{}}}

This is the principal case where the cut formula is $!F$ and it is
promoted, and its dual contracted. This case is solved by using the
rule below that cuts $!F$ with $n$ copies of the formula $?F^\perp$:

\begin{equation}\label{eq-cut-cc}
\infer[mCut]{\vdash \Gamma, \Delta}{
 \deduce{\vdash \Gamma, !F}{}
 &
 \deduce{\vdash \Delta, (?F^\perp)^n}{}
 }
\end{equation}

Hence, the cut-elimination procedure for this system must mutually
eliminate the cut-rules in Theorem \ref{th-cut-mall} and Equation
\ref{eq-cut-cc}.  More precisely, the elimination of one of the cases
of $Cut$ relies on the application of $mCut$ on shorter derivations
and one of the cases in the elimination of $mCut$ requires the
application of $Cut$ on smaller formulas. Here are the two relevant
cases:

\qquad \cutEnv{\cutReduction{\infer[\Cut]{\noheight : \vdash \CTXQ{3}, \CTX{6}}{\infer[!]{\suc \wspace \HEIGHT{1} \wspace : \vdash ! \FORMULA{4}, \CTXQ{3}}{\deduce{\HEIGHT{1} : \vdash \FORMULA{4}, \CTXQ{3}}{}} & \infer[?_C]{\suc \wspace \HEIGHT{5} \wspace : \vdash dual(! \FORMULA{4}), \CTX{6}}{\deduce{\HEIGHT{5} : \vdash \CTX{6}, ? dual(\FORMULA{4}), ? dual(\FORMULA{4})}{}}}}
{\infer[\mCut]{\noheight : \vdash \CTXQ{3}, \CTX{6}}{\infer[\Assumption]{\suc \wspace \HEIGHT{1} \wspace : \vdash \CTXQ{3}, ! \FORMULA{4}}{} & \infer[\Assumption]{\HEIGHT{5} : \vdash \CTX{6}, ? dual(\FORMULA{4}), ? dual(\FORMULA{4})}{}}}}\\

\resizebox{.95\textwidth}{!}{
${\cutReduction{\infer[\mCut]{\noheight : \vdash \CTXQ{3}, \CTX{7}}{\infer[!]{\suc \wspace \HEIGHT{1} \wspace : \vdash ! \FORMULA{4}, \CTXQ{3}}{\deduce{\HEIGHT{1} : \vdash \FORMULA{4}, \CTXQ{3}}{}} & \infer[?]{\suc \wspace \HEIGHT{6} \wspace : \vdash ctr(s \NAT{5}, ? dual(\FORMULA{4})), \CTX{7}}{\deduce{\HEIGHT{6} : \vdash \CTX{7}, dual(\FORMULA{4}), ctr(\NAT{5}, ? dual(\FORMULA{4}))}{}}}}
{\infer[\Cut]{\noheight : \vdash \CTXQ{3}, \CTX{7}}{\infer[\Assumption]{\noheight : \vdash \CTXQ{3}, \FORMULA{4}}{} & \infer[\hCut]{\noheight : \vdash \CTXQ{3}, \CTX{7}, dual(\FORMULA{4})}{\infer[\Assumption]{\suc \wspace \HEIGHT{1} \wspace : \vdash \CTXQ{3}, ! \FORMULA{4}}{} & \infer[\Assumption]{\HEIGHT{6} : \vdash \CTX{7}, dual(\FORMULA{4}), ctr(\NAT{5}, ? dual(\FORMULA{4}))}{}}}}}
$
}

In the last derivation, \lstinline|ctr(s n5, ? dual(F4))| denotes
$(?F_4^\perp)^{s(n_5)}$ and the application of $Cut$ is on a smaller
formula ($F_4$ instead of $!F_4$).

\begin{theorem}[Cut-elim. and id-exp.]
\label{cut-ll-all}
The following rules are admissible in $\LLM$:

$
\infer[Cut]{\vdash \Gamma}{ 
 \deduce{\vdash \Gamma, F}{}
 &
 \deduce{\vdash \Gamma, F^\perp}{}
}
\qquad
\infer[mCut]{\vdash \Gamma, \Delta}{
 \deduce{\vdash \Gamma, !F}{}
 &
 \deduce{\vdash \Delta, (?F^\perp)^n}{}
 }
$. 

\noindent Moreover, for all formulas $F$, the sequent $\vdash F, F^\perp$ is provable. 
\end{theorem}

\begin{proof}
  The procedures using mutual-induction are defined in
  \lstinline{LL/cut-ll.maude} and \lstinline{LL/cut-ll-cc.maude} (for
  $mCut$). The properties are specified in \lstinline{prop-cut.maude}
  and \lstinline{prop-cut-cc.maude}.  Discharging the cases for $Cut$
  (resp., $mCut$) takes 13.2 seconds (resp. 115.6 seconds).  In
  \lstinline{cut-ll.maude}, the definition
\begin{lstlisting}

op mutual-induct : GroundTerm GroundTerm GroundTerm -> RuleSet .
eq mutual-induct('suc[gh], 'suc[gh'], '!_[GTA]) = ... --- specification of mCut
eq mutual-induct('suc[gh], 'suc[gh'], '?_[GTA]) = ... --- specification of mCut
eq mutual-induct(gh, gh',  GTA) = none [owise] .
\end{lstlisting}
receives as parameter the height of the two premises of the cut and
the cut-formula $F$. If $F$ is not a formula marked with $!$ or $?$,
no additional rule is generated (the \lstinline|[owise]| case in the
definition). Also, the patter matching on the heights of the
derivation allows for controlling the application of the cut-rule on
shorter derivations. A similar definition can be found in
\lstinline{cut-ll-cc.maude}.

The elimination of $Cut$ assumes as auxiliary lemmas $\mathsf{H}$ and
also a generalization of $?_W$ on terms of sort
\lstinline|?MSFormula|: $ \infer[?_{GW}]{\vdash \Gamma,
  ?\Delta}{\vdash \Gamma} $.  Note that this rule is not
height-preserving admissible, but it is clearly admissible (and hence,
it can be used only on sequents marked with \lstinline{'inf.INat}).
Currently, the \toolname\ does not support inductive proofs on the
size of lists/multisets as needed for this auxiliary lemma. This rule
is used in the following case (where the height of the derivation is
irrelevant):
\begin{center}
\cutEnv{\cutReduction{\infer[\Cut]{\noheight : \vdash \CTXQ{3}, \CTX{6}}{\infer[!]{\suc \wspace \HEIGHT{1} \wspace : \vdash ! \FORMULA{4}, \CTXQ{3}}{\deduce{\HEIGHT{1} : \vdash \FORMULA{4}, \CTXQ{3}}{}} & \infer[?W]{\suc \wspace \HEIGHT{5} \wspace : \vdash dual(! \FORMULA{4}), \CTX{6}}{\deduce{\HEIGHT{5} : \vdash \CTX{6}}{}}}}
{\infer[?_GW]{\noheight : \vdash \CTXQ{3}, \CTX{6}}{\infer[\Assumption]{\noheight : \vdash \CTX{6}}{}}}}
\end{center}

The elimination of $mCut$ assumes $\mathsf{H}$ and the admissibility
of $\mathsf{W}!$ (Th. \ref{th-adm-wb}), e.g.,

$
\begin{array}{c}
\reduction{\infer[\Cut]{\noheight : \vdash(\CTXQ{4}, ? \FORMULA{3}), \CTX{8}}{\infer[?]{\suc \wspace \HEIGHT{2} \wspace : \vdash ! \FORMULA{5}, \CTXQ{4}, ? \FORMULA{3}}{\deduce{\HEIGHT{2} : \vdash \FORMULA{3}, \CTXQ{4}, ! \FORMULA{5}}{}} & \infer[?]{\suc \wspace \HEIGHT{7} \wspace : \vdash ctr(s \NAT{6}, ? dual(\FORMULA{5})), \CTX{8}}{\deduce{\HEIGHT{7} : \vdash \CTX{8}, dual(\FORMULA{5}), ctr(\NAT{6}, ? dual(\FORMULA{5}))}{}}}}{}
\\\\
\tiny{\infer[\mCut]{\noheight : \vdash \CTXQ{4}, \CTX{8}, ? \FORMULA{3}}{\infer[W!]{\noheight : \vdash \CTXQ{4}, \FORMULA{5}, ? \FORMULA{3}}{\infer[\Assumption]{\noheight : \vdash \CTXQ{4}, ! \FORMULA{5}, ? \FORMULA{3}}{}} & \infer[\hCut]{\noheight : \vdash \CTXQ{4}, \CTX{8}, ? \FORMULA{3}, dual(\FORMULA{5})}{\infer[\Assumption]{\suc \wspace \HEIGHT{2} \wspace : \vdash \CTXQ{4}, ! \FORMULA{5}, ? \FORMULA{3}}{} & \infer[\Assumption]{\HEIGHT{7} : \vdash \CTX{8}, dual(\FORMULA{5}), ctr(\NAT{6}, ? dual(\FORMULA{5}))}{}}}}
\end{array}
$

Also, a generalization of $?_C$ on terms of sort
\lstinline|?MSFormula| is used here:

$
\begin{array}{c}
\reduction{\infer[\Cut]{\noheight : \vdash(\CTXQ{4}, ? \FORMULA{3}), \CTX{10}, \CTX{11}, \FORMULA{8} \tensor \FORMULA{9}}{\infer[?]{\suc \wspace \HEIGHT{2} \wspace : \vdash ! \FORMULA{5}, \CTXQ{4}, ? \FORMULA{3}}{\deduce{\HEIGHT{2} : \vdash \FORMULA{3}, \CTXQ{4}, ! \FORMULA{5}}{}} & \infer[\tensor]{\suc \wspace \HEIGHT{7} \wspace : \vdash ctr(s \NAT{6}, ? dual(\FORMULA{5})), \CTX{10}, \CTX{11}, \FORMULA{8} \tensor \FORMULA{9}}{\deduce{\HEIGHT{7} : \vdash \CTX{10}, \FORMULA{8}, ctr(\NAT{6}, ? dual(\FORMULA{5}))}{} & \deduce{\HEIGHT{7} : \vdash \CTX{11}, \FORMULA{9}, ? dual(\FORMULA{5})}{}}}}{}
\\\\
\tiny{\infer[?_{GC}]{\noheight : \vdash \CTXQ{4}, \CTX{10}, \CTX{11}, ? \FORMULA{3}, \FORMULA{8} \tensor \FORMULA{9}}{\infer[\tensor]{\noheight : \vdash \CTXQ{4}, \CTXQ{4}, \CTX{10}, \CTX{11}, ? \FORMULA{3}, ? \FORMULA{3}, \FORMULA{8} \tensor \FORMULA{9}}{\infer[\hCut]{\noheight : \vdash \CTXQ{4}, \CTX{10}, \FORMULA{8}, ? \FORMULA{3}}{\infer[\Assumption]{\suc \wspace \HEIGHT{2} \wspace : \vdash \CTXQ{4}, ! \FORMULA{5}, ? \FORMULA{3}}{} & \infer[\Assumption]{\HEIGHT{7} : \vdash \CTX{10}, \FORMULA{8}, ctr(\NAT{6}, ? dual(\FORMULA{5}))}{}} & \infer[\hCut]{\noheight : \vdash \CTXQ{4}, \CTX{11}, \FORMULA{9}, ? \FORMULA{3}}{\infer[\Assumption]{\suc \wspace \HEIGHT{2} \wspace : \vdash \CTXQ{4}, ! \FORMULA{5}, ? \FORMULA{3}}{} & \infer[\Assumption]{\HEIGHT{7} : \vdash \CTX{11}, \FORMULA{9}, ? dual(\FORMULA{5})}{}}}}}
\end{array}
$

As explained in Sec. \ref{ex:g3i}, arbitrary applications of
contraction are problematic for proof search. Hence, in
\lstinline{prop-cut-cc.maude}, $?_{GC}$ is introduced as a conditional
rule that can be applied on a given multiset only if there are exactly
one occurrence of it in the current sequent. As in the case of
$?_{GW}$, the admissibility of $?_{GC}$ cannot be proved in \toolname.
Discharging the cases of $Cut$ (resp., $mCut$) takes 30s (resp. 124s).
\qed
\end{proof}

\begin{figure}[t]
\[\small
  \infer[\quest]{\vdash\Gamma : \Delta, \quest F}{\vdash\Gamma, F : \Delta}
  \quad
  \infer[!]{\vdash \Gamma : !F}{\vdash \Gamma : F}
  \quad
  \infer[?_C]{\vdash \Gamma, F : \Delta}{\vdash \Gamma, F : \Delta, F}
  \quad
  \infer[\otimes]{\vdash \Gamma : \Delta_1, \Delta_2, A \tensor B}
        {\vdash \Gamma : \Delta_1, A \quad \vdash \Gamma : \Delta_2, B}
\]
 \caption{Some rules of the dyadic system $\LLD$ (\cite{andreoli92jlc}).}
  \label{fig:Dll}

\end{figure}

\paragraph{Dyadic System for Linear Logic.} 
Since formulas of the form $\quest F$ can be contracted and weakened,
such formulas can be treated as in classical logic. This rationale is
reflected into the syntax of the so called \emph{dyadic sequents} of
the form $\vdash \Gamma : \Delta$, interpreted as the linear logic
sequent $\vdash \quest \Gamma, \Delta$ where $\quest \Gamma = \{\quest
A \mid A \in \Gamma\}$.  It is then possible to define a proof system
without explicit weakening and contraction (system $\LLD$ in
Fig.~\ref{fig:Dll}). The complete dyadic proof system for linear logic
can be found in~\cite{andreoli92jlc}.

\begin{theorem}[Weak-height, Weak-! and invertibility]\label{th-adm-wb-dyll}
  If $\deducn{\LLD}{n}{ \vdash \Gamma : \Delta}$, then
  $\deducn{\LLD}{s(n)}{ \vdash \Gamma:\Delta}$ ($\mathsf{H}$).  The
  following three rules are height-preserving admissible:

$\small 
\infer[\mathsf{C}]{\Gamma,F~:~\Delta}{\Gamma, F, F~:~\Delta}
\qquad
\infer[\mathsf{W}]{\Gamma,F~:~\Delta}{\Gamma~:~\Delta}
\qquad
\infer[\mathsf{W}!]{\vdash \Gamma ~:~  \Delta, F}{\vdash \Gamma ~:~  \Delta, !F}
$.
Moreover, all the rules of the system but $\oplus_i$, $\tensor$ and
$?_C$ are height-preserving invertible.
\end{theorem}

\begin{proof}
  For the admissibility results, see \lstinline{DyLL/prop-H.maude},
  \lstinline{prop-C.maude} \lstinline{prop-W.maude} and
  \lstinline{prop-WB.maude}. The invertibility results are specified
  in \lstinline{DyLL/prop-inv.maude}.  $\mathsf{H}$ and the
  admissibility of $\mathsf{C}$ and $\mathsf{W}$ do not require any
  additional lemma.  The admissibility of $\mathsf{W}!$ depends on
  $\mathsf{H}$.  The invertibility results rely on $\mathsf{H}$,
  $\mathsf{W}$, and $\mathsf{W}!$.\qed
\end{proof}

As in the system $\LLM$, the cut-elimination theorem for $\LLD$
requires mutual induction on two different rules. The rule $Cut!$
below internalizes the storage of formulas marked with $?$ into the
classical context.

\begin{theorem}[Cut-elim. and id-exp.]
  The following rules are admissible in $\LLM$:

$\small
\infer[Cut]{\vdash \Gamma~:~\Delta_1, \Delta_2}{ 
 \deduce{\vdash \Gamma~:~ \Delta_1, F}{}
 &
 \deduce{\vdash \Gamma~:~ \Delta_2,  F^\perp}{}
}
\qquad
\infer[Cut!]{\vdash \Gamma~:~ \Delta}{
 \deduce{\vdash \Gamma ~:~ !F}{}
 &
 \deduce{\vdash \Gamma, F^\perp~:~ \Delta}{}
 }
$.

Moreover, for any formula $F$, the sequent $\vdash \cdot ~:~ F,
F^\perp$ is provable.
\end{theorem}

\begin{proof}
  The procedures are defined in \lstinline{DyLL/cut-dyadic.maude} and
  \lstinline{DyLL/cut-dyadic-cc} (for $Cut!$). The properties are
  specified in \lstinline{prop-cut.maude} and
  \lstinline{prop-cut-cc.maude}.  Discharging the cases for $Cut$
  (resp., $Cut!$) takes 32.4 seconds (resp., 4 seconds).  The
  admissibility of $Cut$ (resp., $Cut!$) relies on $\mathsf{H}$ and
  $\mathsf{W}$ (resp., $\mathsf{H}$, $\mathsf{W}$ and $\mathsf{W}!$).
  \qed
\end{proof}

\paragraph{Intuitionistic Linear Logic.}
The directory \lstinline{ILL} contains the specification and analyses
for intuitionistic linear logic ($\ILL$). In this system, the
multiplicative disjunction $\lpar$ is not present and the linear
implication $\limp$ needs to be added (in classical $\LLM$, $F\limp G$ is a
shorthand for $F^\perp \lpar G$).  The resulting system is two-sided
and single-conclusion. Formulas marked with $!$ on the left of the
sequent can be weakened and contracted.  The proof of cut-elimination
follows the same principles that the one for $\LLM$ and the multicut
rule below is required where
$!\Gamma$ is a multiset of formulas marked with $!$ (specified
as the sort \lstinline{!MSFormula})

\[
\infer[mCut]{!\Gamma, \Delta\vdash G}{
 \deduce{!\Gamma \vdash  !F}{}
 &
 \deduce{\Delta, (!F)^n \vdash G}{}
 }
\] 

\subsection{Normal modal logics:  $\kRule$ and $\Sfour$}\label{case:modal}
Modal logics extend classical logic with the addition of modal
connectives (e.g., the unary modal connective $\Box$), used to qualify
the truth of a judgment, widening the scope of logical-conceptual
analysis.  The alethic interpretation of $\Box A$ is ``the formula $A$
is necessarily true''.  A modal logic is {\em normal} if it contains
the axiom
\[\begin{array}{ll}
(\mathsf{k}) & \Box(A\iimp B)\iimp\Box A\iimp\Box B
\end{array}
\]
and it is closed under {\em generalization}: if $A$ is a theorem then
so it is $\Box A$.

The smallest normal modal logic is called $\kRule$, and $\Sfour$
extends $\kRule$ by assuming the axioms $(\mathsf{T}) \Box A\iimp A$
and $(\mathsf{4}) \Box A\iimp \Box\Box A$.  The sequent systems
considered here for the modal logics $\kSystem$ and $\Sfour$ are
extensions of $\Gtcp$ with the additional rules for the modalities
depicted in Fig.~\ref{fig:modal}.

\paragraph{Structural rules.}
The admissibility of the structural rules follows as in Section
\ref{case:gtcp}.

\begin{figure}[t]
\[\small
    \infer[\mathsf{k}]{\Gamma', \Box\Gamma\vdash \Box A,\Delta}{\Gamma\vdash A} \qquad
    \infer[\mathsf{T}]{\Gamma,\Box A\vdash \Delta}{\Gamma, \Box A,A\vdash \Delta}\qquad
    \infer[\mathsf{4}]{\Gamma', \Box\Gamma\vdash \Box A,\Delta}{\Box\Gamma\vdash A}
\]
  \caption{The modal sequent rules for $\kRule$ ($\mathsf{k}$) and  $\Sfour$
  ($\mathsf{T}+\mathsf{4}$)}
  \label{fig:modal}
\end{figure}

\begin{theorem}[Structural rules]\label{th:weak-K}
  If $\deducn{\kSystem}{n}{\Gamma \vdash \Delta}$, then
  $\deducn{\kSystem}{s(n)}{\Gamma \vdash \Delta}$
  ($\mathsf{H}$). Similarly for $\Sfour$.  Moreover, the rules
  $\mathsf{W}_L$, $\mathsf{W}_R$, $\mathsf{C}_L$, and $\mathsf{C}_R$
  are height-preserving admissible in both $\kSystem$ and $\Sfour$.
\end{theorem}

\paragraph{Invertibility.}
As in $\Gtcp$, all the rules for the propositional connectives are
invertible. Furthermore, $\mathsf{T}$ is invertible, but $\mathsf{k}$
and $\mathsf{4}$ are not (due to the implicit weakening).

\begin{theorem}[Invertibility]
  Only the rule $\mathsf{T}$ in Fig. \ref{fig:modal} is invertible.
\end{theorem}

\begin{proof}
  See the specification in \lstinline{K/prop-inv.maude} and
  \lstinline{S4/prop-inv.maude}.  $\mathsf{H}$ is required in this
  proof.  Below, one of the trivial cases showing that the rule
  $\mathsf{k}$ is not invertible:
  \reductionFail{\infer[\top_R]{\suc \wspace \HEIGHT{2} \wspace :
      \MSBF{4}, \CTX{5} \vdash \top, \CTX{3}, \Box
      \FORMULA{1}}{}}{\textcolor{red}{\infer[\failTK]{\suc \wspace
        \HEIGHT{2} \wspace : unbox(\MSBF{4}) \vdash \FORMULA{1}}{}}}
  Multisets of boxed formulas belong to the sort
  \lstinline{MSBFormula} and the operation

\noindent \lstinline{op unbox : MSBFormula -> MSFormula .} 
removes the boxes. 
\qed
\end{proof}

\paragraph{Cut-elimination and Identity Expansion.} 
The cut-elimination 
procedure for  $\kSystem$ and $\Sfour$
uses the same infrastructure developed for 
$\Gtcp$. 

\begin{theorem}[Cut-elimination and ID-expansion]\label{th:cut-s4}
  The cut-rule in Theorem \ref{th:cut-mlj} is admissible in both
  $\kSystem$ and $\Sfour$. Moreover, for any $F$, the sequent $F
  \vdash F$ is provable in $\kSystem$ and $\Sfour$.
\end{theorem}

\begin{proof}
  The specifications are in \lstinline{K/prop-cut.maude} and
  \lstinline{S4/prop-cut.maude}.  In both cases, $\mathsf{H}$ is
  required as well as the invertibility of the propositional
  rules. \qed
\end{proof}

\paragraph{Modal Logic $\Sfive$. }
Some extensions of the modal logic $\kRule$ do not have (known)
cut-free sequent systems. In particular, consider the system $\Sfive$,
obtained by extending $\kRule$ with the axiom $\mathsf{T}$ and the
rule below:
\[\small
\infer[\mathsf{45}]{\Gamma', \Box\Gamma \vdash \Delta', \Box A, \Box \Delta}{
\Box\Gamma \vdash  A, \Box \Delta
}
\]
(see \lstinline{KT45/KT45.maude}).  Using the same strategy as in
Theorem \ref{th:cut-s4}, the tool is able to discard some of the proof
obligations for cut-elimination. However, some subcases involving
$\mathsf{45}$ and the other two modal rules ($\mathsf{T}$ and
$\mathsf{k}$) fail.  Here, one example: 

\reductionFail{\infer[\Cut]{\noheight : \Box\Delta_{13}, \CTX{14} \vdash \Box\Delta_{11}, \CTX{12},\Box \FORMULA{8}}{\infer[A45]{\suc \wspace \HEIGHT{1} \wspace : \Box\Delta_{13}, \CTX{14} \vdash(\Box\Delta_{11}, \CTX{12},\Box \FORMULA{8}),\Box \FORMULA{10}}{\deduce{\HEIGHT{1} : \Box\Delta_{13} \vdash \Box\Delta_{11}, \FORMULA{8},\Box \FORMULA{10}}{}} & \infer[AT]{\suc \wspace \HEIGHT{9} \wspace :(\Box\Delta_{13}, \CTX{14}),\Box \FORMULA{10} \vdash \Box\Delta_{11}, \CTX{12},\Box \FORMULA{8}}{\deduce{\HEIGHT{9} : \Box\Delta_{13}, \FORMULA{10}, \CTX{14},\Box \FORMULA{10} \vdash \Box\Delta_{11}, \CTX{12},\Box \FORMULA{8}}{}}}}{\textcolor{red}{\infer[\failTK]{\noheight : \Box\Delta_{13}, \CTX{14} \vdash \Box\Delta_{11}, \CTX{12},\Box \FORMULA{8}}{}}}

This case is clearly not provable: the cut-formula $\Box F_{10}$ is
not decomposed in the left premise. Hence, cutting with $F_{10}$ will
not finish the proof.  The only alternative is to reduce the height of
the cut. The rule $\mathsf{T}$ cannot be applied (on $F_{10}$) on the
last sequent. If $\mathsf{45}$ is applied on the last sequent, either
$F_8$ or one of the formulas in the boxed context $\Delta_{11}$ will
lose the box. In both cases, none of the leaves of the left derivation
can be used.

The failing cases are interesting, because they spot the point where
the cut-elimination fails: if the goal is to propose a cut-free
sequent system for a logic, certain shapes of rules should be avoided.


\section{Concluding Remarks}\label{sec.concl}

Checking structural properties of proof systems demands trustful
methods. Usually, proving such properties is done via case-by-case
analysis, where all the possible combinations of rule applications in
a system are exhausted.  The advent of automated reasoning has changed
completely the landscape, since theorems can nowadays be proved
automatically in meta-logical frameworks (see e.g.~\cite{DBLP:conf/lpar/DawsonG10}).  This approach has brought a
fresh perspective to the field of proof theory: useless proof search
steps, usually singular to a specific logic, have been disregarded in
favor of developing universal methods for providing general automation
strategies. These developments have ultimately helped in abstracting
out conceptual characteristics of logical systems, as well as in
identifying effective frameworks that can capture (and help in
reasoning about) them in a natural way.

This work moves forward towards such a direction: it proposes a
general, natural, and uniform way of proving key structural properties
of sequent systems by using the rewriting logic logical and
meta-logical framework~\cite{basin-reflmetlogframe-2004}. It ultimately
enables modular and incremental proofs of meta-level properties of
propositional sequent systems, both specified and mechanized in the
language of Maude~\cite{clavel-maudebook-2007}.  The approach builds
on top of core algorithms that are combined and used for proving
admissibility and invertibility of rules, as well as cut-elimination
and identity expansion of sequent systems.

The choice of rewriting logic as a meta-logical framework brings key
advantages. Indeed, as detailed below, while approaches using logical
frameworks depend heavily on the specification method and/or the
implicit properties of the meta and object logics, rewriting logic
enables the specification of the rules as they are actually written in
text and figures
\cite{marti-oliet-rlframework-2002,basin-reflmetlogframe-2004}. Moreover,
notions as derivability and admissibility of rules can be neatly
formulated in the context of rewriting systems
\cite{DBLP:journals/corr/abs-0911-1412}.

Consider, for instance, the LF framework~\cite{pfenning00ic} based on
intuitionistic logic, where the left context is handled by the
framework as a set. Specifying sequent systems based on multisets
requires elaborated mechanisms in most logical frameworks, which makes
the encoding of the system of interest far from being natural and
straightforward.  Moving from intuitionistic to linear logic solves
this problem~\cite{cervesato02ic,DBLP:journals/tcs/MillerP13}, but
still several sequent systems cannot be naturally specified in
frameworks based on $\LLM$, as it is the case of $\mLJ$. This latter
situation can be partially fixed by adding subexponentials
\cite{danos93kgc} to linear logic
($\SELL$)~\cite{nigam14jlc,DBLP:conf/cade/NigamRL14}. However, the
resulting encodings are often non-trivial and difficult to automate.
Moreover, several logical systems cannot be naturally specified in
$\SELL$, such as the modal sequent system $\kRule$
\cite{OLARTE2020143}. 

A completely different approach is presented in~\cite{Zach20}, where cut-elimination is obtained by considering the relation between the cut-rule in Gentzen's system $\LK$ and the resolution rule for (propositional) classical logic. But, again, this method is restricted to systems
having a (classical) semantics as a starting point.

All in all, this paper presents yet more evidence of the fact that
rewriting logic is an innovative and elegant framework for both
specifying and reasoning in and about logical systems, with the
further benefit of easy modular extension. In fact, a strong
conjecture is that mild adjustments to the proposed approach and
algorithms are needed to reason about other systems, including normal
(multi-)modal~\cite{DBLP:conf/lpar/LellmannP15} and
paraconsistent~\cite{DBLP:journals/lu/LahavMZ17} sequent systems.  It
seems also possible to have an extension to handle variants of sequent
systems themselves, like nested~\cite{Brunnler:2009kx} or linear
nested~\cite{DBLP:conf/tableaux/Lellmann15} systems. This ultimately
widens the scope of logics that can be analyzed in the \toolname, such
as normal and non-normal modal
logics~\cite{DBLP:journals/tocl/LellmannP19}. It is also an
interesting future research path to consider first-order sequent
systems. Previous work on mechanizing first-order sequent systems in
rewriting logic, including proof-search heuristics at the meta-level,
has been presented before in~\cite{rocha-modulorelmics-2008}.

A usual concern when a new sequent system is proposed is to implement
it.  Few implementational efforts have provided tools for emerging
sequent systems for logics such as epistemic, lax, deontic, knotted,
linear-modal, etc., logics.  It would not require much effort to reuse
some of the algorithms presented here to implement a procedure that,
given the inference rules of a sequent system, outputs an
implementation of such a system in Maude.  Also, implementing new
induction principles on lists/multisets may help in obtaining new
automatic proofs (see Theorem~\ref{cut-ll-all}). More interestingly,
using invertibility lemmas would also enable the generation of a
weak-focus~\cite{andreoli92jlc} version of the original systems, thus
eliminating part of the non-determinism during proof-search.  It
should be noted, however, that this depends on a deeper investigation
of the role of invertible rules as equational rules in rewriting
logic. While this idea sounds more than reasonable, it is necessary to
check whether promoting invertible rules to equations preserves
completeness of the system (e.g., the resulting equational theory
needs to be Church-Rosser and terminating modulo the structural axioms
of the operators). If the answer to this question is in the
affirmative for a large class of systems, then the approach presented
here also opens the possibility to the automatic generation of focused
systems.

Finally, a word about cut-elimination.  The usual Gentzen's cut-elimination
proof strategy can be summarized by the following steps: (i)
transforming a proof with cuts into a proof with principal cuts; (ii)
transforming a proof with principal cuts into a proof with atomic
cuts; and (iii) transforming a proof with atomic cuts into a cut-free
proof.  While step (ii) is not difficult to solve (see,
e.g.,~\cite{DBLP:journals/tcs/MillerP13}), steps (i) and (iii) can be
problematic to mechanize. The results presented in this work suggest
that such techniques can be fully automated to a certain degree, as
showcased with various proof systems in Section~\ref{sec.case}.

\bibliographystyle{abbrv}
\bibliography{main}

\end{document}